\documentclass[12pt]{article}

\usepackage{cite}                      
\usepackage[dvips]{color}                 
\usepackage{graphicx}                     
\usepackage{amssymb}
\usepackage{amsmath}

\usepackage{xspace}                       

\newcommand{\mx   }[3]{\langle\,{#1} \mid {#2} \mid {#3}\,\rangle}
 
\newcommand{\mxemp}[2]{\langle\,{#1} \mid {#2}          \,\rangle}

\newcommand{\ket}[1] {\mid     {#1}\,\rangle}

\newcommand{\Mfi}[1]{{\mathcal{M}}_{fi}^{#1}}
\newcommand{\eps}  {{\hat{\epsilon}}_{{\lambda}_i}}
\newcommand{\epspr}{{\hat{\epsilon}}_{{\lambda}_f}^{\ast}}
\newcommand{\ki}   {{\vec{k}}_i}
\newcommand{\kf}   {{\vec{k}}_f}

\newcommand{\vsh}[3]{ {\vec{T}}_{ {#1}\,{#2}\,{#3} }(\hat{r})}
\newcommand{\li}{{\lambda}_i}
\newcommand{\lf}{{\lambda}_f}

\newcommand{\wignerd}[3]{d^{#1}_{{#2},{#3}}(\theta)}
\newcommand{\green}  {G_{\hat{C}}(r,r';E_0) }

\newcommand{\calO}{ \mathcal{O} }
\newcommand{\ofxi}{(\vec{\xi}\,)}

\newcommand{\mpi}{m_\pi}

\newcommand{\MeV}{\mathrm{MeV}}
\newcommand{\fm}{\mathrm{fm}}

\newcommand{\Hint}{H^\mathrm{int}}
\newcommand{\w}{\omega}
\newcommand{\PC}{\vec{P}_{C}}
\newcommand{\PCsq}{\!\left.\vec{P}_C\!\!\!\right.^2}

\newcommand{\EC}{E_{C}}
\newcommand{\md}{m_{d}}
\newcommand{\mC}{m_{C}}
\newcommand{\denoms}{ \omega+\frac{\w^2}{2m_d}-B-E_C}
\newcommand{\denomu}{-\omega+\frac{\w^2}{2m_d}-\frac{\PCsq}{2m_C}-B-E_C}
\newcommand{\e}{\mathrm{e}}
\newcommand{\nab}{\vec{\nabla}}
\newcommand{\phiihat}{\hat{\phi}_i}
\newcommand{\phifhat}{\hat{\phi}_f}

\newcommand{\Aone}{\vec{A}^{(1)}}
\newcommand{\Atwo}{\vec{A}^{(2)}}
\newcommand{\Jsigma}{\vec{J}^{(\sigma)}}
\newcommand{\Jp}{\vec{J}^{(p)}}

\newcommand{\doubleint}{\int\!\!\!\int}

\newcommand{\HIGS}{HI$\gamma$S\xspace}

\newcommand{\mytitle}[1]{\begin{center}\LARGE{\textbf{#1}}\end{center}}
\newcommand{\myauthor}[1]{\textbf{#1}}
\newcommand{\myaddress}[1]{\textit{#1}}
\newcommand{\mypreprint}[1]{\begin{flushright}#1\end{flushright}}

\begin{document}
%

\begin{titlepage}
  
  \mypreprint{
        \hfill
    nucl-th/0512063; 
    TUM-T39-05-18;
    FAU-TP3-05/9\\
Revised version 3rd August 2010.}

  \vspace*{0.5cm}
  
  \mytitle{Nucleon Polarizabilities from Deuteron Compton Scattering 
           within  a Green's-Function Hybrid Approach}

  \vspace*{0.5cm}

  \begin{center}
    \myauthor{Robert P. Hildebrandt$^{a}$}, \myauthor{Harald W.\
      Grie{\ss}hammer$^{a,b,c,}$}\footnote{
      Corresponding author; email: hgrie@gwu.edu; permanent address: c}\\
    and \myauthor{Thomas R.~Hemmert$^{a}$}

  
    \myaddress{$^a$
      Institut f\"ur Theoretische Physik (T39), Physik-Department,\\
      Technische Universit{\"a}t M{\"u}nchen, D-85747 Garching, Germany}
    \\[2ex]
    \myaddress{$^b$ Institut f\"ur Theoretische Physik III,
      Naturwissenschaftliche Fakult\"at I, Universit\"at~Erlangen-N\"urnberg,
      D-91058 Erlangen, Germany}
    \\[2ex]
    \myaddress{$^c$ Center for Nuclear Studies, Department of Physics,
      The~George~Washington~University,~Washington DC 20052, USA}
  \end{center}


\begin{abstract}
  We examine elastic Compton scattering from the deuteron for photon energies
  ranging from zero to 100~MeV, using state-of-the-art deuteron wave functions
  and $NN$-potentials.  Nucleon-nucleon rescattering between emission and
  absorption of the two photons is treated by Green's functions in order to
  ensure gauge invariance and the correct Thomson limit.  With this
  Green's-function hybrid approach, we fulfill the low-energy theorem of
  deuteron Compton scattering and there is no significant dependence on the
  deuteron wave function used.  Concerning the nucleon structure, we use
  Chiral Effective Field Theory with explicit $\Delta(1232)$ degrees of
  freedom within the Small Scale Expansion up to leading-one-loop order.
  Agreement with available data is good at all energies.  Our 2-parameter fit
  to all elastic $\gamma d$ data leads to values for the static isoscalar
  dipole polarizabilities which are in excellent agreement with the isoscalar
  Baldin sum rule. Taking this value as additional input, we find $\alpha_E^s=
  (11.3\pm0.7\,(\mathrm{stat})\pm0.6\,(\mathrm{Baldin})
  \pm1\,(\mathrm{theory}) )\cdot10^{-4}\;\fm^3$ and $\beta_M^s = (
  3.2\mp0.7\,(\mathrm{stat})\pm0.6\,(\mathrm{Baldin}) \pm1\,(\mathrm{theory})
  )\cdot10^{-4}\;\fm^3$ and conclude by comparison to the proton numbers that
  neutron and proton polarizabilities are 
  the same within rather small errors.
\end{abstract}
\vskip 1.0cm
\noindent
\begin{tabular}{rl}
Suggested PACS numbers:& 13.40.-f, 13.60.Fz, 25.20.-x, 21.45.+v
\\[1ex]
Suggested Keywords: &\begin{minipage}[t]{11cm}
                    Effective Field Theory, Compton Scattering,\\
                    Deuteron, Delta Resonance, Nucleon Polarizabilities. 
                    \end{minipage}
\end{tabular}

\vskip 1.0cm

\end{titlepage}

\setcounter{page}{2} \setcounter{footnote}{0} \newpage

%

\section{Introduction}
\setcounter{equation}{0}
\label{sec:introduction}

Analyzing protons and neutrons with electromagnetic probes 
has a long history in the field of nucleon-structure studies.
In Compton scattering off a single nucleon, 
the electromagnetic field of the scattered photon
attempts to deform the nucleon. The global resistance of the
nucleon's internal degrees of freedom against this deformation is measured 
in terms of the electromagnetic polarizabilities, which makes 
them an excellent tool to study the structure of the nucleon. 
In principle, each polarizability is a function of the frequency of the 
electromagnetic wave. Therefore, energy-dependent 
polarizabilities have been introduced in Refs.~\cite{GH01,HGHP}. 
In this work, we 
determine the \textit{static} values of the nucleon polarizabilities 
from experiment, i.e. the values in the limit of vanishing photon 
energy, which we therefore denote as \textit{the} polarizabilities for 
simplicity. 
The two most prominent nucleon polarizabilities are the static 
electric and magnetic dipole polarizabilities $\alpha_E$ and $\beta_M$.
For the proton, rather precise experimental values for these two 
quantities exist, e.g. 
those given in a recent review~\cite{Schumacher}, which were obtained 
as the weighted average over several experiments,
\begin{align}
  \alpha_{E}^p=(12.0\pm0.6)\cdot 10^{-4}\;\mathrm{fm}^3,\qquad \beta_{M}^p =(
  1.9\mp0.6)\cdot 10^{-4}\;\mathrm{fm}^3.
  \label{eq:reviewp}
\end{align}
Our recent fit to the proton Compton data yielded~\cite{HGHP}
\begin{align}
  \alpha_E^p&=(11.04\pm1.36(\mathrm{stat})\pm0.4(\mathrm{Baldin}))
  \cdot 10^{-4}\;\mathrm{fm}^3,\qquad\nonumber\\
  \beta_M^p &=( 2.76\mp1.36(\mathrm{stat})\pm0.4(\mathrm{Baldin})) \cdot
  10^{-4}\;\mathrm{fm}^3,
  \label{eq:exppHGHP}
\end{align}
in agreement with Eq.~(\ref{eq:reviewp}), and with the central value of the
Baldin sum rule $\alpha_E^p+\beta_M^p=(13.8\pm0.4)\cdot
10^{-4}\;\mathrm{fm}^3$~\cite{Olmos}.  The framework that has been chosen for
this extraction is third-order ($\mathcal{O}(\epsilon^3)$) Small Scale
Expansion (SSE), which is one possible way to include the $\Delta(1232)$
resonance explicitly in Chiral Effective Field Theory, cf.
Ref.~\cite{HHKLett1,HHKLett2}. It is this very framework on which the
one-nucleon sector of our present work is built.

On the other hand, neutron polarizabilities are much harder to access
experimentally, as there is no stable single-neutron target for Compton
scattering. Therefore, one has to rely on other methods in order to
investigate these quantities. One approach is quasi-free Compton scattering
off the neutron bound in the deuteron, or the scattering of neutrons from a
lead target. The weighted average over several experiments investigating these
two processes gives the result~\cite{Schumacher}
\begin{align}
  \alpha_{E}^n=(12.5\pm1.7)\cdot 10^{-4}\;\mathrm{fm}^3,\qquad \beta_{M}^n =(
  2.7\mp1.8)\cdot 10^{-4}\;\mathrm{fm}^3.
  \label{eq:reviewn}
\end{align}

The numbers given in Eq.~(\ref{eq:reviewn}) do not include information from
elastic low-energy Compton scattering from light nuclei, as these processes so
far led to values inconsistent with Eq.~(\ref{eq:reviewn}).  However, several
such experiments have already been performed~\cite{Lucas,Lund,Hornidge} and
further proposals exist, e.g.  from the proton, deuteron or $^3\!\mathrm{He}$
at
TUNL/HI$\gamma$S~\cite{Weller:2009zza,Miskimen,Miskimentalk,Weller,Gao,Ahmed}
and on deuteron targets at MAXlab~\cite{Feldman:2008zz,Feldman2} as well as at
the S-DALINAC~\cite{Richter}, and for the proton at
MAMI~\cite{AhrensBeckINT08}.  They promise an extensive study of elastic
deuteron Compton scattering below the pion-production threshold.  From a
theorist's point of view, extracting the neutron polarizabilities requires an
accurate description both of the structure of the nucleon \textit{and} of the
dynamics of the low-energy degrees of freedom within the deuteron, as one has
to account for the nucleon polarizabilities as well as for meson-exchange
currents.  We remind the reader that the deuteron as the proton-neutron bound
state is an isoscalar object, so that only
$\alpha_E^s\equiv\frac{1}{2}\,(\alpha_E^p+\alpha_E^n)$, $\beta_
M^s\equiv\frac{1}{2}\,(\beta _M^p+\beta _M^n)$ is measured.  The isovector
polarizabilities are defined as
$\alpha_E^v\equiv\frac{1}{2}\,(\alpha_E^p-\alpha_E^n)$, $\beta_
M^v\equiv\frac{1}{2}\,(\beta _M^p-\beta _M^n)$.  A first attempt to fit the
isoscalar polarizabilities to the elastic deuteron Compton-scattering data
from Illinois~\cite{Lucas} and SAL~\cite{Hornidge} was made in~\cite{Lvov}.
The extracted neutron polarizabilities $\alpha_E^n=( 9.0\pm3.0)\cdot
10^{-4}\;\mathrm{fm}^3$, $\beta_ M^n=(11.0\pm3.0)\cdot 10^{-4}\;\mathrm{fm}^3$
indicated the possibility of a rather \textit{large} isovector part,
especially in the magnetic dipole polarizability, in contrast to the
quasi-elastic result from \cite{Kossert} contained in Eq.~(\ref{eq:reviewn}).
The enhancement of $\beta_ M^n$ in~\cite{Lvov} is in our opinion due to an
insufficient description of the SAL-data around 94~MeV, in particular in the
backward direction, which have been a puzzle for several years.  It was
finally resolved by the authors of~\cite{deuteronpaper} in an
Effective-Field-Theory calculation by observing that dynamical effects from
explicit $\Delta(1232)$ resonance degrees of freedom are large at these
energies.  The calculation presented in~\cite{deuteronpaper} agrees reasonably
with data but only works above some lower energy limit of the order 50-60~MeV.
The ``best'' (Baldin-constrained) fit results for the isoscalar
polarizabilities given in~\cite{deuteronpaper} are in excellent agreement with
Eqs.~(\ref{eq:reviewp}) and~(\ref{eq:reviewn}):
\begin{align}
  \alpha_E^s&=(12.6\pm0.8(\mathrm{stat})\pm0.7(\mathrm{wf})
  \pm0.6(\mathrm{Baldin}))\cdot 10^{-4}\;\mathrm{fm}^3,\qquad\nonumber\\
  \beta_M^s &=( 1.9\mp0.8(\mathrm{stat})\mp0.7(\mathrm{wf})
  \pm0.6(\mathrm{Baldin}))\cdot 10^{-4}\;\mathrm{fm}^3,
  \label{eq:deuteron1fits}
\end{align}
with 'wf' denoting the uncertainty arising from the residual wave-function
dependence.

On the theory side, Heavy Baryon Chiral Perturbation Theory (HB$\chi$PT)
predicts that the proton and neutron polarizabilities are equal at
leading-one-loop order~\cite{BKKM}.  Comparing Eq.~(\ref{eq:reviewp})
to~(\ref{eq:reviewn}) or~(\ref{eq:deuteron1fits}), respectively, suggests that
$\alpha_E$ and $\beta_M$ have indeed only small isovectorial components.  A
thorough discussion and review of the chiral aspects of Compton scattering can
for example be found in Ref.~\cite{danielreview} and references therein.

Let us first sketch where our approach agrees or differs from other
calculations, and which problems we hope to solve by such modifications. In
the present work, we investigate elastic deuteron Compton scattering including
photon couplings to the one-pion exchange.  Diagrams which are characterized
by the propagation of the nucleons between the two photon interactions are
calculated using Green's-function methods in analogy to
Ref.~\cite{Karakowski1,Karakowski2}. The single-nucleon structure is included
within Chiral Effective Field Theory ($\chi$EFT) with explicit $\Delta$
degrees of freedom up to leading-one-loop order like in
Ref.~\cite{deuteronpaper}, where isovectorial components are zero.  The
$\Delta(1232)$ is treated according to the power-counting rules of the Small
Scale Expansion~\cite{HHKLett1,HHKLett2}.  We aim for a consistent calculation
at photon energies $\w$ below $100$~MeV, i.e. we
\begin{itemize}
\item[1)]give an improved description of the high-energy
  ($\w\sim95\;\mathrm{MeV}$) data from~\cite{Hornidge} with respect to the
  calculations presented in \cite{Lvov,Karakowski1,Karakowski2}.
\item[2)]present an alternative approach that does not suffer from the
  shortcomings of the EFT calculations of
  Refs.~\cite{deuteronpaper,Phillips,McGPhil1,McGPhil2}, which are not applicable below
  some lower energy limit. Stated differently, we demand that our calculation
  reaches the correct static limit, i.e. the Thomson limit for deuteron
  Compton scattering.
\item[3)]contribute to the ongoing discussion of the neutron polarizabilities
  via fits of the isoscalar polarizabilities $\alpha_E^s$ and $\beta_M^s$ to
  all existing elastic deuteron Compton data.
\end{itemize}
Effective Field Theory has already been used for the latter purpose
in~\cite{deuteronpaper} and~\cite{McGPhil1,McGPhil2}.  The authors of
\cite{deuteronpaper,McGPhil1,McGPhil2}, as well as those of
Ref.~\cite{Phillips}, followed Weinberg's proposal~\cite{Weinberg1,Weinberg2} to
calculate the irreducible kernel for the $\gamma N N\rightarrow \gamma N N$
process in Effective Field Theory and then folded this with external deuteron
wave functions, derived from high-precision $NN$-potentials such as
Nijm93~\cite{Nijm}, CD-Bonn~\cite{Bonn} or AV18~\cite{AV18}. This ``hybrid''
approach has proven quite successful in describing scattering reactions like
$\pi d$~\cite{Beane}, $e^-d$~\cite{Danieled} and other processes, see e.g.
also~\cite{Rholecture}.  However, the $\calO(p^4)$-HB$\chi$PT calculation of
Ref.~\cite{McGPhil1,McGPhil2} in the Weinberg hybrid approach gives an
insufficient description of the SAL-data~\cite{Hornidge} measured in the
backward direction. Therefore, the authors of~\cite{McGPhil1,McGPhil2}
excluded these (two) data points in some of their fits.  Those of
Ref.~\cite{deuteronpaper} had to restrict themselves to the published data
above 60~MeV.  It is one of the central results of the approach presented in
this work that we do not need any such constraints, in particular in view of
forthcoming data even beyond $\w=100$~MeV~\cite{Feldman:2008zz,Feldman2}.

In this work, we present an extended hybrid approach, called the \emph{Green's
  function hybrid approach}. It includes the full two-nucleon Green's function
in all diagrams with an $NN$-intermediate state, which was only treated
perturbatively in Ref.~\cite{deuteronpaper}, according to the power-counting
rules of third-order SSE for high-energy external probes.  Besides the
single-nucleon current, we couple the photon field also to the meson-exchange
currents of the two-nucleon system.  The calculations
of~\cite{deuteronpaper,Phillips,McGPhil1,McGPhil2} are strictly perturbative
in the interaction kernel and therefore have the disadvantage that they become
inapplicable below $\w\sim50$-60~MeV, see discussion in
Section~\ref{sec:shortnote}.  For example, they would over-predict the
deuteron Thomson limit by more than a factor 2~\cite{PHD}.  This is the more
damaging as the Thomson limit is a simple consequence of gauge invariance, so
that the calculation in a too-simple application of Weinberg's counting rules
obviously violates gauge invariance.  The calculation presented in this
publication does not suffer from that limitation. Instead, we implement
Weinberg's original suggestion that chiral counting can safely only be applied
to the \emph{two-nucleon irreducible} pieces of the amplitudes. The
non-perturbative nature of two-nucleon rescattering must for energies well
below the pion mass be implemented by iteration, no matter whether in initial,
final or intermediate states. Since this approach reaches the exact Thomson
limit, we are able to describe the low-energy ($\w<60$~MeV) data well.

It is worth noting that a fully self-consistent $\chi$EFT of the two-nucleon
system is not yet available. To take the counting rules of the one-nucleon
system simply over to the potential of the two-nucleon system was Weinberg's
original proposal~\cite{Weinberg1,Weinberg2}. However, Nogga et al.~and
Birse~\cite{Nogga,Birse} showed recently that this approach is not
self-consistent in $NN$-scattering, but that additional short-distance
interactions are needed in higher partial waves to recover phase shifts which
are insensitive to details of short-distance physics. While this problem does
not affect the ${}^3\mathrm{S}_1$ and ${}^3\mathrm{D}_1$ channels, it may be
of importance in the $NN$-intermediate state. However, we use here a so-called
``high-precision'' $NN$ potential, AV18~\cite{AV18}, which reproduces the $NN$
partial waves reasonably well.

The various cross-checks we perform in this presentation make us confident
that the cross sections and polarizabilities we obtain in this paper will be
essentially unchanged once the fully systematic theory of few-nucleon systems
with pion-nucleon interactions is found. Its relation to a more rigorous EFT
approach will be elaborated on in a future publication on power-counting in
two-nucleon $\chi$EFT~\cite{hg1,chidyn2006,menu07}. Weinberg's
suggestion~\cite{Weinberg1,Weinberg2} of a hybrid approach seems still to be applicable
in deuteron Compton scattering.

The dependence of deuteron Compton scattering on the nucleon polarizabilities
has also been explored to next-to-next-to leading order in the EFT variant in
which pions are integrated out~\cite{Rupak1,Rupak2,Rupak3}.  While the Thomson
limit is in this approach implemented automatically, the range of
applicability of this ``pion-less'' EFT is limited to typical momenta well
below the pion mass and thus to typical photon energies
$\omega\lesssim\mpi^2/M\approx20\;\MeV$.  The available deuteron Compton data
lie well above this scale.

Comparing to Refs.~\cite{Lvov,Karakowski1,Karakowski2}, we see the main
advantage of our calculation in an improved treatment of the single-nucleon
Compton multipoles, which in those works are only included via the
leading~\cite{Karakowski1,Karakowski2} and subleading~\cite{Lvov} terms of a
Taylor expansion in $\w$, whereas we keep the full energy dependence of the
Compton multipoles, including the explicit $\Delta(1232)$ following
third-order SSE as worked out in~\cite{HGHP}.  The huge influence of the
$\Delta(1232)$ in single-nucleon Compton backscattering is well-known. It is
due to the strong paramagnetic $M1$ coupling of the photon to the
$N\rightarrow\Delta$ transition.  A similarly strong influence of the $\Delta$
resonance has been found in deuteron Compton scattering in the backward
direction in~\cite{deuteronpaper}. Therefore, we advocate to retain this
degree of freedom explicitly in any Compton calculation.

This paper is organized as follows: In Section~\ref{sec:theory}, we give a
brief survey of the theoretical formalism applied, focusing on those diagrams
which have a two-nucleon intermediate state. We close the section by
demonstrating that our approach is gauge invariant and therefore fulfill the
low-energy theorem~\cite{Friar}, i.e. it generates the correct static limit.
In Section~\ref{sec:results}, we discuss our predictions for four different
photon energies, ranging from 50~MeV to 100~MeV, and we compare those to data
and to the $\mathcal{O}(\epsilon^3)$ SSE calculation of
Ref.~\cite{deuteronpaper}.  The subsequent Section~\ref{sec:fits} contains our
fits of the isoscalar polarizabilities to all existing elastic deuteron
Compton data.  We show that the data are self-consistent and in good agreement
with the theoretical expectation that isovector components are small. We
conclude in Section~\ref{sec:conclusion}, having shifted the most technical
parts to Appendices~\ref{app:multipoleexp} and~\ref{app:resonant} and to
Ref.~\cite{PHD}. In Appendix~\ref{app:photodisintegration}, we derive the
total deuteron-photodisintegration cross section from our Compton amplitude in
order to cross-check our calculation. We find perfect agreement with data and
predictions from Effective Range Theory. We also published summaries of these
findings previously in conference proceedings~\cite{chidyn2006,menu07}.

\section{Theoretical Framework}
\setcounter{equation}{0}
\label{sec:theory}
In this work, elastic deuteron Compton scattering is examined, including
explicit $\Delta$ resonance degrees of freedom.  In general, the $T$-matrix
for Compton scattering off the deuteron is derived as the matrix element of
the interaction kernel, evaluated between initial- and final-state deuteron
wave functions, as explained in great detail in~\cite{Phillips}:
\begin{equation}
  T=\left<\Psi_f|K_{\gamma\gamma}+K_\gamma\,G\,K_\gamma|\Psi_i\right>
  \label{eq:Tgammagamma}
\end{equation}
The first piece in Eq.~(\ref{eq:Tgammagamma}) is called the ``two-nucleon
irreducible'' part and the second is the ``two-nucleon reducible'' part.
Two-nucleon reducible diagrams are those which contain an intermediate state
with only the two nucleons as particle content.  $G$ is the two-particle
Green's function, constructed from the two-nucleon irreducible interaction $V$
and the free two-nucleon Green's function.  $K_{\gamma}$ denotes the coupling
of one photon to the two-nucleon system, $K_{\gamma \gamma}$ is the
two-nucleon irreducible kernel for the coupling of the two photons.

\begin{figure}[!htb]
\begin{center} 
\parbox[m]{.11\linewidth}{
\includegraphics*[width=\linewidth]{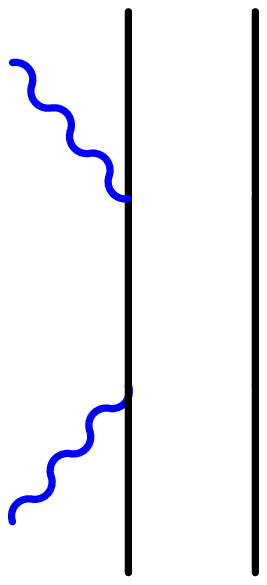}}
\hspace{.5cm}
\parbox[m]{.114\linewidth}{
\includegraphics*[width=\linewidth]{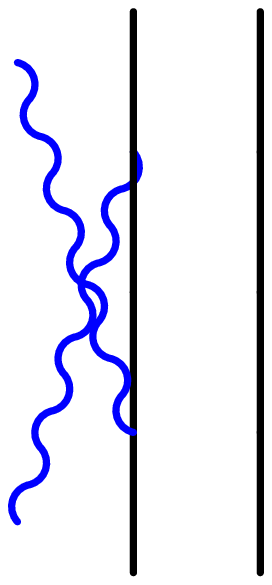}}
\parbox{1.\textwidth}{
\caption{Nucleon-pole terms without rescattering of the nucleons between the 
two photon interactions. The photon-nucleon vertex is given by minimal 
substitution or magnetic-moment interactions.}
\label{fig:pole}}
\end{center}
\end{figure}

The main difference between this work and Ref.~\cite{deuteronpaper} is the
treatment of those diagrams which are characterized by the propagation of the
two nucleons in the intermediate state between the two photon interactions.
In \cite{deuteronpaper,Phillips,McGPhil1,McGPhil2}, such diagrams have been calculated
following the power-counting rules of Effective Field Theory for
``high-energy'' external probes, i.e. $\omega\sim m_\pi$. For large photon
energies, a perturbative treatment is possible as is easily understood
heuristically: The absorption of a high-energy photon immediately tears the
two nucleons apart, so the deuteron would be destroyed if the second photon
was not emitted near-instantaneously.  It turns out that up to
leading-one-loop order in HB$\chi$PT, as well as in SSE, the only diagrams
with a two-nucleon intermediate state are the nucleon-pole diagrams, sketched
in Fig.~\ref{fig:pole}. Note that the $s$-channel diagram, i.e. the left
diagram in Fig.~\ref{fig:pole}, is part of $K_\gamma\,G\,K_\gamma$, as it has
an intermediate state with only the two nucleons propagating. While the
nucleons can in general interact with each other between the two photon
vertices, the authors of Refs.~\cite{deuteronpaper,Phillips} calculated only
up to leading-one-loop order, where no such interactions are possible.
Therefore, we sketch the propagation of two free nucleons in
Fig.~\ref{fig:pole}.

\begin{figure}[!htb]
  \begin{center}
\parbox[m]{.118\linewidth}{
  \includegraphics*[width=\linewidth]{spole.eps}} \hspace{.5cm}
\parbox[m]{.118\linewidth}{
  \includegraphics*[width=\linewidth]{upole.eps}}
$\;\;\;\;\;\longrightarrow\;$
\parbox[m]{.1\linewidth}{
  \includegraphics*[width=\linewidth]{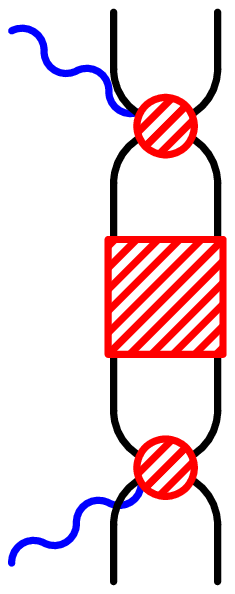}} \hspace{.5cm}
\parbox[m]{.112\linewidth}{
  \includegraphics*[width=\linewidth]{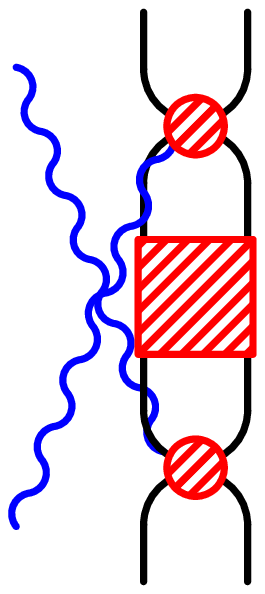}}

\vspace{.5cm}
\includegraphics*[width=.6\linewidth]{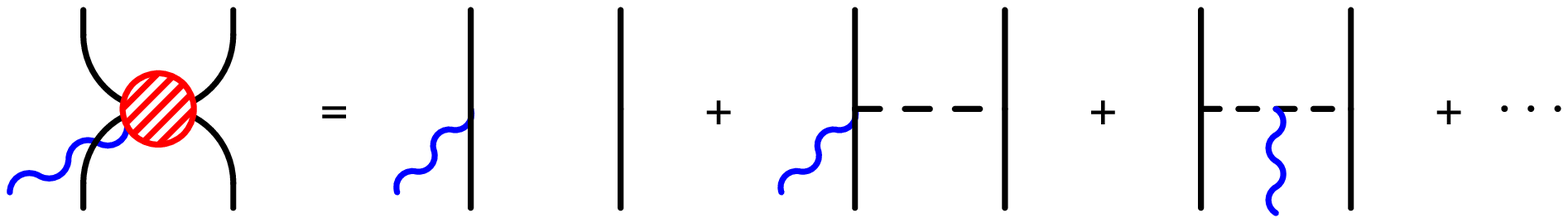}
\parbox{1.\textwidth}{
\caption{Upper line: 
Sketch of the different  
treatment of diagrams with two-nucleon intermediate state in the present work
with respect to Refs.~\cite{deuteronpaper,Phillips}. 
The square symbolizes the $NN$-$S$-matrix. The blobs denote photon coupling 
to the one-body current and possible one-pion exchange, as 
indicated in the lower line.}
\label{fig:disp}}
\end{center}
\end{figure}

As a consequence of the power counting applied, the calculations of 
Refs.~\cite{deuteronpaper,Phillips,McGPhil1,McGPhil2} break down in the low-energy 
regime, manifested in an incorrect Thomson limit. Their lower energy limit 
was found to be of the order of 50-60~MeV.
The reason for the mismatch at low energies is that one has to treat 
the $np$-interaction non-perturbatively.
Therefore, we now include the full $np$-$S$-matrix in the intermediate 
state. The possible rescattering between the two nucleons is denoted by a 
square in Fig.~\ref{fig:disp}, where 
we sketch the differences between Ref.~\cite{deuteronpaper} and this work. 
Foremost, we do not only include the free propagation of the two nucleons in
the nucleon-pole diagrams, Fig.~\ref{fig:pole}, but we construct the full
$np$-Green's function whenever a two-nucleon intermediate state is involved.
Furthermore, we allow for more ways of coupling the photon field to the two
nucleons with respect to Refs.~\cite{deuteronpaper,Phillips}: Besides coupling
to the single-nucleon current, like in Fig.~\ref{fig:pole}, we also include
pion-exchange currents as shown in the lower line of Fig.~\ref{fig:disp},
making use of Siegert's theorem~\cite{Siegert}, cf.
Section~\ref{sec:resonant}.  Such diagrams appear only at $\calO(p^4)$ in
HB$\chi$PT and are therefore not included in~\cite{deuteronpaper}. In
Ref.~\cite{McGPhil1,McGPhil2}, the $np$-rescattering in the intermediate state is not
fully included. It is well-known that only full inclusion of
$np$-rescattering, together with the appropriate explicit pion-exchange
diagrams, generates the Thomson limit of deuteron Compton
scattering~\cite{Lvov,Karakowski1,Karakowski2,Arenhoevel,ArenhoevelII} as direct
consequence of only demanding gauge invariance~\cite{Friar}.

In this publication, we follow closely the work of Karakowski and
Miller~\cite{Karakowski1,Karakowski2} to construct the rescattering \textit{and} the photon
coupling to the meson-exchange currents via Siegert's theorem, cf.
Fig.~\ref{fig:disp}. However, before we turn to the calculation of the
diagrams with an intermediate $np$-state, we recall all other contributions,
which therefore are part of $K_{\gamma\gamma}$, see
Eq.~(\ref{eq:Tgammagamma}). As those terms have already been discussed in
Ref.~\cite{deuteronpaper} and partly in earlier references therein, we shall
be brief in the upcoming section.

\subsection{Diagrams without Intermediate $NN$-Scattering}
\label{sec:nonresonant}
Except for the diagrams with two-nucleon intermediate states, see
Fig.~\ref{fig:disp}, we apply the power-counting rules of the Small Scale
Expansion, where the expansion parameter is $\epsilon$, denoting either a
small momentum, the pion mass or the mass difference $\Delta_0$ between the
real part of the $\Delta$ mass and the nucleon mass.  We refer the interested
reader to~\cite{BKM} for the $N\pi$ Lagrangean and to~\cite{HHKK}
and~\cite{HGHP} for the relevant pieces of the $\Delta\pi$ Lagrangean.  The
power-counting scheme that we use for the nucleon-structure part of our
calculation is motivated by Weinberg's idea to count powers only in the
interaction kernel~\cite{Weinberg1,Weinberg2}.  This hybrid approach is a
well-established tool by now. While the kernel is power counted according to
the rules of the Effective Field Theory, a deuteron wave function from
state-of-the-art $NN$-potentials is used.  Unfortunately, a fully
self-consistent $\chi$EFT of the two-nucleon system is not yet
available~\cite{Nogga,Birse}.  We apply therefore here an extension of the
hybrid approach which uses basic facts of nuclear phenomenology in the
two-nucleon sector such as $np$-rescattering and meson-exchange currents. Its
relation to a more rigorous EFT approach will be given in a forthcoming work
on power counting in two-nucleon $\chi$EFT~\cite{hg1,chidyn2006,menu07}.
Still, the deuteron wave functions that we use are derived from modern
$NN$-potentials: the AV18-potential~\cite{AV18} and the ``NNLO chiral''
potential~\cite{Epelbaum}.  This last potential also follows Weinberg's
suggestion and is derived by applying HB$\chi$PT power counting to the
$NN$-potential $V$ itself. It therefore contains some terms which are strictly
speaking of higher order. Except for the un-resolved power-counting issues
mentioned in the Introduction, one does not expect this to be detrimental for
the accuracy of our calculation. A NLO potential would actually suffice in our
calculation since the photo-nuclear interaction kernel is expanded only up to
NLO as well. We will demonstrate in Sec.~\ref{sec:potentialdep} that the
differences to a LO chiral potential are very small and can be used to
estimate residual theoretical uncertainties.

Considering only contributions without intermediate two-nucleon states, there
is another possibility to classify diagrams.  It is the separation into
one-body and two-body pieces, i.e. into diagrams where only one of the two
nucleons or both of them are involved in the Compton-scattering process.
Obviously, such a rigorous separation is no longer possible when we calculate
the diagrams including $NN$-rescattering between the photon interactions, cf.
Fig.~\ref{fig:disp}.  As in Ref.~\cite{deuteronpaper}, we calculate up to
$\mathcal{O}(\epsilon^3)$, i.e. there is no difference in the diagrams without
an intermediate $np$-state between this work and~\cite{deuteronpaper}.
Therefore, we refer to Ref.~\cite{deuteronpaper} for further details on these
contributions and only list again the various diagrams for completeness:
\begin{itemize}
\item One-body contributions without explicit $\Delta(1232)$ degrees of
  freedom. These are the single-nucleon seagull,
  Fig.~\ref{fig:chiPTsingle}(a), and the contributions from the leading chiral
  dynamics of the pion cloud around the nucleon
  (Figs.~\ref{fig:chiPTsingle}(c)-(f)).  The pion pole
  (Fig.~\ref{fig:chiPTsingle}(b)), i.e. the $\pi^0$-exchange in the
  $t$-channel, does not contribute to deuteron Compton scattering at this
  order, as it is isovectorial.  Note that the nucleon-pole terms,
  Fig.~\ref{fig:pole}, also contribute at
  $\mathcal{O}(\epsilon^3)$~\cite{deuteronpaper}. In this work, however, these
  diagrams are not only included perturbatively, like in
  Ref.~\cite{deuteronpaper}, but we calculate them using the full $np$-Green's
  function $G$, cf. Eq.~(\ref{eq:Tgammagamma}). Their evaluation is postponed
  to Section~\ref{sec:resonant}.
  \begin{figure}[!htb]
    \begin{center}
      \includegraphics*[width=.121\linewidth]{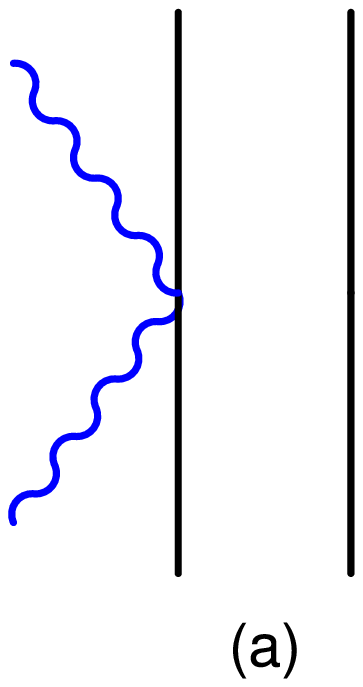}
      \includegraphics*[width=.121\linewidth]{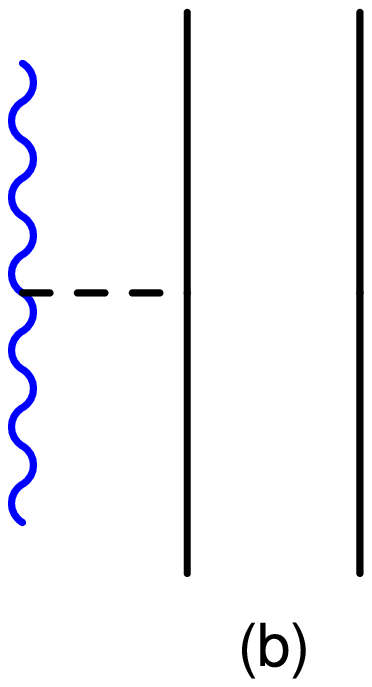}
      \includegraphics*[width=.121\linewidth]{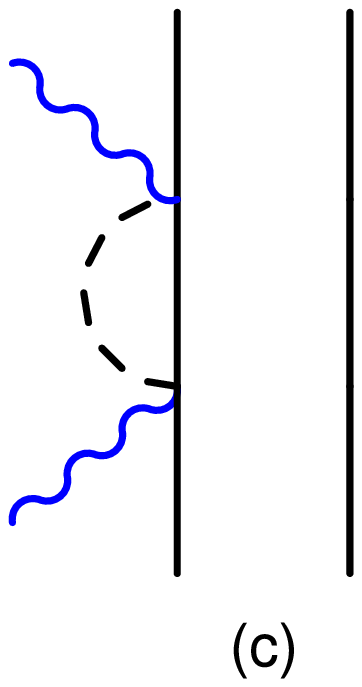}
      \includegraphics*[width=.121\linewidth]{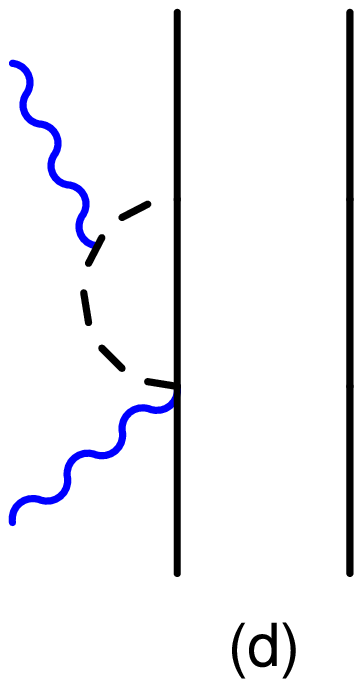}
      \includegraphics*[width=.121\linewidth]{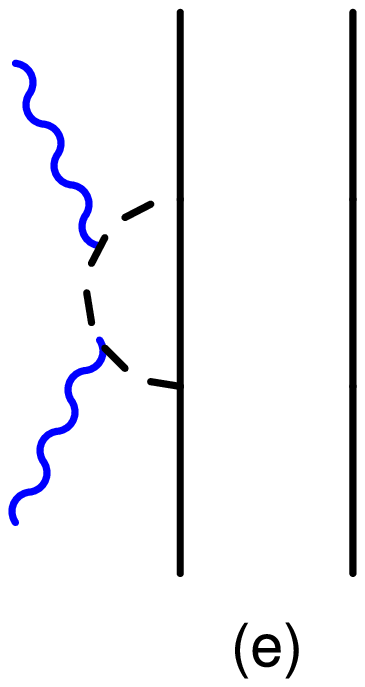}
      \includegraphics*[width=.121\linewidth]{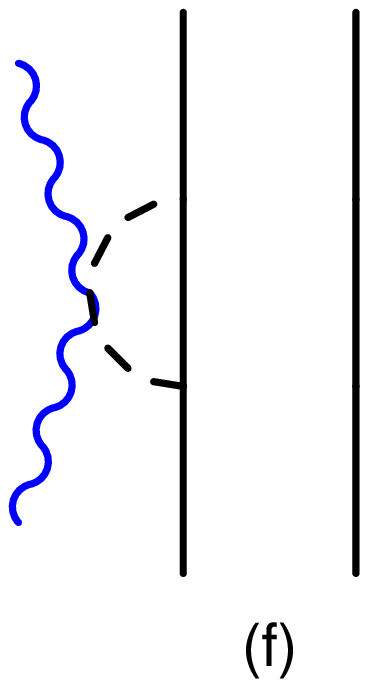}
\parbox{1.\textwidth}{
\caption{One-body interactions without $\Delta(1232)$ propagator 
contributing to deuteron Compton scattering  up to $\mathcal{O}(\epsilon^3)$ 
in SSE. Permutations and crossed graphs are not shown.}
\label{fig:chiPTsingle}}
\end{center}
\end{figure}
\item One-body diagrams with explicit $\Delta$ degrees of freedom, as shown in
  Fig.~\ref{fig:SSEsingle}: The $\Delta$-pole diagrams
  (Fig.~\ref{fig:SSEsingle}(a)) and the contributions from the pion cloud
  around the $\Delta(1232)$ (Figs.~\ref{fig:SSEsingle}(b)-(e)).
\item Two isoscalar short-distance one-body operators
  (Fig.~\ref{fig:SSEsingle}(f)), whose coupling constants are denoted as
  $g_{117}$, $g_{118}$ in Table~\ref{tab:const}. These operators, which we
  determine via fits to either proton or deuteron Compton cross-section data,
  contribute energy-independently to the dipole polarizabilities $\alpha_E^s$
  and $\beta_M^s$.  They are formally of $\mathcal{O}(\epsilon^4)$ but turn
  out to give an anomalously large contribution to the single-nucleon Compton
  amplitude. Therefore, we promote them to next-to-leading order as discussed
  in detail in~\cite{HGHP}. As pointed out in Refs.~\cite{deuteronpaper,HGHP},
  two isoscalar parameters suffice for an accurate description of all Compton
  data on the proton and deuteron. We will confirm later that isovectorial
  short-distance effects seem to be suppressed. We therefore do not augment
  the number of short-distance parameters which are promoted to lower orders
  beyond the necessary minimal set consisting of these two.
  \begin{figure}[!htb]
    \begin{center}
      \includegraphics*[width=.121\linewidth]{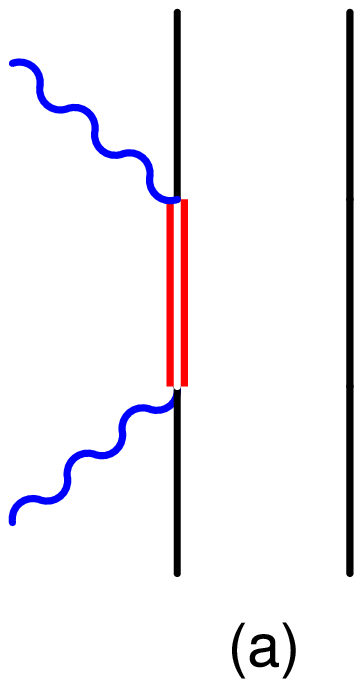}
      \includegraphics*[width=.121\linewidth]{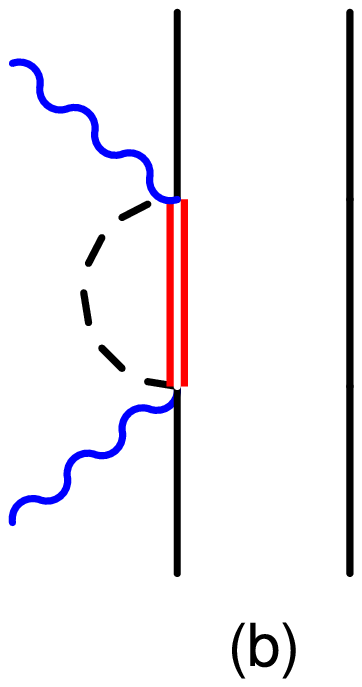}
      \includegraphics*[width=.121\linewidth]{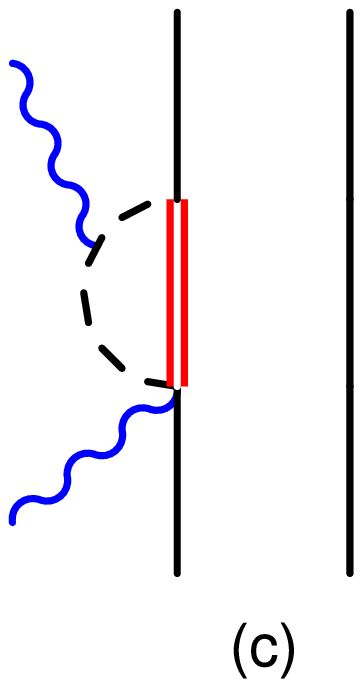}
      \includegraphics*[width=.121\linewidth]{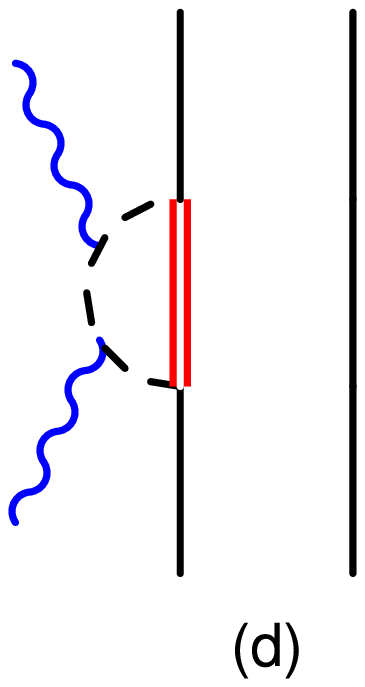}
      \includegraphics*[width=.121\linewidth]{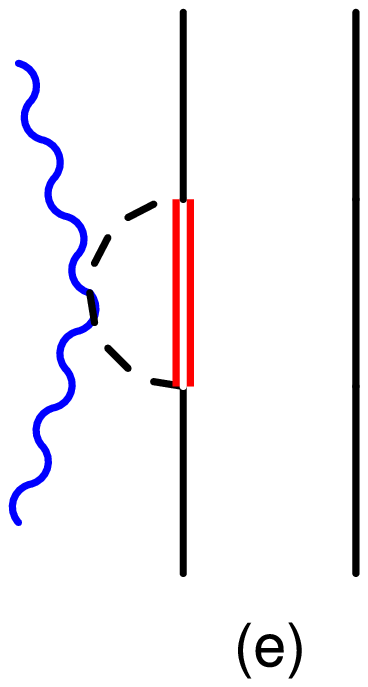}
      \includegraphics*[width=.121\linewidth]{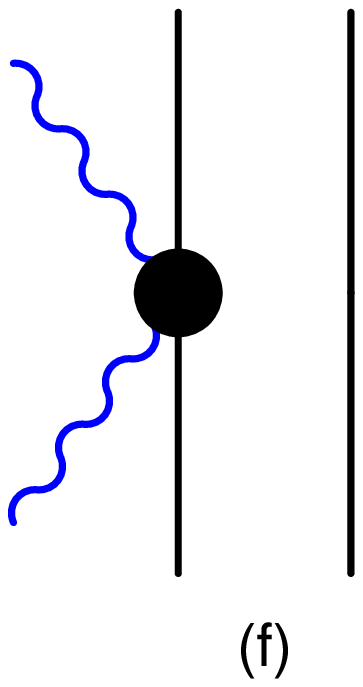}
\parbox{1.\textwidth}{
  \caption{Additional one-body interactions which contribute to deuteron
    Compton scattering at $\mathcal{O}(\epsilon^3)$ in SSE compared to
    third-order HB$\chi$PT.  Permutations and crossed graphs are not shown.}
\label{fig:SSEsingle}}
\end{center}
\end{figure}
\item Two-body contributions with one pion exchanged between the two nucleons
  (Fig.~\ref{fig:chiPTdouble}).  As discussed in Ref.~\cite{deuteronpaper},
  the meson-exchange diagrams are identical in third-order HB$\chi$PT and SSE.
  These diagrams, together with those given in Figs.~\ref{fig:disp}
  and~\ref{fig:chiPTsingle}(a) are responsible for complying with the
  low-energy theorem, as discussed in Section~\ref{sec:Thomson}.
  \begin{figure}[!htb]
    \begin{center}
      \includegraphics*[width=.75\textwidth]{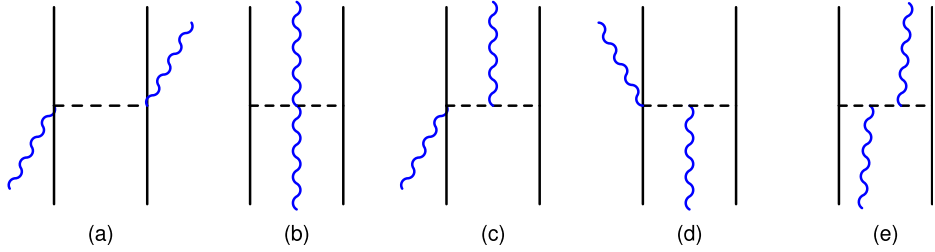}
\parbox{1.\textwidth}{
\caption{Two-body interactions 
contributing to the kernel for deuteron Compton scattering  
at $\mathcal{O}(\epsilon^3)$ in SSE. Diagrams which differ only by 
nucleon interchange are not shown.}
\label{fig:chiPTdouble}}
\end{center}
\end{figure}
\end{itemize}

All these diagrams (Figs.~\ref{fig:chiPTsingle}--\ref{fig:chiPTdouble}) are
included in our interaction kernel. The SSE single-nucleon amplitudes can be
found in~\cite{HGHP}, while the two-body contributions are given explicitly in
\cite{Phillips}.  The only diagrams that remain to be calculated are the
diagrams with an intermediate $np$-state, cf. Fig.~\ref{fig:disp}.

\subsection {Diagrams with Intermediate $NN$-Scattering}
\label{sec:resonant}

\subsubsection{A Short Note on the Thomson Limit \label{sec:shortnote}}
Before we describe how we calculate the $NN$-rescattering diagrams 
sketched in Fig.~\ref{fig:disp}, we 
consider the well-known Thomson limit, i.e. 
the deuteron Compton amplitude in the limit of vanishing photon energy:
\begin{equation}
  A^\mathrm{Thomson}_d=\frac{e^2}{m_d }\,\eps\cdot\epspr\approx
  \frac{e^2}{2m_N}\,\eps\cdot\epspr.
  \label{eq:Thomsond}
\end{equation}
Friar showed that Eq.~(\ref{eq:Thomsond}) is a consequence of current
conservation and -- as long as there are no photons in internal loops -- gauge
invariance \cite{Friar}.  The EFT
calculations~\cite{deuteronpaper,Phillips,McGPhil1,McGPhil2} investigate the r\'egime
$\omega\sim\mpi$ and thus are by construction inapplicable in that limit. If
taken at face-value, they would there also violate gauge invariance, albeit
their interaction kernel is gauge invariant by construction. The violation
appears when evaluating the kernel between the deuteron wave functions,
without allowing the two nucleons in the intermediate state to interact with
each other. The reason is that the deuteron wave function implies this
interaction, which can be interpreted as the exchange of mesons, e.g. of
pions, between the two nucleons.  In order to achieve current conservation and
gauge invariance, it is therefore mandatory to include rescattering of the two
nucleons on one hand \textit{and} to couple the photons to these
meson-exchange currents, cf.  Fig.~\ref{fig:disp} and
Refs.~\cite{Lvov,Karakowski1,Karakowski2}.  It is one of the main advantages of this work
with respect to \cite{deuteronpaper,Phillips,McGPhil1,McGPhil2} that our calculation
does fulfill Eq.~(\ref{eq:Thomsond}).

Reaching the Thomson limit is a non-trivial check because the deuteron mass is
involved, whereas the Thomson seagull for Compton scattering from the proton,
Fig.~\ref{fig:chiPTsingle}(a), yields 
\begin{equation}
  A^\mathrm{Thomson}_p=\frac{e^2}{m_p}\,\eps\cdot\epspr
  \label{eq:Thomsonp}
\end{equation}
with $m_p$ the proton mass.  The single-neutron amplitude vanishes in the
static limit. Therefore, all other contributions to deuteron Compton
scattering in the limit $\w\rightarrow 0$ have to cancel half of the proton
amplitude~(\ref{eq:Thomsonp}).  In Section~\ref{sec:Thomson}, we discuss how
this cancellation comes about, after sketching in Sections~\ref{sec:dominant}
and~\ref{sec:subleading} our way to calculate diagrams with intermediate
$np$-scattering.

\subsubsection{Dominant Terms \label{sec:dominant}}

In the following, we briefly explain the ``Green's-function method'' to
include the $NN$-rescattering in the diagrams given in Fig.~\ref{fig:disp},
using second-order time-ordered perturbation theory in the two-photon
interaction.  For further details see Appendix~\ref{app:resonant} and
Refs.~\cite{Karakowski1,Karakowski2,PHD}.  In general, the scattering amplitude for these
processes can be written as
\begin{align}
  \Mfi{}&=\sum_C\left \{\frac{\mx{d_f,\gamma_f}{\Hint}{C}
      \mx{C}{\Hint}{d_i,\gamma_i}}
    { \w+\frac{\w^2}{2\md}-B-\EC}\right.\nonumber\\
  &+\left.  \frac{\mx{d_f,\gamma_f}{\Hint}{C,\gamma_f,\gamma_i}
      \mx{C,\gamma_f,\gamma_i}{\Hint}{d_i,\gamma_i}}
    {-\w+\frac{\w^2}{2\md}-B-\EC-\frac{\PCsq}{2\mC}}\right\},
  \label{eq:disp}
\end{align}
where we sum over all possible intermediate eigenstates $C$ of the
$NN$-interaction Hamiltonian.  Recall that the intermediate rescattering state
\eqref{eq:disp} is the off-shell $S$-matrix of $NN$-scattering, as
pointed out at the beginning of Sec.~\ref{sec:theory}. It contains also the
case that the two nucleons do not interact with each other between
photon emission and absorption.

\begin{figure}[!htb]
  \begin{center}
\parbox[m]{.35\linewidth}{
  \includegraphics*[width=\linewidth]{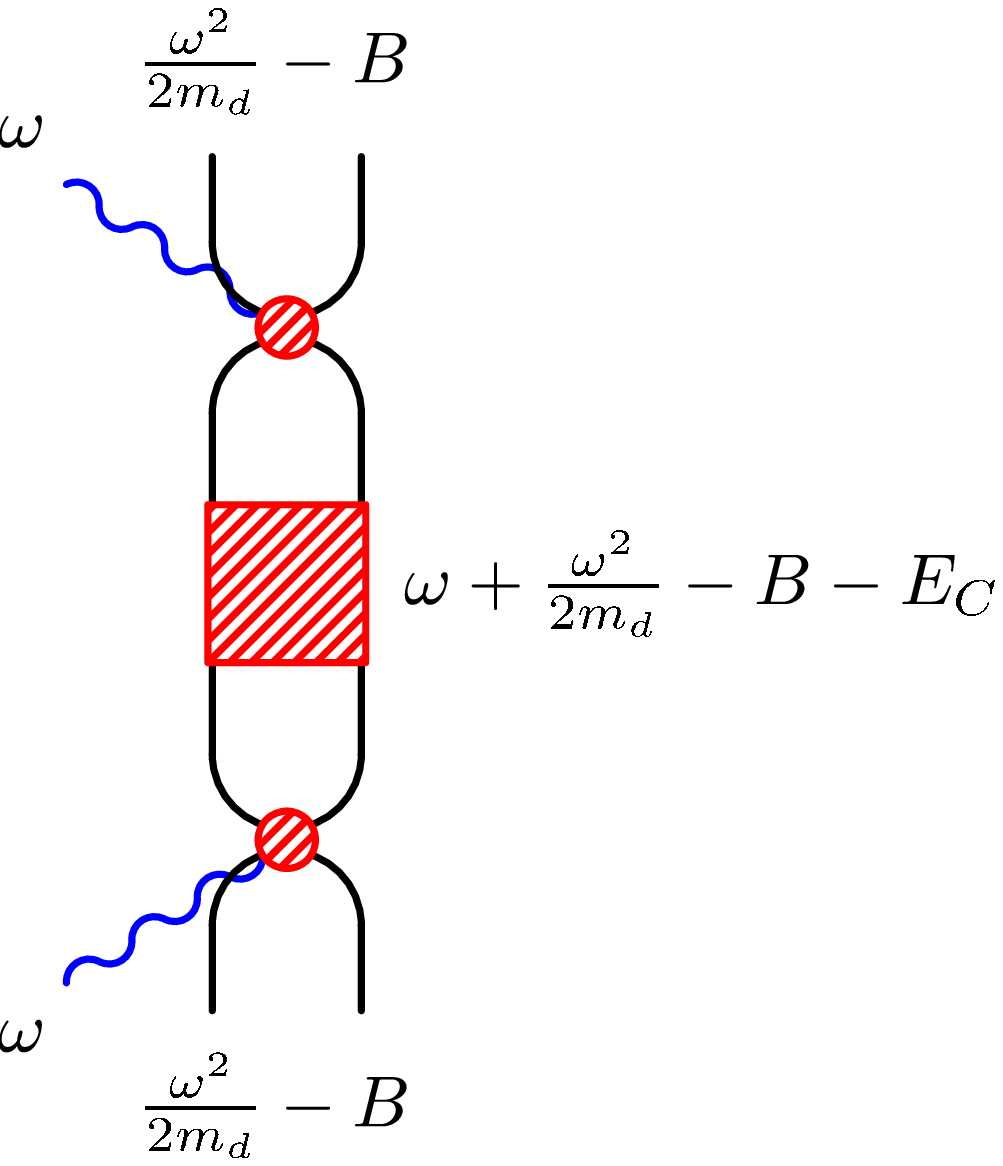}} \hspace{.5cm}
\parbox[m]{.46\linewidth}{
  \includegraphics*[width=\linewidth]{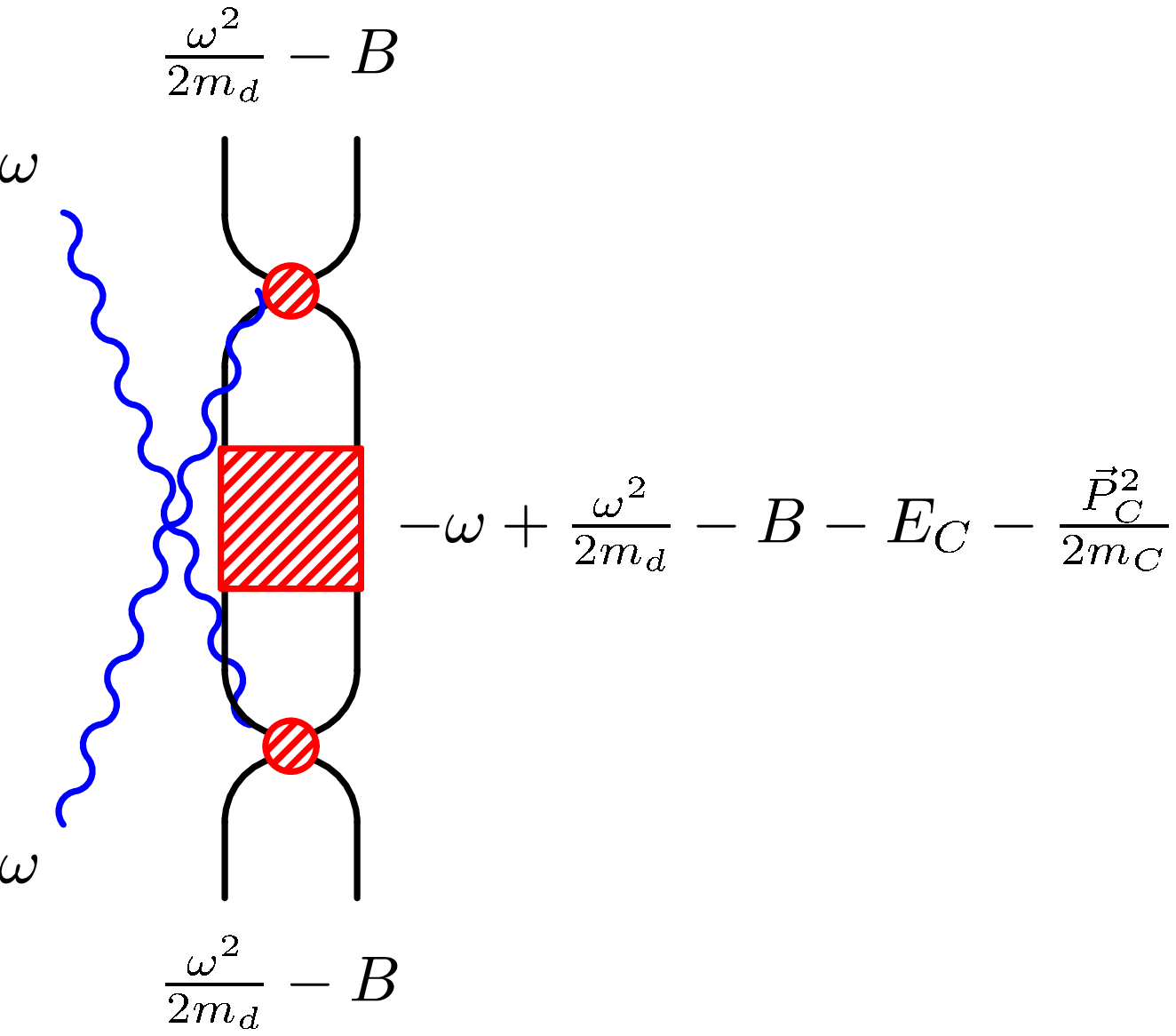}}
\parbox{1.\textwidth}{
  \caption{Diagrams with two-nucleon intermediate states (hatched square) in
    the $s$-channel (left) and in the $u$-channel (right). The labels denote
    the energies of the various particles and states.}
\label{fig:displabel}}
\end{center}
\end{figure}

The two terms in Eq.~(\ref{eq:disp}) correspond to the two diagrams shown on
the right hand side of Fig.~\ref{fig:disp}. We sketch these diagrams once
again in Fig.~\ref{fig:displabel} in order to explain the various terms in the
denominators of Eq.~(\ref{eq:disp}), constituting the energy of the
intermediate nucleons.
In the $\gamma d$ cm frame, the incoming and outgoing photons have the same
energy $\w$. $B$ denotes the deuteron binding energy, $\frac{\w^2}{2\md}$ is
the kinetic energy of the incoming deuteron, $\frac{\PCsq}{2\mC}$ that of the
intermediate two-nucleon system.  For our numerical evaluations we use the
masses $m_C=2m_N$, $m_d=2m_N-B$ and neglect isospin-breaking effects,
$m_p\equiv m_n\equiv m_N$. As we calculate in the cm frame of the $\gamma d$
system, $\frac{\PCsq}{2\mC}=0$ in the $s$-channel diagram, whereas
$\PC=-\ki-\kf$ in the $u$-channel, i.e.
$\frac{\PCsq}{2\mC}=\frac{\w^2}{m_C}\,(1+\cos\theta)$.  $-\EC$ denotes the
excitation energy of the intermediate state $C$.

The interaction Hamiltonian in Eq.~(\ref{eq:disp}) is
\begin{equation}
  \Hint=-\int\vec{J}\ofxi\cdot\vec{A}\ofxi\,d^3\xi.
  \label{eq:Hint}
\end{equation}
An explicit expression for the photon field, expanded into multipoles, has
been derived in~\cite{Karakowski1,Karakowski2} in analogy to Ref.~\cite{Rose} and is given
in Eq.~(\ref{eq:multipoleexp}).  It consists of three parts and can
schematically be written as
\begin{equation}
  \vec{A}=\nab\phi+\Aone+\Atwo
  \label{eq:schematically}
\end{equation} 
with $\Aone$ denoting the magnetic part of $\vec{A}$, $\nab\phi+\Atwo$ the
electric part. The definitions of the function $\phi\ofxi$, as well as of
$\Aone$ and $\Atwo$, are given in Appendix~\ref{app:multipoleexp}.  Now we
systematically replace the photon field in the interaction
Hamiltonian~(\ref{eq:Hint}) by the three terms contained in
Eq.~(\ref{eq:schematically}).  Therefore, when we only distinguish between the
various possibilities for $\vec{A}$, we find nine different combinations in
Eq.~(\ref{eq:disp}).  The largest contributions are those where we substitute
$\vec{A}\ofxi\rightarrow\nab\phi\ofxi$ at both vertices, which is the only
part of the photon field that contributes for $\w=0$, cf.
Appendix~\ref{app:multipoleexp}. Further terms, where this replacement is made
only once, are discussed in Section~\ref{sec:subleading}. Only a few
combinations of interactions without the gradient part of $\vec{A}$ give
visible contributions.  These are also taken care of in
Section~\ref{sec:subleading}.  

Contributions to the elastic cross sections of the order of $\leq2\%$ are
neglected throughout this article since the theoretical uncertainty from
effects which are higher than $\calO(\epsilon^3)$ is larger than that. This is
discussed and quantified in detail in the following Sections and summarized in
the Conclusions. Most of these contributions are treated in detail in
Hildebrandt's PhD thesis~\cite{PHD}. The numerical error is $\leq2\%$ at very
low energies and less for $\omega\sim\mpi$, see e.g.~the discussion of the
Thomson limit in Sect.~\ref{sec:Thomson}.

In this section, we calculate Eq.~(\ref{eq:disp}) with 
$\Hint\rightarrow-\int\vec{J}\ofxi\cdot\nab\phi\ofxi\,d^3\xi$ simultaneously 
at both vertices, i.e. we restrict ourselves to the terms arising from 
minimal coupling. 
In order to simplify the calculation on the one hand, and to ensure gauge 
invariance and the correct Thomson limit on the other, we integrate by parts 
and use current conservation:
\begin{align}
  -\int\vec{J}\ofxi\cdot\nab\phi\ofxi\,d^3\xi&=
  \int\nab\cdot\vec{J}\ofxi\,\phi\ofxi\,d^3\xi\label{eq:pI}\\
  \nab\cdot\vec{J}\ofxi&=-\frac{\partial\rho\ofxi}{\partial t}=
  -i\left[H,\rho\ofxi\right]
  \label{eq:continuity}
\end{align}
The fact that one only needs to know the charge density $\rho$ in order to
calculate the amplitude in the long-wavelength limit is referred to as
``Siegert's theorem''~\cite{Siegert}.  For $\rho(\vec{\xi})$ one can find
in~\cite{Ericson} the general decomposition
\begin{equation}
  \rho\ofxi=\rho^{(0)}\ofxi+\rho^{\mathrm{ex}}(\vec{\xi};\vec{x}_p,\vec{x}_n)
  \label{eq:chargedensity}
\end{equation}
with $\rho^{(0)}$ the charge density of the two nucleons and
$\vec{x}_p,\;\vec{x}_n$ the position of proton and neutron, respectively.
$\rho^{\mathrm{ex}}(\vec{\xi};\vec{x}_p,\vec{x}_n)$ is the charge density
arising from meson-exchange currents, e.g. from those given in the lower line
of Fig.~\ref{fig:disp}.  The dominant term in Eq.~(\ref{eq:chargedensity}) is
\begin{equation}
  \rho^{(0)}\ofxi=\sum_{j=n,p}e_j\,\delta(\vec{\xi}-\vec{x}_j)=
  e\,\delta(\vec{\xi}-\vec{x}_p),
  \label{eq:rhonull}
\end{equation}
which is the only non-vanishing contribution to $\rho\ofxi$ in the static
limit (``Siegert's hypothesis''~\cite{Siegert}).  Note that the
$\delta$-functions in Eq.~(\ref{eq:rhonull}) indicate that the two nucleons
are treated as pointlike particles, i.e. unlike the authors of e.g.
Ref.~\cite{Lvov} we do not introduce any nucleon form factors.  We also
performed calculations including
$\rho^{\mathrm{ex}}(\vec{\xi};\vec{x}_p,\vec{x}_n)$.  From these
investigations, which are reported in Ref.~\cite{PHD}, we conclude that
$\rho^{\mathrm{ex}}$ is well negligible in the energy range considered.
Indeed, such terms are suppressed by three orders in the EFT power-counting,
as shown in Refs.~\cite{Danieled,danielpaper}. Therefore, we are only
concerned with $\rho^{(0)}\ofxi$.

\begin{figure}[!htb]
  \begin{center}
    \includegraphics*[width=.4\linewidth]{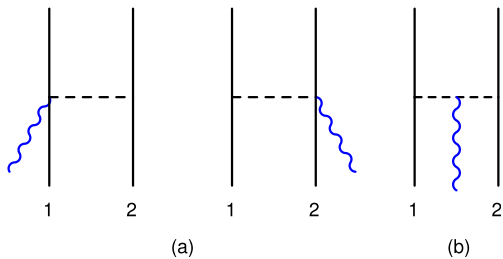}
    \caption{One-pion-exchange currents contributing to our calculation: the
      ``Kroll-Ruderman current'' (a) and the ``pion-pole current'' (b).}
    \label{fig:mesonexchange}
  \end{center}
\end{figure}

Albeit it is not obvious, the use of current conservation in
Eq.~(\ref{eq:continuity}) causes that meson-exchange currents (cf.
Fig.~\ref{fig:mesonexchange}) are also implicitly included in the calculation,
as
\begin{equation}
  \nab\cdot\vec{J}^{\mathrm{ex}}=
  -i\left[V^{\mathrm{ex}}\,\vec{\tau}_1\cdot\vec{\tau}_2,\rho^{(0)}\right]
  -i\left[H,\rho^{\mathrm{ex}}\right].
  \label{eq:implicit}
\end{equation}
$V^{\mathrm{ex}}$ is the $np$-potential from one-pion exchange~\cite{Ericson},
which is part of the Hamiltonian $H$. $\vec{\tau}_i$ is the isospin operator
of the $i$th nucleon.  These contributions, which go beyond coupling the
photon field to the single-nucleon current, like in Fig.~\ref{fig:pole}, were
already indicated in Fig.~\ref{fig:disp}.

Substituting $\nab\cdot\vec{J}$ by $-i\left[H,\rho^{(0)}\ofxi\right]$ in 
Eq.~(\ref{eq:pI}), the integral 
over the dummy variable $\vec{\xi}$  can easily be performed to yield
\begin{equation}
  \Hint=-\int\vec{J}\ofxi\cdot\nab\phi\ofxi\,d^3\xi=
  -i\left[H,e\,\phi(\vec{x}_p)\right],
  \label{eq:commutator}
\end{equation}
where $H$ is the full Hamiltonian of the $np$-system
\begin{equation}
  H=\frac{\vec{p}_p^{\,\,2}}{2m_p}+\frac{\vec{p}_n^{\,\,2}}{2m_n}+V
  \label{eq:fullHamiltonian}
\end{equation}
with the $np$-potential $V$.  To evaluate the commutator
(\ref{eq:commutator}), we switch to cm variables, i.e.
\begin{equation}
  \vec{p}=\frac{\vec{p}_p-\vec{p}_n}{2},\;\;\vec{P}=\vec{p}_p+\vec{p}_n,\;\;
  \vec{r}=\vec{x}_p-\vec{x}_n,\;\;\vec{R}=\frac{\vec{x}_p+\vec{x}_n}{2}.
  \label{eq:cmvariables}
\end{equation}
Our analysis shows that recoil corrections, which arise from the cm motion of
the deuteron, are well negligible. This observation agrees with
Ref.~\cite{Karakowski1,Karakowski2}, where such corrections have been evaluated as well.
The net effect is that we may set $\vec{R}=\vec{0}$, i.e. we neglect the cm
velocity of the two nucleons\footnote{There is one exception to
  this rule: We do include the kinetic energy of the two-nucleon system in
  energy denominators, cf. Eq.~(\ref{eq:disp}).}. As a consequence, we find
$\vec{x}_p=\vec{r}/2$ and the Hamiltonian~(\ref{eq:fullHamiltonian})
simplifies to the ``internal'' Hamiltonian
\begin{equation}
  H^{np}=\frac{\vec{p}^{\,2}}{m_N}+V.
  \label{eq:Hinternal}
\end{equation}
Therefore, we can rewrite Eq.~(\ref{eq:commutator}) as
\begin{equation}
  \Hint=-i\left[H^{np},e\,\phi(\vec{r}/2)\right].
  \label{eq:equivalent}
\end{equation}
Inserting the commutator~(\ref{eq:equivalent}) into Eq.~(\ref{eq:disp}) and
defining $\phiihat=e\,\phi_i(\vec{r}/2)$, $\phifhat=e\,\phi_f(\vec{r}/2)$ in
analogy to Ref.~\cite{Karakowski1,Karakowski2} we get
\begin{align}
  \Mfi{\phi\phi}&=-\sum_C\left
    \{\frac{\mx{d_f}{\left[H^{np},\phifhat\right]}{C}
      \mx{C}{\left[H^{np},\phiihat\right]}{d_i}}
    { \w+\frac{\w^2}{2\md}-B-\EC}\right.\nonumber\\
  &+\left.  \frac{\mx{d_f}{\left[H^{np},\phiihat\right]}{C}
      \mx{C}{\left[H^{np},\phifhat\right]}{d_i}}
    {-\w+\frac{\w^2}{2\md}-B-\EC-\frac{\PCsq}{2\mC}}\right\}.
\end{align}
In order to keep track of the various combinations of interaction Hamiltonians
we have labeled the double-$\phi$ transition matrix '$\phi\phi$'; the photon
states have been skipped for brevity.  Now the commutators are expanded and,
as $\ket{d_{i,f}}$, $\ket{C}$ are eigenstates of $H^{np}$, we can act with
$H^{np}$ on these states.  We end up with four amplitudes, which have already
been derived (in the lab system) in Ref.~\cite{Karakowski1,Karakowski2}:
\begin{align}
  \label{eq:Mfiphiphi1added}
  \Mfi{\phi\phi 1}&=\left(\w+\frac{\w^2}{2\md}\right)^2
  \sum_C\frac{\mx{d_f}{\phifhat}{C}\mx{C}{\phiihat}{d_i}}
  {\w+\frac{\w^2}{2\md}-B-\EC}\\
  \label{eq:Mfiphiphi2added}
  \Mfi{\phi\phi 2}&=\left(\w+\frac{\PCsq}{2\mC}-\frac{\w^2}{2\md}\right)^2
  \sum_C\frac{\mx{d_f}{\phiihat}{C}\mx{C}{\phifhat}{d_i}}
  {-\w-\frac{\PCsq}{2\mC}+\frac{\w^2}{2\md}-B-\EC}\\
  \label{eq:Mfiphiphi3added}
  \Mfi{\phi\phi 3}&=\left(\frac{\PCsq}{2\md}-\frac{\w^2}{\md}\right)
  \mx{d_f}{\phifhat\,\phiihat}{d_i}\\
  \label{eq:Mfiphiphi4added}
  \Mfi{\phi\phi 4}&=\frac{1}{2}
  \mx{d_f}{\left[\left[H^{np},\phiihat\right],\phifhat\right]+
    \left[\left[H^{np},\phifhat\right],\phiihat\right]}{d_i}
\end{align}
$\Mfi{\phi\phi 4}$ is the only one of these amplitudes which contributes in
the static limit.  It is responsible for the correct low-energy behavior of
the calculation, as will be discussed in detail in Section~\ref{sec:Thomson}.

We defer the evaluation of these amplitudes to Appendix~\ref{app:resonant} and
turn now to those contributions, where the substitution
$\vec{A}\ofxi\rightarrow\nab\phi\ofxi$ (cf. the beginning of this section) is
made at most once.

\subsubsection{Subleading Terms \label{sec:subleading}}

So far we only considered contributions arising from minimal coupling of the
photon field to the two-nucleon system at both vertices. In the following, we
describe how to calculate the amplitudes given in Eq.~(\ref{eq:disp}), when
the replacement
\begin{equation}
  \Hint=-\int\vec{J}\ofxi\cdot\vec{A}\ofxi\, d^3\xi\rightarrow 
  -\int\vec{J}\ofxi\cdot\nab\phi\ofxi\,d^3\xi
\end{equation}
is made only once. The term 'subleading' refers to the fact that the resulting
amplitude is numerically less important than that of
Section~\ref{sec:dominant}~-- its contribution to the differential cross
section is suppressed with respect to the dominant terms from
Section~\ref{sec:dominant} by at least one order of magnitude for all energies
and angles considered, see Fig.~\ref{fig:separation}.  The amplitude is
denoted by $\Mfi{\phi}$ and follows immediately from Eq.~(\ref{eq:disp}):
\begin{align}
  \Mfi{\phi}&=\sum_C\left\{
    \frac{\mx{d_f}{\int\vec{J}\ofxi\cdot\nab\phi_f\ofxi\,d^3\xi}{C}
      \mx{C}{\int\vec{J}\ofxi\cdot\vec{A}\ofxi\,d^3\xi}{d_i}}{\denoms}\right.
  \nonumber\\&+ \frac{\mx{d_f}{\int\vec{J}\ofxi\cdot\vec{A}\ofxi\,d^3\xi}{C}
    \mx{C}{\int\vec{J}\ofxi\cdot\nab\phi_i\ofxi\,d^3\xi}{d_i}}{\denoms}
  \nonumber\\&+
  \frac{\mx{d_f}{\int\vec{J}\ofxi\cdot\nab\phi_i\ofxi\,d^3\xi}{C}
    \mx{C}{\int\vec{J}\ofxi\cdot\vec{A}\ofxi\,d^3\xi}{d_i}}{\denomu}
  \nonumber\\&+
  \left.\frac{\mx{d_f}{\int\vec{J}\ofxi\cdot\vec{A}\ofxi\,d^3\xi}{C}
      \mx{C}{\int\vec{J}\ofxi\cdot\nab\phi_f\ofxi\,d^3\xi}{d_i}}{\denomu}\right\}
\end{align}
Now we perform the same steps as described in
Eqs.~(\ref{eq:commutator}-\ref{eq:Mfiphiphi4added}), i.e. we first replace
$\int\vec{J}\ofxi\cdot\nab\phi\ofxi\,d^3\xi$ by
$i\left[H^{np},e\,\phi(\vec{r}/2)\right]$, then act with $H^{np}$ on $\ket{d}$
and $\ket{C}$, respectively, and finally add and subtract terms in order to
perform some cancellations against the denominator. We find, again neglecting
recoil terms and the deuteron velocity,
\begin{align}
  \label{eq:Mfiphi}
  \Mfi{\phi}&=i\sum_C\bigg\{
  \mx{d_f}{\phifhat}{C}\mx{C}{\int\vec{J}\ofxi\cdot\vec{A}\ofxi\,d^3\xi}{d_i}
  \nonumber\\&- \bigg(\w+\frac{\w^2}{2m_d}\bigg)\frac{\mx{d_f}{\phifhat}{C}
    \mx{C}{\int\vec{J}\ofxi\cdot\vec{A}\ofxi\,d^3\xi}{d_i}}{\denoms}
  \nonumber\\&-
  \mx{d_f}{\int\vec{J}\ofxi\cdot\vec{A}\ofxi\,d^3\xi}{C}\mx{C}{\phiihat}{d_i}
  \nonumber\\&+ \bigg(\w+\frac{\w^2}{2m_d}\bigg)\frac{\mx{d_f}
    {\int\vec{J}\ofxi\cdot\vec{A}\ofxi\,d^3\xi}{C}\mx{C}{\phiihat}{d_i}}{\denoms}
  \nonumber\\&+
  \mx{d_f}{\phiihat}{C}\mx{C}{\int\vec{J}\ofxi\cdot\vec{A}\ofxi\,d^3\xi}{d_i}
  \nonumber\\&+
  \bigg(\w-\frac{\w^2}{2m_d}+\frac{\PCsq}{2m_C}\bigg)\frac{\mx{d_f}{\phiihat}{C}
    \mx{C}{\int\vec{J}\ofxi\cdot\vec{A}\ofxi\,d^3\xi}{d_i}}{\denomu}
  \nonumber\\&-
  \mx{d_f}{\int\vec{J}\ofxi\cdot\vec{A}\ofxi\,d^3\xi}{C}\mx{C}{\phifhat}{d_i}
  \nonumber\\&-
  \bigg(\w-\frac{\w^2}{2m_d}+\frac{\PCsq}{2m_C}\bigg)\frac{\mx{d_f}
    {\int\vec{J}\ofxi\cdot\vec{A}\ofxi\,d^3\xi}{C}\mx{C}{\phifhat}{d_i}}{\denomu}
  \bigg\}.
\end{align}
Whenever the energy denominator has been canceled, the sum over $C$ may be
collapsed. As
$\hat{\phi}\left(\int\vec{J}\ofxi\cdot\vec{A}\ofxi\,d^3\xi\right)=
\left(\int\vec{J}\ofxi\cdot\vec{A}\ofxi\,d^3\xi\right)\hat{\phi}$, these four
terms cancel exactly, and only the terms including an energy denominator
remain.

The relevant parts of the photon field $\vec{A}\ofxi$ are the non-gradient
terms in Eq.~(\ref{eq:schematically}). The current $\vec{J}\ofxi$ includes
one-body and two-body currents. The latter are the pion-exchange currents of
Fig.~\ref{fig:mesonexchange}, however we found that only the Kroll-Ruderman
current, Fig.~\ref{fig:mesonexchange}(a), gives visible contributions. The
single-nucleon current consists of two parts, which we call $\Jsigma$ and
$\Jp$, with
\begin{align}
  \label{eq:Jsigma}
  \Jsigma\ofxi&=\frac{e}{2m_N}\sum_{j=n,p}\left[\nab_\xi\times\mu_j\,
    \vec{\sigma}_j\,\delta(\vec{\xi}-\vec{x}_j)\right],\\
  \Jp \ofxi&=\frac{1}{2m_N}\sum_{j=n,p}\left\{
    e_j\,\delta(\vec{\xi}-\vec{x}_j),\vec{p}_j\right\},
  \label{eq:Jp}
\end{align}
cf. e.g.~\cite{Ericson}. So we may schematically write
\begin{equation}
  \vec{J}\ofxi=\Jsigma\ofxi+\Jp\ofxi+\vec{J}\,^\mathrm{KR}\ofxi.
\end{equation}
$\mu_j$ is the magnetic moment, $\vec{\sigma}_j$ the spin operator and
$\vec{p}_j$ the momentum of the $j$th nucleon.  We observe sizeable
contributions only from the magnetic moment interaction encoded in the spin
current $\Jsigma$, Eq.~(\ref{eq:Jsigma}), whereas $\Jp$ turned out negligibly
small.  Our notation for the amplitudes is $\Mfi{\phi\,\sigma}$ when we
replace $\vec{J}\ofxi$ by $\Jsigma\ofxi$ in Eq.~(\ref{eq:Mfiphi}) and
$\Mfi{\phi\,\mathrm{KR}}$ when we use the Kroll-Ruderman current instead.  As
this part is rather technical, we shift it to Appendix~\ref{app:resonant}.
There, we also discuss those amplitudes which do not contain the gradient part
of the photon field but nevertheless give sizeable contributions, i.e.
$\Mfi{\sigma\,\sigma}$ and $\Mfi{\mathrm{KR}\,\sigma}$.

Some of these contributions have been considered before, namely diagrams with
one photon coupling to one-pion-exchange and one photon coupling to a nucleon,
see~\cite{McGPhil1} and \cite[Fig.~9]{McGPhil2}. As demonstrated there, they
extend the region of validity $\omega\sim\mpi$ of the kernel ``without''
rescattering to lower energies, but are not sufficient to restore the Thomson
limit.

\begin{figure}[!htb]
  \begin{center}
    \includegraphics*[width=.4\linewidth]{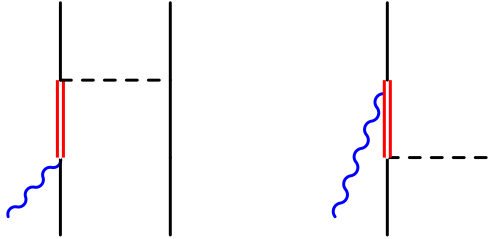}
    \caption{Exemplary one-pion-exchange currents with explicit $\Delta(1232)$
      degrees of freedom.}
    \label{fig:mesonexchangeDelta}
  \end{center}
\end{figure}

Two-body currents with explicit $\Delta(1232)$ degrees of freedom, as
displayed in Fig.~\ref{fig:mesonexchangeDelta}, are suppressed by one order in
$\epsilon$ with respect to the Kroll-Ruderman current,
Fig.~\ref{fig:mesonexchange}(a), due to the $\gamma N \Delta$~vertex being
part of $\mathcal{L}_{N\Delta}^{(2)}$~\cite{HHKLett1,HHKLett2}.  This agrees
with the findings of \cite{Lvov}, where such contributions to elastic deuteron
Compton scattering below 100~MeV were claimed to be of the order of 2\%. A
similar size is reported in~\cite{Ericson} for the process $np\rightarrow
d\gamma$, where the contributions from the $\Delta$(1232) current turn out to
be considerably smaller than those from pionic exchange currents. Therefore,
and due to the excellent agreement of the total deuteron-photodisintegration
cross section, extracted from our elastic Compton amplitude, with data, cf.
Appendix~\ref{app:photodisintegration}, we so far refrain from including these
terms into our calculation. It would be an interesting future task to perform
a detailed investigation of their size.

Now we have prepared all ingredients of our deuteron Compton calculation. In
the next section we demonstrate that it fulfills the well-known low-energy
theorem, Eq.~(\ref{eq:Thomsond}), i.e. we obtain the correct low-energy limit
within our approach. In Section~\ref{sec:results}, we present our results at
non-zero energies and compare them to those from Ref.~\cite{deuteronpaper} and
to data.

\subsubsection{Low-Energy (Thomson) Limit}
\label{sec:Thomson}

This work constitutes an extended hybrid approach to the calculations of
Refs.~\cite{deuteronpaper,Phillips,McGPhil1,McGPhil2} which are not applicable
for photon energies below 50~MeV.  In this section, we prove that we have
indeed \textit{removed} the limitations of these works at low energies, i.e.
we reach the correct limit of vanishing photon energy,
Eq.~(\ref{eq:Thomsond}).

The only non-vanishing amplitudes in the static limit~-- except for the proton
seagull, Fig.~\ref{fig:chiPTsingle}(a)~-- are the explicit pion-exchange
diagrams, Fig.~\ref{fig:chiPTdouble}, and the double-commutator term,
Eq.~(\ref{eq:Mfiphiphi4added}).  This double-commutator involves the internal
Hamiltonian $H^{np}=\frac{\vec{p}\,^2}{m_N}+V$, cf. Eq.~(\ref{eq:Hinternal}),
and therefore can be separated into a kinetic and a potential part.
Arenh\"ovel showed analytically that in the static limit, the potential energy
part, using the one-pion-exchange potential, cancels exactly the contributions
from explicit pion exchange, Fig.~\ref{fig:chiPTdouble} \cite{Arenhoevel}.
Therefore, the kinetic energy part has to cancel half of the proton seagull.
This can easily be shown to be true, cf.  Ref.~\cite{Karakowski1,Karakowski2}
or~\cite{PHD}. Note that the Thomson amplitude~(\ref{eq:Thomsond}) is
independent of the deuteron wave function and the $np$-potential chosen.

\begin{figure}[!htb]
  \begin{center}
    \includegraphics*[width=.48\linewidth]{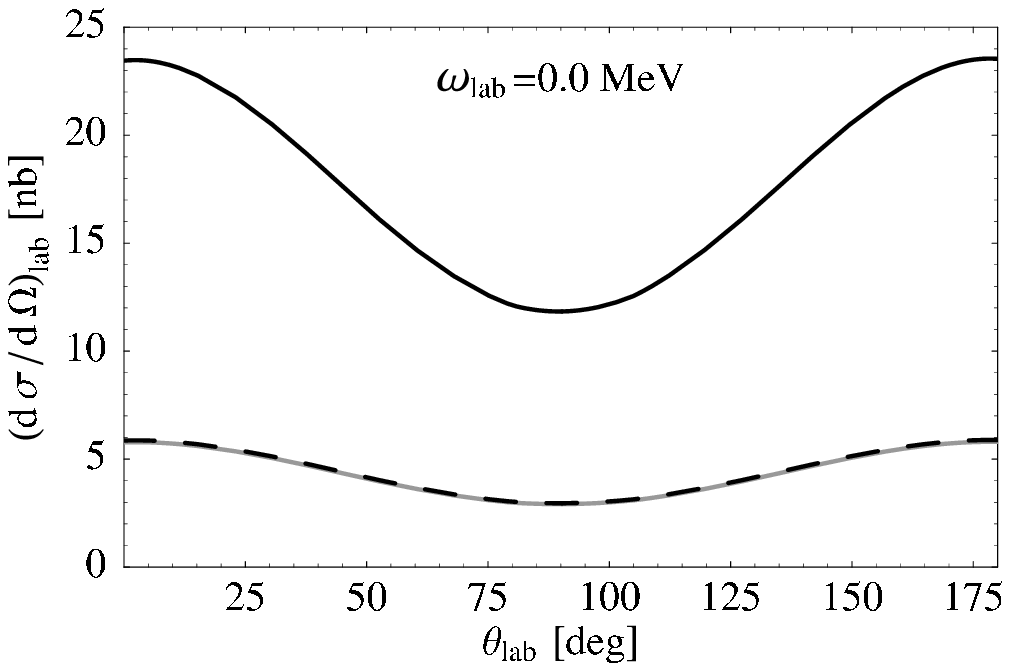}
    \hfill
    \includegraphics*[width=.48\linewidth]{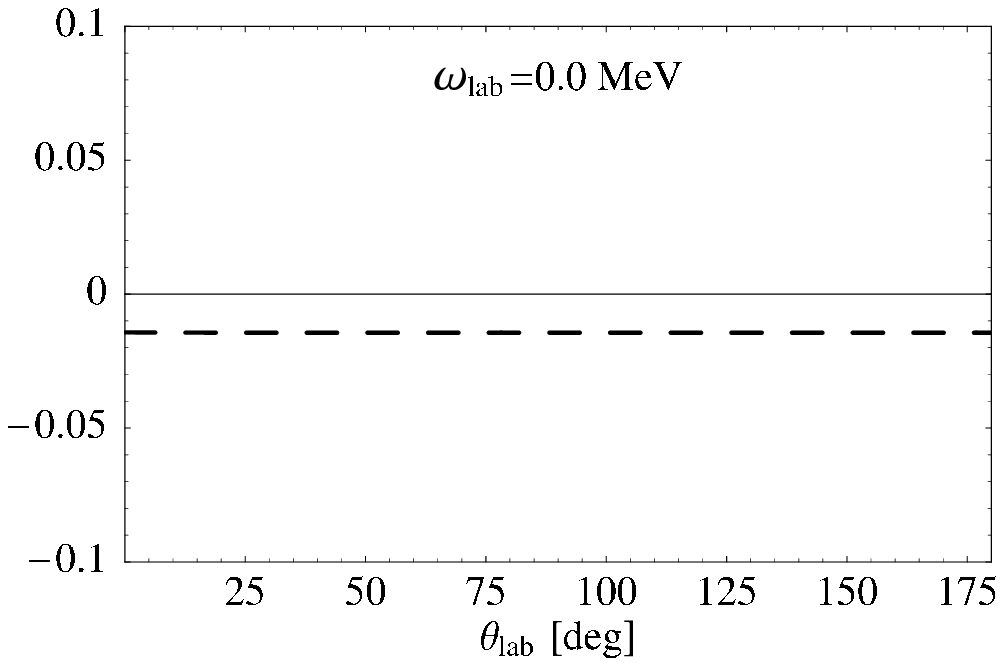}
    \caption {Left panel: Comparison of the proton (black, solid), deuteron
      (grey, solid) and $1/4$ of the proton (black, dashed) Compton cross
      section in the static limit. The function plotted in the right panel is
      $\left( \frac{d\sigma}{d\Omega}\right)_d/
      \left(\frac{1}{4}\frac{d\sigma}{d\Omega}\right)_p-1$.}
    \label{fig:Thomson2}
  \end{center}
\end{figure}

Our numerical evaluation agrees well with the Thomson limit
(\ref{eq:Thomsond}), as demonstrated in a comparison (Fig.~\ref{fig:Thomson2})
between the proton Compton cross section,
$\left(\frac{1}{2}\right)^2=\frac{1}{4}$ of this cross section and the
deuteron Compton cross section at zero photon energy (we remind the reader
that $\frac{d\sigma}{d\Omega}\propto|\Mfi{}|^2$). The latter two curves are
nearly indistinguishable. In the right panel of Fig.~\ref{fig:Thomson2}, we
see that the relative error $\left( \frac{d\sigma}{d\Omega}\right)_d/
\left(\frac{1}{4}\frac{d\sigma}{d\Omega}\right)_p-1$ is constant in $\theta$
and less than 2\%. Therefore it can be accounted for by a constant factor.
The main part of this discrepancy is due to numerical uncertainties in the
normalization of the wave function within our code.

In this section, we showed that our Green's-function hybrid approach to
deuteron Compton scattering fulfills the low-energy theorem and therefore
guarantees gauge invariance of the calculation.  In the next section, we
present our results for non-zero photon energies.  From the good description
of all data available we conclude that we have achieved a consistent
description of $\gamma d$ scattering for photon energies ranging from 0~MeV up
to $\w\sim100$~MeV.

\section{Predictions and Discussion}
\setcounter{equation}{0}
\label{sec:results}
  
\subsection{Comparison to Previous Work 
\label{sec:comparison}}

\begin{figure}[!htb]
\begin{center} 
\includegraphics*[width=.48\linewidth]{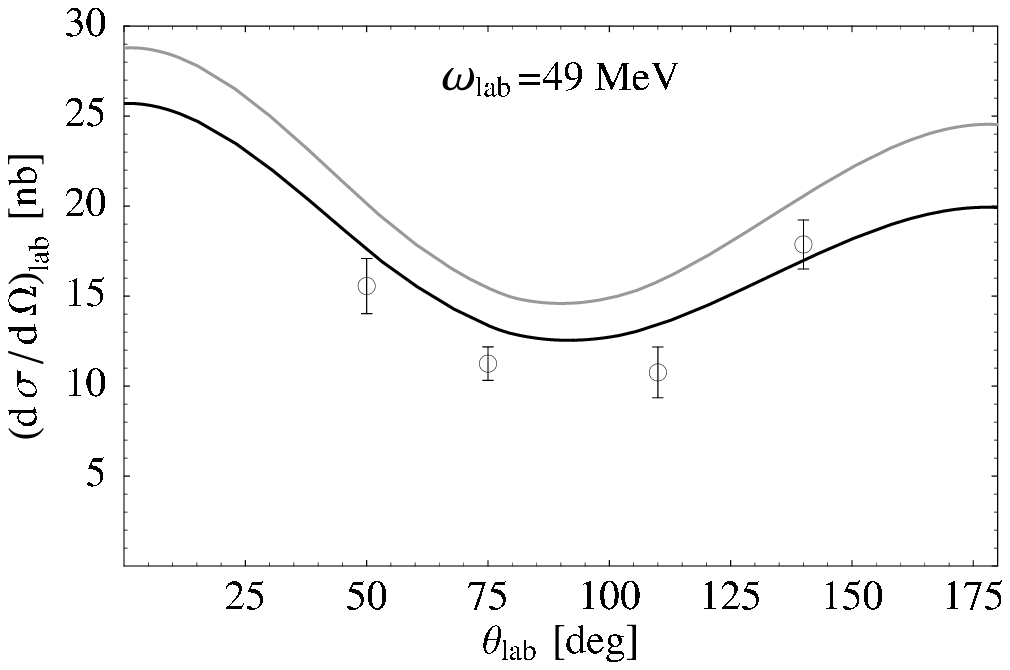}
\hfill
\includegraphics*[width=.48\linewidth]{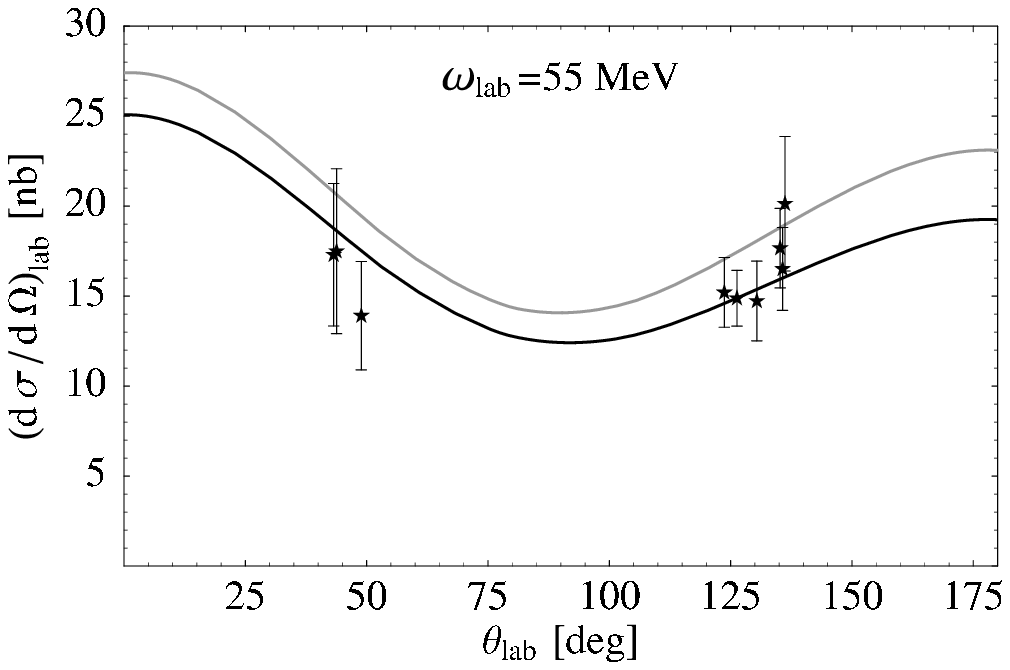}\\
\includegraphics*[width=.48\linewidth]{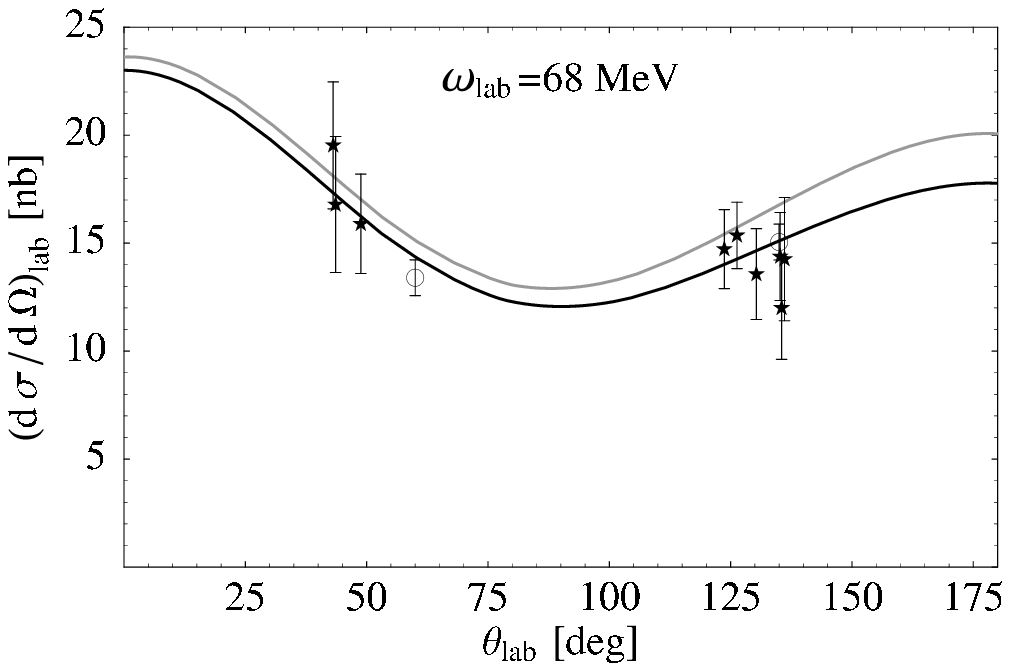}
\hfill
\includegraphics*[width=.48\linewidth]{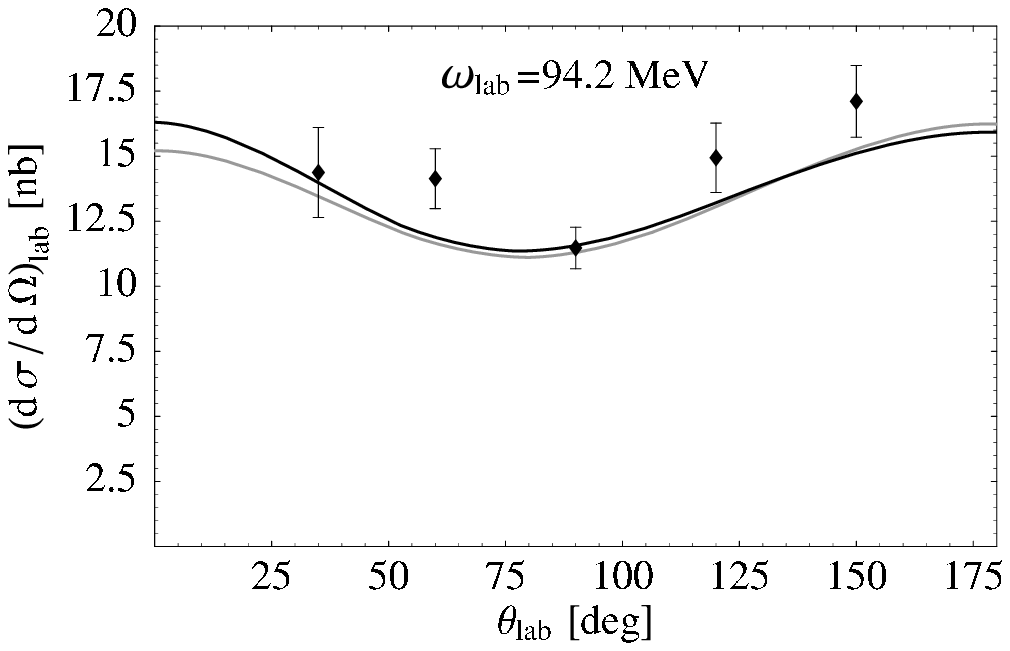}
\parbox{1.\textwidth}{
  \caption {Comparison of our predictions for deuteron Compton scattering
    (black) with those from Ref.~\cite{deuteronpaper} (grey). In both
    calculations the numbers from the 2-parameter SSE fit to proton data of
    Ref.~\cite{HGHP} are used for the isoscalar polarizabilities. The data are
    from \cite{Lucas}~(circle), \cite{Lund}~(star) and
    \cite{Hornidge}~(diamond).}
  \label{fig:comparisonpertnonpert}}
\end{center}
\end{figure}
\noindent
In Fig.~\ref{fig:comparisonpertnonpert}, we show our predictions for the
elastic deuteron Compton cross sections, compared to those from
Ref.~\cite{deuteronpaper} and to the data from Illinois~\cite{Lucas},
Lund~\cite{Lund} and SAL~\cite{Hornidge}.  Our results are parameter-free, as
we use the values obtained from proton Compton scattering in~\cite{HGHP} for
the two a priori unknown parameters $g_{117}$ and $g_{118}$, cf.
Fig.~\ref{fig:SSEsingle}(f).  Obviously we have reached our final goal: a
(chirally) consistent calculation of elastic deuteron Compton scattering which
describes all existing data reasonably well and also satisfies the low-energy
theorem exactly, cf.~Section~\ref{sec:Thomson}.  If not stated differently,
all curves throughout this work have been derived using the NNLO chiral wave
function from Ref.~\cite{Epelbaum} with cutoff $\Lambda=650$~MeV. However, as
we demonstrate in Section~\ref{sec:wavefunctiondep}, we achieve very similar
results with other state-of-the-art wave functions.  The values we use for the
various input parameters are given in Table~\ref{tab:const}. The numbers for
the short-distance couplings $g_{117}$, $g_{118}$ are taken from the
Baldin-constrained fit~\cite{HGHP} of $\alpha_E$ and $\beta_M$ to the proton
Compton data, cf.  Eq.~(\ref{eq:exppHGHP}).  We use the resulting proton
polarizabilities for the neutron analogues as well, as there are no isovector
contributions up to third order in the SSE scheme.
\begin{table}[!htb] 
\begin{center}
\begin{tabular}{|c|c|c|}
\hline 
Parameter         & Value       & Comment \\
\hline 
$m_\pi$           & $139.6$~MeV   & charged pion mass \\
$m_N$             & $938.9$~MeV   & isoscalar nucleon mass \\
$f^2$             & $0.075$       & pion-nucleon coupling constant \\
$f_\pi$           & $92.4$~MeV    & pion-decay constant \\
$g_A$             & $1.267$       & axial coupling constant \\
$\alpha$          & $1/137$       & QED fine-structure constant \\
$\mu_p$           & $2.795 $      & magnetic moment of the proton \\
$\mu_n$           & $-1.913$      & magnetic moment of the neutron \\
\hline
$\Delta_0$        & $271.1$~MeV   & $N\Delta$ mass splitting \\
$g_{\pi N\Delta}$ & $1.125$       & $\pi N\Delta$ coupling constant \\
$b_1$             & $4.67$        & $\gamma N\Delta$ coupling constant\\
$g_{117}$         & $18.82$       & short-distance coupling constant\\
$g_{118}$         & $-6.05$       & short-distance coupling constant\\
\hline
$m_d$             & $1875.58$~MeV & deuteron mass \\
$B$               & $2.2246$~MeV  & deuteron binding energy\\
\hline
\end{tabular}
\end{center}
\caption{$\chi$EFT parameters determined independently of this work. 
Magnetic moments are given in nuclear magnetons.}
\label{tab:const}
\end{table}

There are still minor deviations from the experiments, e.g. 
our calculation lies slightly above the three 49~MeV data from 
\cite{Lucas} which have been measured at 
angles below $120^\circ$. However, this is a feature that our calculation has 
in common with other approaches which also reach the correct static limit, 
e.g. \cite{Lvov,Karakowski1,Karakowski2,Rupak1,Rupak2,Rupak3}. 
At higher energies, the two calculations of Ref.~\cite{deuteronpaper} and
of this work approach each other. This is another important 
cross-check as the power counting of the calculations of 
Ref.~\cite{deuteronpaper}, as well as of Refs.~\cite{Phillips,McGPhil1,McGPhil2},
was designed for photon energies of $\w\sim100$~MeV. Consequently, both 
curves in Fig.~\ref{fig:comparisonpertnonpert} describe the 94.2~MeV data from 
\cite{Hornidge} equally well~-- in fact they nearly lie on top of each other.

\begin{figure}[!htb]
  \begin{center}
    \includegraphics[width=.48\linewidth]{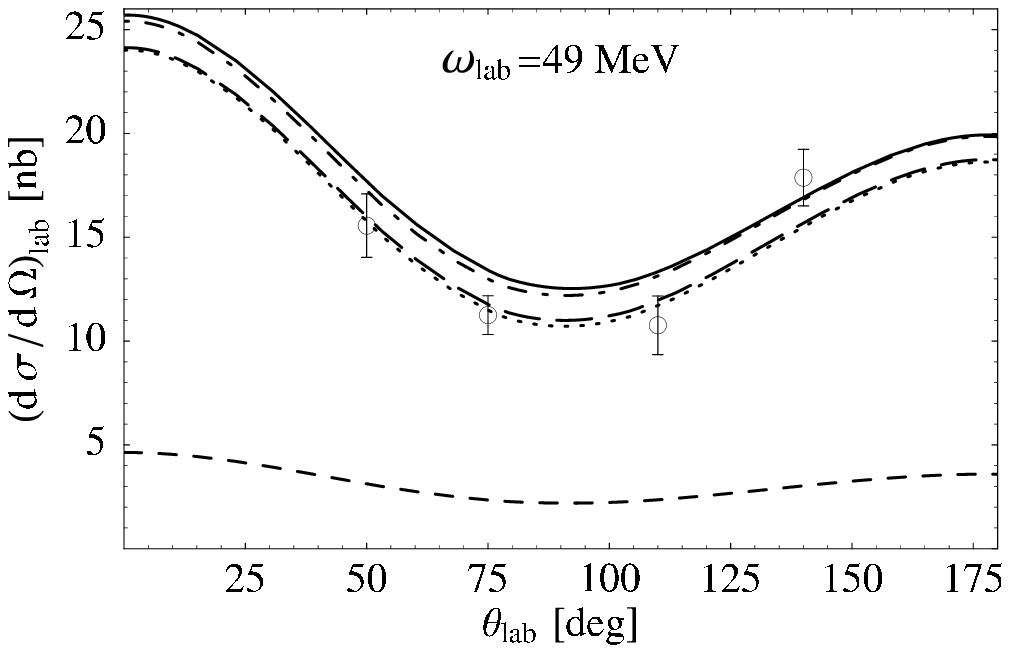}
    \hfill
    \includegraphics[width=.48\linewidth]{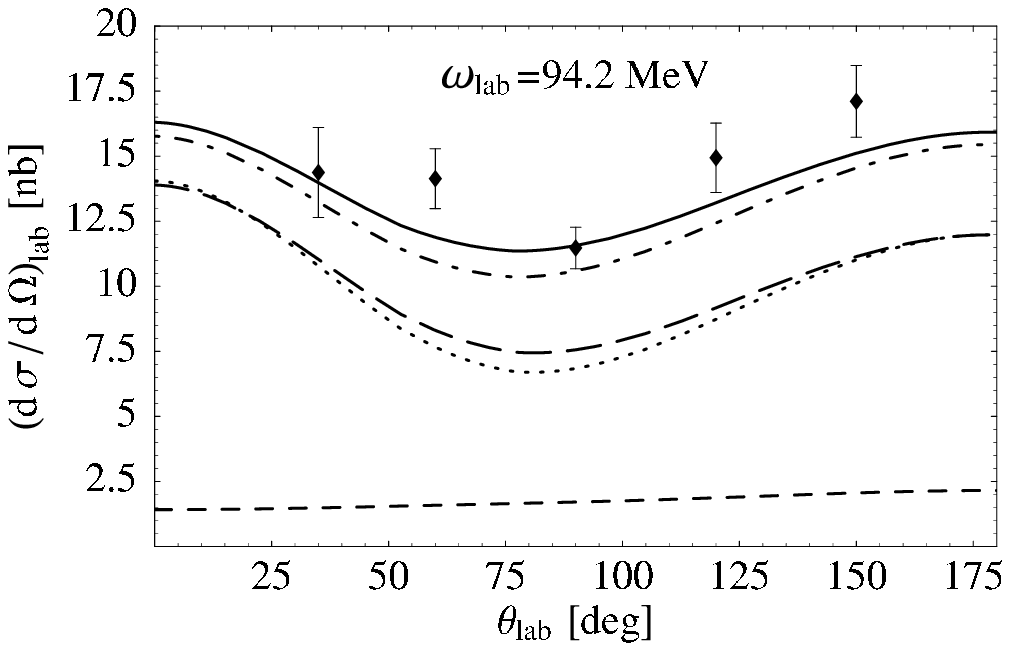}
    \\
    \includegraphics[width=.48\linewidth]{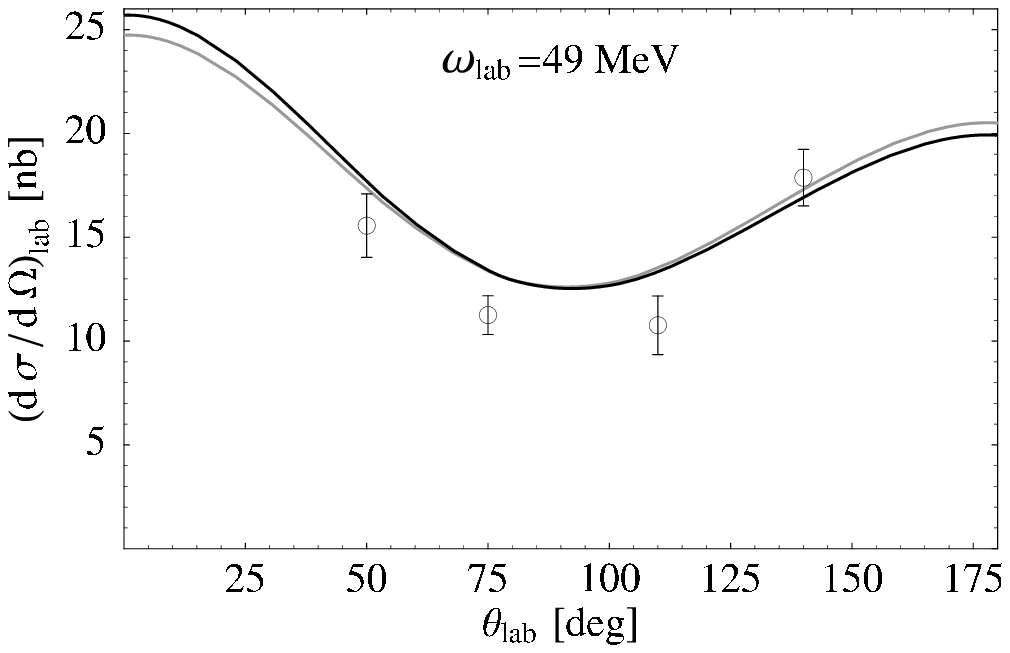}
    \hfill
    \includegraphics[width=.48\linewidth]{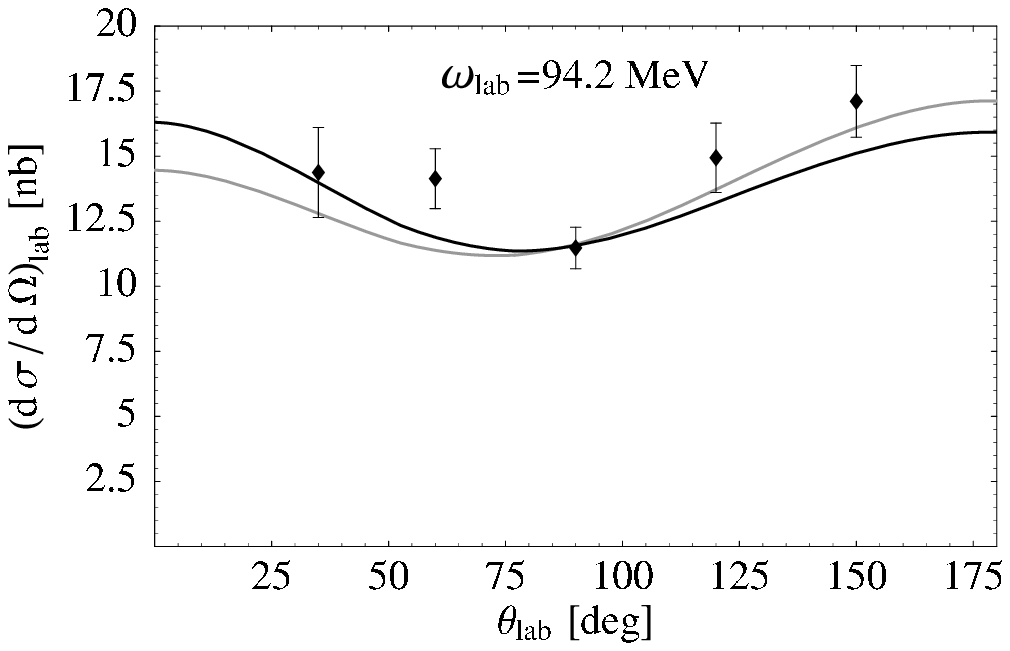}
\parbox{1.\textwidth}{
\caption
{Comparison of separate contributions to our deuteron Compton-scattering 
results. In the upper panels we compare the full result (solid) to curves with 
several amplitudes subtracted; the subtracted amplitudes are:
$\Mfi{\mathrm{KR}}$ (dotdashed), 
$\Mfi{\mathrm{KR}}+\Mfi{\sigma\sigma}$ (dotted),
$\Mfi{\mathrm{KR}}+\Mfi{\sigma\sigma}+\Mfi{\phi\sigma}$ (longdashed),
$\Mfi{\mathrm{KR}}+\Mfi{\sigma\sigma}+\Mfi{\phi\sigma}+\Mfi{\phi\phi1,2,3}$ 
(shortdashed).
In the lower panels, we compare our full result (black) to a curve where 
the amplitudes $\Mfi{\sigma\sigma}$ and $\Mfi{\phi\phi}$ have been replaced 
by their $L=L'=1$-approximations (grey).
The data are from \cite{Lucas}~(circle) and \cite{Hornidge}~(diamond).}
\label{fig:separation}}
\end{center}
\end{figure}

Apart from the total result, we discuss the strength of several contributions
separately.  There are however certain amplitudes which are closely related to
each other, e.g. the kinetic energy part of the double commutator,
Eq.~(\ref{eq:Mfiphiphi4added}), cancels half of the proton seagull in the
static limit, cf. Section~\ref{sec:Thomson}.  The sum of the potential energy
part of the commutator and the nine two-body contributions from
Fig.~\ref{fig:chiPTdouble} is zero in the limit of vanishing photon energy. It
stays small in the whole energy range considered in this work, as already
observed in Refs.~\cite{ArenhoevelII,Karakowski1,Karakowski2}.  Therefore we do not
separate these contributions from each other.  Nevertheless, there are a few
issues worth investigating in more detail:
\begin{itemize}
\item[1)] The prominent role of the amplitudes $\Mfi{\phi\phi1,2}$, cf.
  Eqs.~(\ref{eq:Mfiphiphi1added},~\ref{eq:Mfiphiphi2added}), which include an
  $E1$-interaction at both vertices.
\item[2)] The importance of the amplitudes $\Mfi{\phi\,\sigma}$ and
  $\Mfi{\sigma\sigma}$, with $\sigma$ denoting the coupling to the spin
  current.
\item[3)] The strength of the amplitudes with the explicit Kroll-Ruderman
  current at one vertex, $\Mfi{\mathrm{KR}}$.
\end{itemize}
In the upper two panels of Fig.~\ref{fig:separation}~-- we investigate the two
extreme energies of Fig.~\ref{fig:comparisonpertnonpert}~-- these
contributions are successively added to the remaining terms: the
single-nucleon amplitudes from Figs.~\ref{fig:chiPTsingle} and
\ref{fig:SSEsingle}, the two-nucleon diagrams from Fig.~\ref{fig:chiPTdouble}
and the double-commutator amplitude $\Mfi{\phi\phi4}$, cf.
Eq.~(\ref{eq:Mfiphiphi4added}). The amplitude $\Mfi{\phi\phi3}$
(Eq.~(\ref{eq:Mfiphiphi3added})) is a small correction and has been added to
the leading amplitudes $\Mfi{\phi\phi1,2}$.

Obviously, the amplitudes $\Mfi{\phi\phi}$ are the dominant ones in
Fig.~\ref{fig:separation}.  This observation holds for both energies
considered. However, the amplitudes $\Mfi{\sigma\sigma}$ give also important
contributions.  The same pattern occurs in the calculation of total
deuteron-photodisintegration cross sections, cf.
Appendix~\ref{app:photodisintegration}. The contributions from
$\Mfi{\phi\sigma}$ are nearly negligible. The small size of these terms is due
to the fact that the two amplitudes which arise from coupling the two
non-gradient parts of $\vec{A}$ to the spin current, cf.
Eqs.~(\ref{eq:schematically},~\ref{eq:Jsigma}), largely cancel each other. The
diagrams with one photon explicitly coupling to the Kroll-Ruderman current are
tiny for low energies, but give a sizeable correction at 94.2~MeV. This
contribution is stronger in our calculation than it appears in
\cite{Karakowski1,Karakowski2}, which one may partly attribute to the fact that we do not
neglect the photon energy in the denominator of the pion propagator of the
Kroll-Ruderman current, in contradistinction to~\cite{Karakowski1,Karakowski2}, see
Eqs.~(\ref{eq:fKRs},~\ref{eq:fKRu}).

We also give an estimate of the strength of contributions from photons with 
multipolarity $L=2$, cf.~Eq.~(\ref{eq:multipoleexp}). 
In Ref.~\cite{Karakowski1,Karakowski2}, these next-to-leading terms
in the multipole expansion of the photon field are claimed to be  
small and therefore have been neglected. However, we slightly disagree from
this statement, as can be seen in the lower two panels of 
Fig.~\ref{fig:separation}, where we compare our full results 
to curves which only include the $L=L'=1$-approximation of the dominant 
amplitudes $\Mfi{\phi\phi1,2,3}$ and $\Mfi{\sigma\sigma}$. For low energies,
these corrections are certainly negligible, but they are of the order of 10\% 
in the high-energy regime in the forward and backward direction.

Comparing to Ref.~\cite{Karakowski1,Karakowski2}, we see the main difference to this work
in our systematic $\chi$EFT description of the single-nucleon structure.
In~\cite{Karakowski1,Karakowski2}, the structure of the nucleon is included only via the
static polarizabilities $\alpha_{E}$ and $\beta_{M}$, i.e. via the leading
terms of a Taylor expansion of the single-nucleon Compton amplitudes. In our
work, these amplitudes have been calculated up to third order in the Small
Scale Expansion, as explained in detail in Refs.~\cite{HGHP,deuteronpaper},
and are included with their full energy dependence. Another advantage of our
approach is the treatment of the pion propagator in the pion-exchange diagrams
of Fig.~\ref{fig:chiPTdouble}.  We calculate these diagrams using the full
pion propagator, whereas the authors of \cite{Karakowski1,Karakowski2} always assume that
the photon energy is small compared to the energy of the virtual pion and
therefore may be neglected. This, however, is no longer a good approximation
as the photon energy approaches the pion mass. A similar difference occurs in
the Kroll-Ruderman amplitudes, as discussed above. Finally, we do not agree
with the statement of \cite{Karakowski1,Karakowski2} that $L=2$-contributions are
negligible for all amplitudes and energies considered, see
Fig.~\ref{fig:separation}.

\begin{figure}[!htb]
  \begin{center}
    \includegraphics*[width=.6\linewidth]{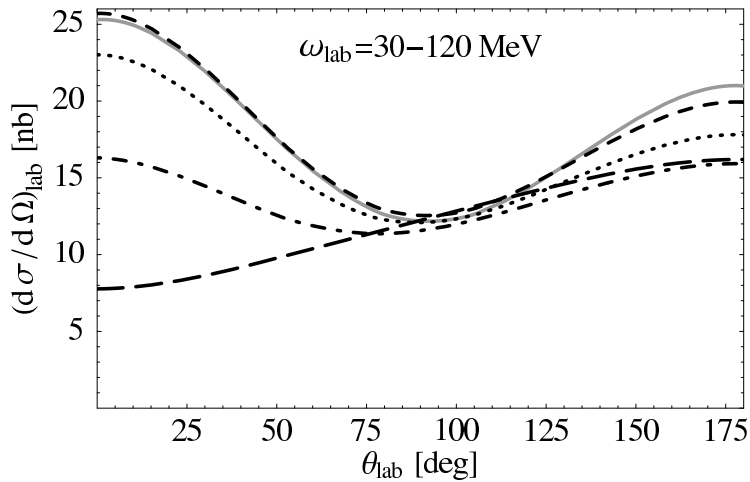}
\parbox{1.\textwidth}{
\caption{Comparison of our results for differential Compton cross sections at 
various energies: 30~MeV (grey), 49~MeV (shortdashed), 68~MeV (dotted), 
94.2~MeV (dotdashed), 120~MeV (longdashed).}
\label{fig:energies}}
\end{center}
\end{figure}

We showed in the last two sections that our calculation provides a consistent
description of elastic deuteron Compton scattering below $100$~MeV.  It will
be interesting to see whether data from a forthcoming experiment at MAXlab at
$\w\sim120$~MeV are closer to the Green's-function hybrid approach or
Weinberg's hybrid approach.  There, however, corrections due to the
kinematically correct position of the pion-production threshold in analogy to
Ref.~\cite{HGHP}, which are not yet included in our calculation, should not be
neglected.  Therefore, the 120~MeV curve in the direct comparison of our
results at various energies in Fig.~\ref{fig:energies} is only a qualitative
statement about the behavior of the differential cross section for
$\w\rightarrow m_\pi$.  Hildebrandt et al.~\cite{deuteronpaper} estimate this
effect to be negligible below 100~MeV.  We also show our prediction at
$\w_\mathrm{lab}=30$~MeV, which might be considered as a guideline to
forthcoming experiments in this energy region~\cite{Feldman:2008zz,Feldman2}.  It is
comparable in magnitude to the 49~MeV curve. Its shape, however, is less
asymmetric between the forward and the backward direction, due to the exact
forward-backward symmetry of the static cross section, cf.
Fig.~\ref{fig:Thomson2}.

In the next two sections, we investigate the sensitivity of our results to the
deuteron wave function and the $NN$-potential used.

\subsection{Dependence on the Deuteron Wave Function 
\label{sec:wavefunctiondep}}

\begin{figure}[!htb]
  \begin{center}
    \includegraphics*[width=.48\linewidth]{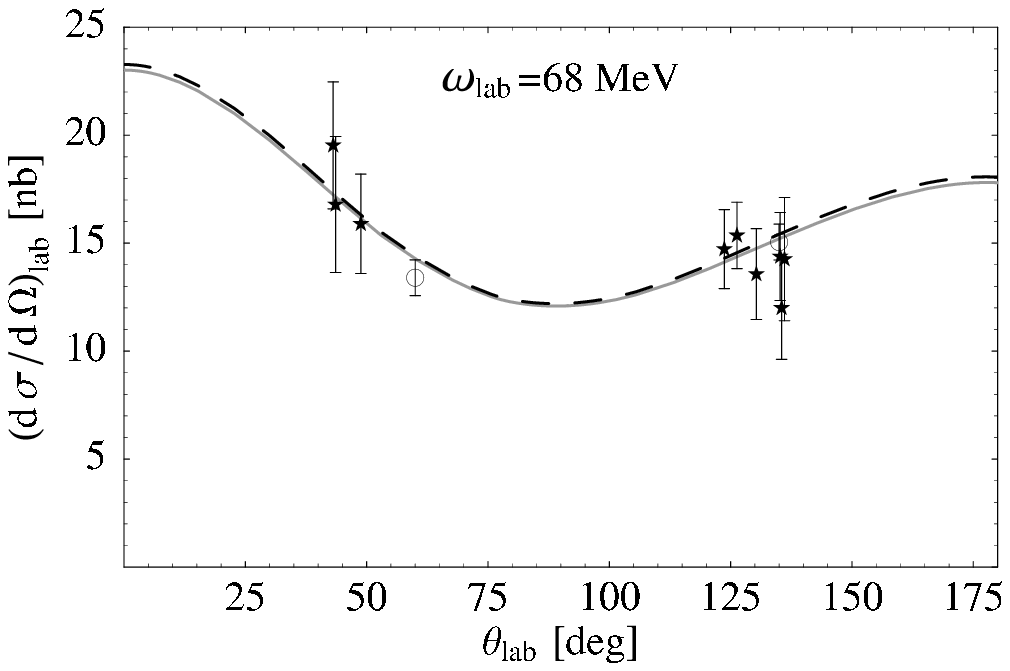}
    \hfill
    \includegraphics*[width=.48\linewidth]{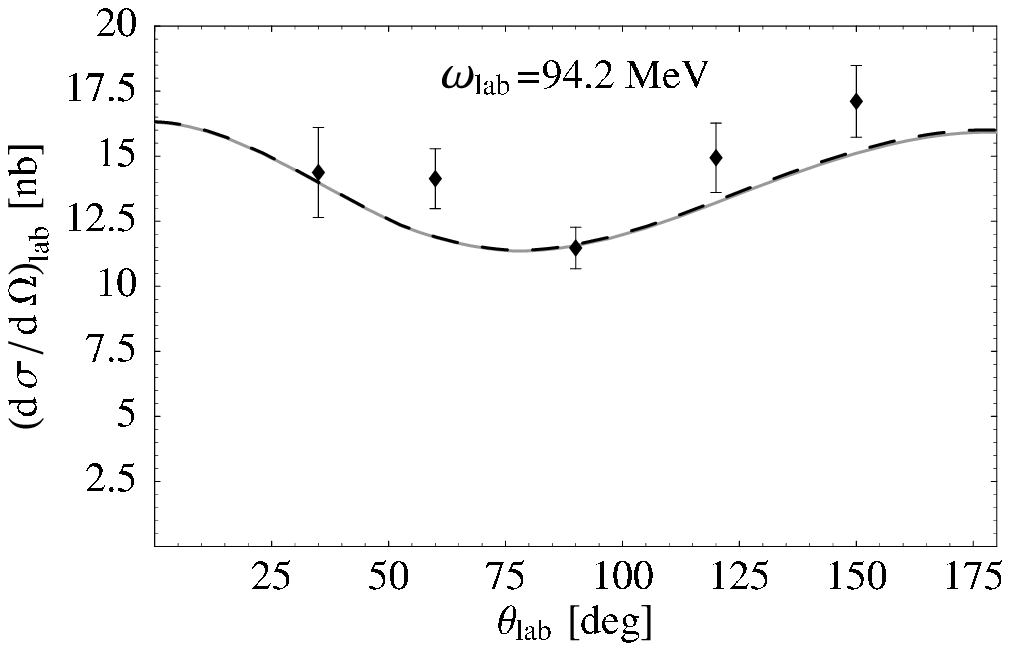}
    \includegraphics*[width=.48\linewidth]{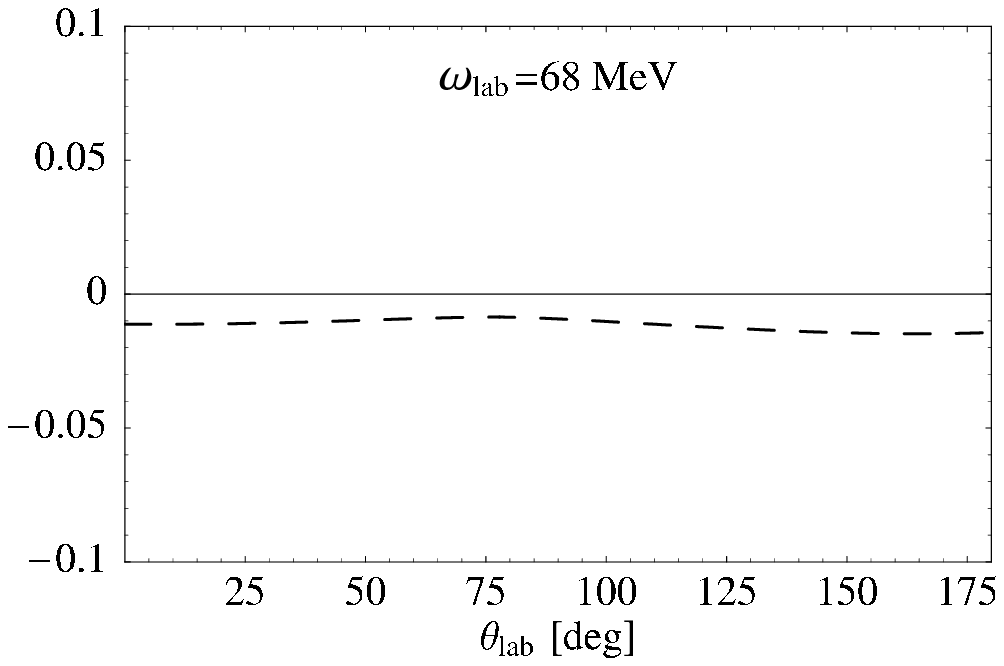}
    \hfill
    \includegraphics*[width=.48\linewidth]{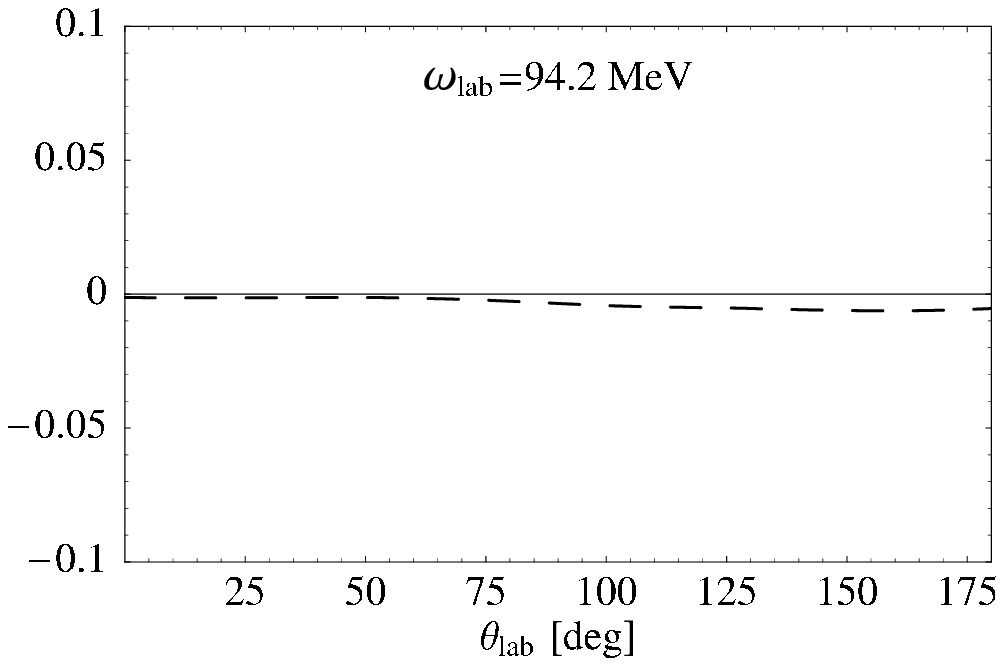}
    \caption{Comparison of our deuteron Compton cross-section results for 68
      and 94.2~MeV, using two different wave functions: NNLO $\chi$PT
      (grey)~\cite{Epelbaum} and AV18 (dashed)~\cite{AV18}. In the lower two
      panels we show $\left(\frac{d\sigma}{d\Omega}\right)_\mathrm{NNLO}/
      \left(\frac{d\sigma}{d\Omega}\right)_\mathrm{AV18}-1$.}
    \label{fig:wavefunctiondep}
  \end{center}
\end{figure}

As demonstrated in Section~\ref{sec:Thomson}, our calculation fulfills 
the low-energy theorem, Eq.~(\ref{eq:Thomsond}), which in turn is independent 
of the deuteron wave function. Therefore and because of the nearly 
energy-independent offset between the cross sections calculated with the 
chiral~\cite{Epelbaum} and the AV18-wave function~\cite{AV18}, observed in 
Refs.~\cite{deuteronpaper,PHD}, it is 
not surprising that the wave-function dependence of our present calculation 
is also at non-zero energies considerably reduced with respect to 
Refs.~\cite{deuteronpaper,McGPhil1,McGPhil2}. In fact, the remaining 
dependence is of the order of 1\% and therefore nearly invisible, 
cf. Fig.~\ref{fig:wavefunctiondep}, where we compare our cross sections 
with the two wave functions that turned out as the extreme ones in 
Ref.~\cite{deuteronpaper}: 
the AV18 \cite{AV18} and the NNLO $\chi$PT \cite{Epelbaum} wave function 
(the same observation holds for other state-of-the-art deuteron wave 
functions). 
This is another important success of our present approach to deuteron Compton 
scattering, as it demonstrates that our calculation is not sensitive to details
of high-energy physics, i.e. short-distance contributions of the wave 
function, whereas the 10\%-effect observed in 
Refs.~\cite{deuteronpaper,McGPhil1,McGPhil2} manifests a much stronger dependence on 
short-distance dynamics than would be expected from a low-energy Effective 
Field Theory.

\subsection{Dependence on the Potential 
\label{sec:potentialdep}}

In this section, we investigate briefly the sensitivity 
to the $np$-potential used for the rescattering part. Usually, we use the 
AV18-potential~\cite{AV18} which provides an excellent theoretical 
description of the Nijmegen partial-wave analysis. We compare here
our results achieved with this modern 'high-precision' potential to those of
the leading-order chiral potential, which includes only the one-pion
exchange and a simple parameterization of short-distance effects via two
point-like, momentum-independent contact operators. This potential is given 
e.g. in Ref.~\cite{Rho}, using a
Gaussian regulator in order to render the pion-exchange potential finite at the
origin. 

At leading order, there are two free parameters, $C_0^{^1\!S_0}$ and $C_0^d$,
with $d$ denoting the (deuteron) $^3\!S_1$-$^3\!D_1$ channel. $C_0^{^1\!S_0}$
has been fixed in~\cite{Rho} via the $^1\!S_0$ scattering length,
$a_0\approx-23.75$~fm, $C_0^d$ to the deuteron binding energy, parameterizing
any short-distance physics in the spin-singlet and -triplet channel,
respectively.  $C_0^{^1\!S_0}$ and $C_0^d$, as given in Ref.~\cite{Rho} for
the cutoff-value $\Lambda=600$~MeV, are reported in Table~\ref{tab:Rho}.
\begin{table}[!htb]
  \begin{center}
    \begin{tabular}{|l|l|}
      \hline 
      $C_0^{^1\!S_0}$&$C_0^d$\\
      \hline
      $-0.422$~fm&$0.795$~fm\\
      \hline 
    \end{tabular}
  \end{center}
  \caption{Parameters of the LO chiral potential as determined in~\cite{Rho}
    for $\Lambda=600$~MeV.}
  \label{tab:Rho}
\end{table}

\begin{figure}[!htb]
  \begin{center}
    \includegraphics*[width=.48\linewidth]{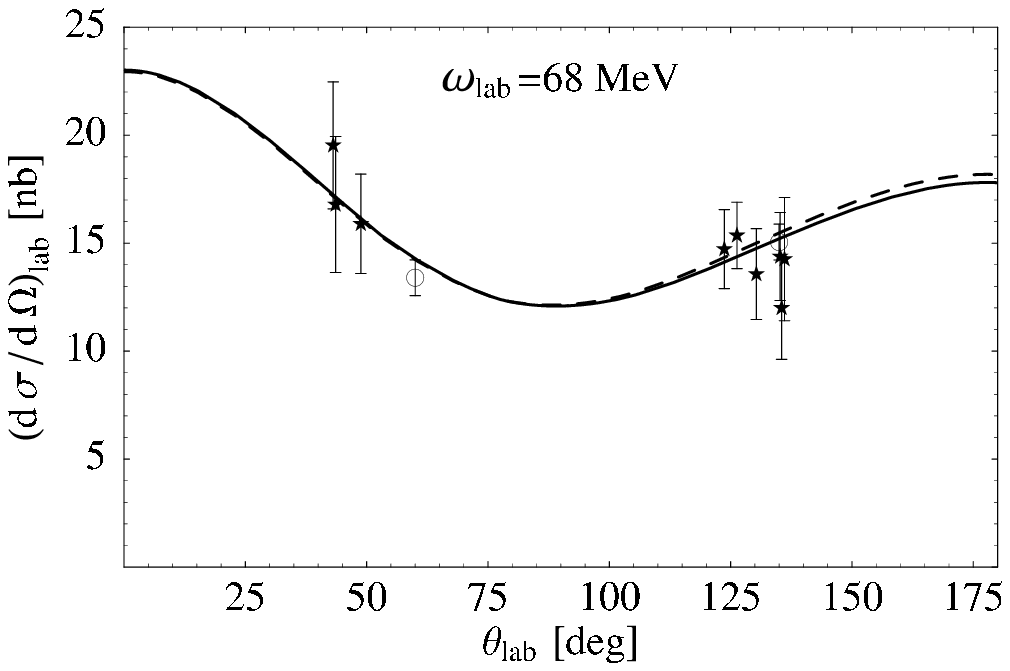}
    \hfill
    \includegraphics*[width=.48\linewidth]{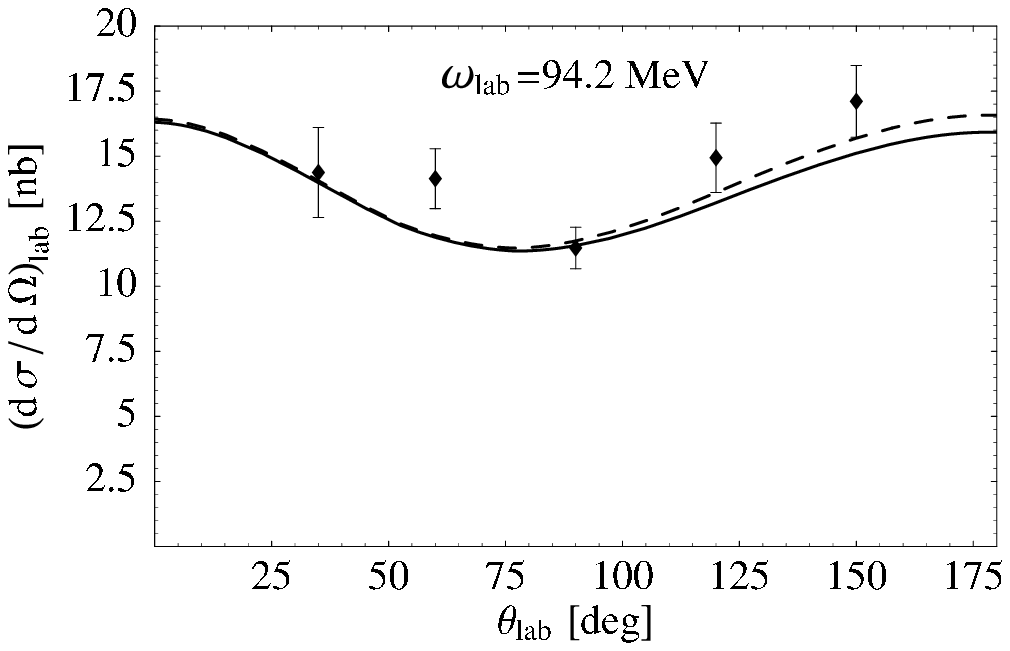}
    \includegraphics*[width=.48\linewidth]{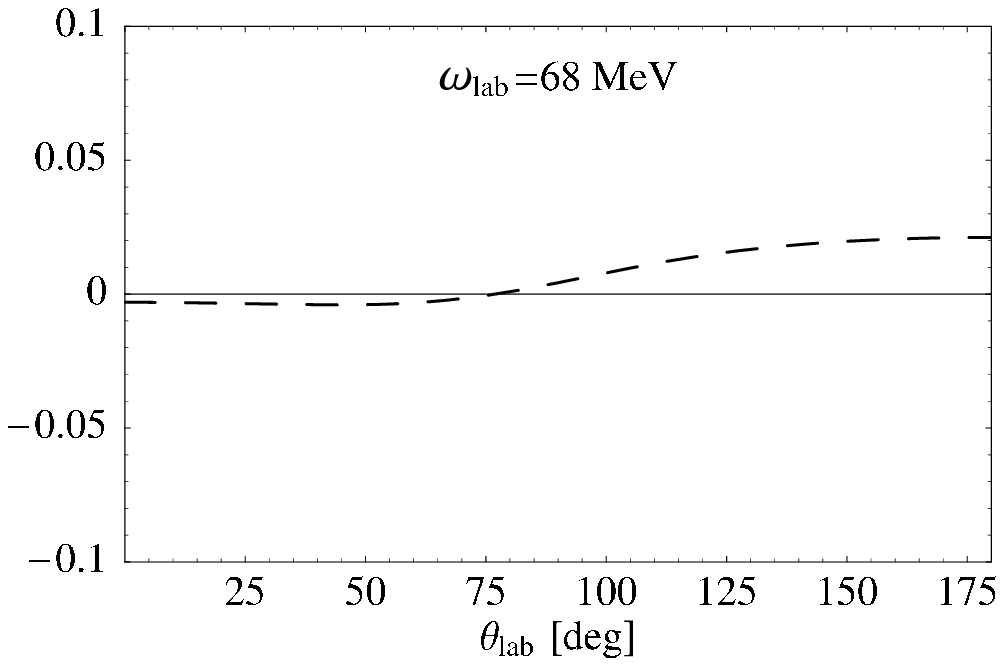}
    \hfill
    \includegraphics*[width=.48\linewidth]{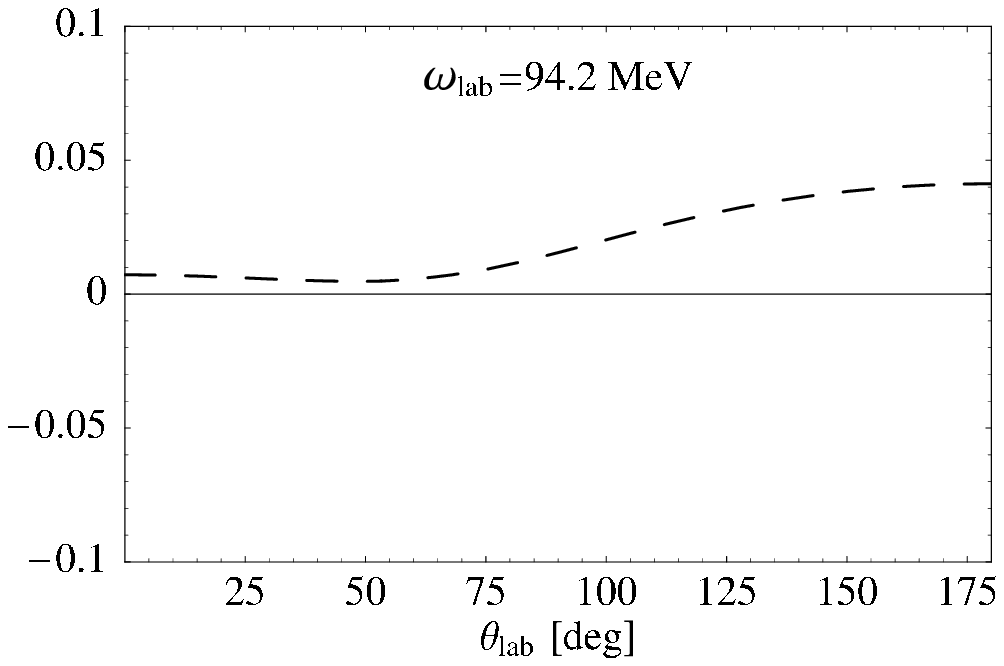}
    \caption{Upper panels: Comparison of our result at 68~MeV and 94.2~MeV
      using two different $np$-potentials: the AV18-potential~\cite{AV18}
      (solid) and the LO chiral potential~\cite{Rho} (dashed). For both curves
      the chiral wave function~\cite{Epelbaum} has been used. Lower panels:
      Corresponding error plots, calculated in analogy to
      Fig.~\ref{fig:Thomson2}.}
    \label{fig:LOchiPT}
  \end{center}
\end{figure}

Even with this rather crude approximation of the neutron-proton interaction,
we obtain results close to those of the AV18-potential, cf.
Fig.~\ref{fig:LOchiPT}. Obviously, the one-pion-exchange potential, adequately
regulated for $r\rightarrow 0$, together with a reasonable parameterization of
the hard core gives an approximation of the potential which is well sufficient
for the process under consideration. We conclude that we are mainly sensitive
to the long-range part of the potential.  Nevertheless, there are small
deviations (of the order of $\leq4\%$) visible in Fig.~\ref{fig:LOchiPT},
which justify the application of a more sophisticated potential.  These
deviations increase with increasing photon energy and scattering angle.
However, the correct Thomson limit is reached with any combination of
wave-function and two-nucleon rescattering potential, as shown by
Arenh\"ovel~\cite{Arenhoevel} and recalled in Sec.~\ref{sec:Thomson}.
Therefore, deviations coming from using different wave-functions or potentials
must necessarily increase with increasing energy.  One can therefore not
simply determine if the seed of discrepancies is a difference between LO
potential and phase shifts at low energies, and/or if the different curves
arise from the well-explored poor LO-description of the ${}^1\text{S}_0$,
${}^1\text{P}_1$ and ${}^3\text{P}_{0,2}$ partial waves at energies $\sim\mpi$
for some cutoffs, see e.g.~\cite{Rho,Nogga,Birse}.  The situation is
complicated by the fact that rescattering contributions are small at high
energies, see Sec.~\ref{sec:comparison}.  The error induced by neglecting them
is comparable to that from using the cruder LO potential in
Fig.~\ref{fig:LOchiPT} but has a different angular dependence.

Having proven that our calculation is rather insensitive to the choice of the
deuteron wave function and the $np$-potential, we turn now to fits of the
isoscalar nucleon polarizabilities (or equivalently the short-distance
contributions $g_{117}$ and $g_{118}$) to all existing elastic $\gamma d$
data.

\section{Fits of the Isoscalar Polarizabilities}
\setcounter{equation}{0}
\label{sec:fits}

We saw in Section~\ref{sec:comparison} that
our results for the elastic deuteron Compton cross sections give a good 
description of all existing data. Furthermore, the cross-check described in 
Appendix~\ref{app:photodisintegration}, i.e. extracting the total 
deuteron-photodisintegration cross section from our
Compton amplitude, together with the exact 
reproduction of the low-energy theorem, cf. Section~\ref{sec:Thomson}, gives 
a strong hint that the numerically most important amplitudes have been 
calculated correctly.
Therefore, we now use our deuteron Compton cross sections to \textit{fit}
the static isoscalar nucleon polarizabilities $\alpha_{E}^s$ and $\beta_{M}^s$
to elastic $\gamma d$ experiments. This corresponds to fitting the coupling 
constants $g_{117}$ and $g_{118}$
of the two short-distance operators, Fig.~\ref{fig:SSEsingle}(f), to elastic 
$\gamma d$ rather than $\gamma p$ data. We can  use \textit{all} data 
for the fits, whereas the authors of Ref.~\cite{deuteronpaper} had to restrict 
themselves to the experiments performed around 68 and 94.2~MeV and those
of Ref.~\cite{McGPhil1,McGPhil2} excluded the two 94.2~MeV data in the backward 
direction in certain fits. 

\begin{figure}[!htb]
  \begin{center}
    \includegraphics*[width=.32\linewidth]{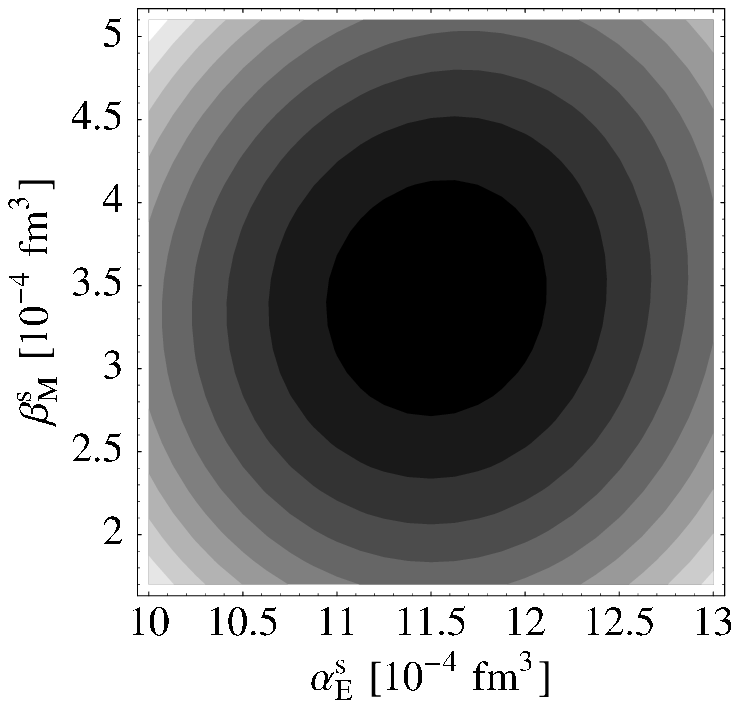}
    \hspace{.1\linewidth}
    \includegraphics*[width=.32\linewidth]{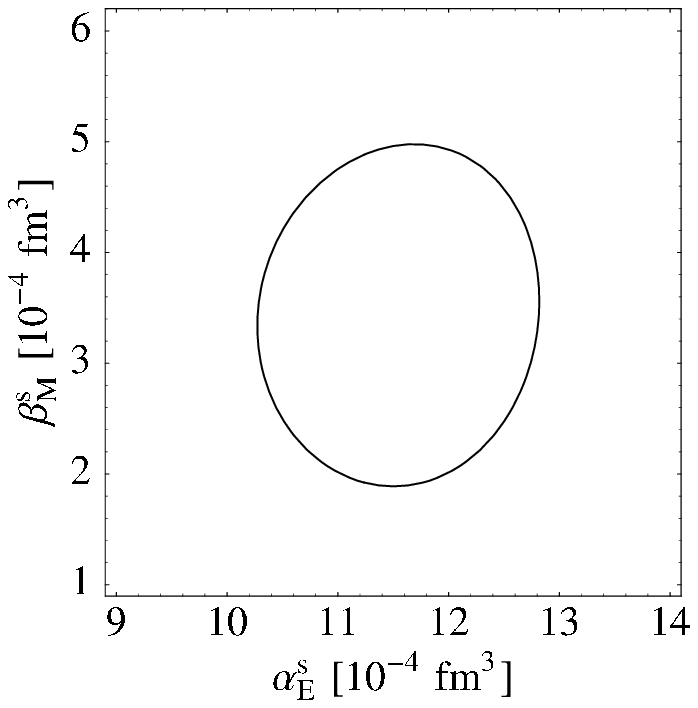}
    \caption{Contour plot of the $\chi^2$ achieved in our global 2-parameter
      fit for $\alpha_{E}^s$ and $\beta_{M}^s$ (left panel) and the
      70\%-confidence ellipse (right panel). The shadings in the left panel
      are separated by steps of $\Delta\chi^2=0.67$.}
    \label{fig:chisquareplot}
  \end{center}
\end{figure}

We do a least-$\chi^2$ fit, using the chiral NNLO wave function with
$\Lambda=650$~MeV~\cite{Epelbaum}.  Our results for the isoscalar
polarizabilities from the global fit to all data read
\begin{align}
  \alpha_{E}^s&=(11.5\pm1.4\,(\mathrm{stat}))\cdot10^{-4}\;\fm^3,\nonumber\\
  \beta_{M}^s &=( 3.4\pm1.6\,(\mathrm{stat}))\cdot10^{-4}\;\fm^3.
  \label{eq:globalfit}
\end{align}
For now, we only give the statistical error and postpone a discussion
of sources and sizes of theoretical errors to the Conclusions. The
corresponding $\chi^2$ per degree of freedom is
\begin{equation}
  \frac{\chi^2}{\text{d.o.f.}}=0.98
\end{equation}
with 27 degrees of freedom (4 data points from \cite{Lucas} at 49~MeV, 9 from
\cite{Lund} at 55~MeV, 2 from \cite{Lucas}, 9 from \cite{Lund} around 68~MeV
and 5 from \cite{Hornidge} at 94.2~MeV, along with two fit parameters). In
Fig.~\ref{fig:chisquareplot}, we give a contour plot of the achieved $\chi^2$,
together with the 70\%-confidence ellipse.

\begin{figure}[!htb]
  \begin{center}
    \includegraphics*[width=.48\linewidth]{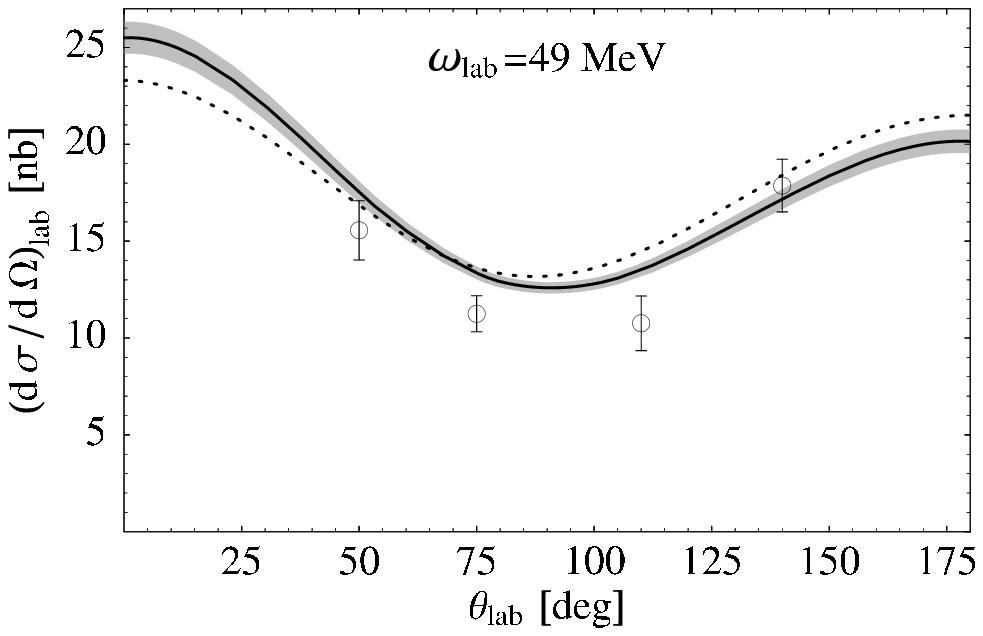}
    \hfill
    \includegraphics*[width=.48\linewidth]{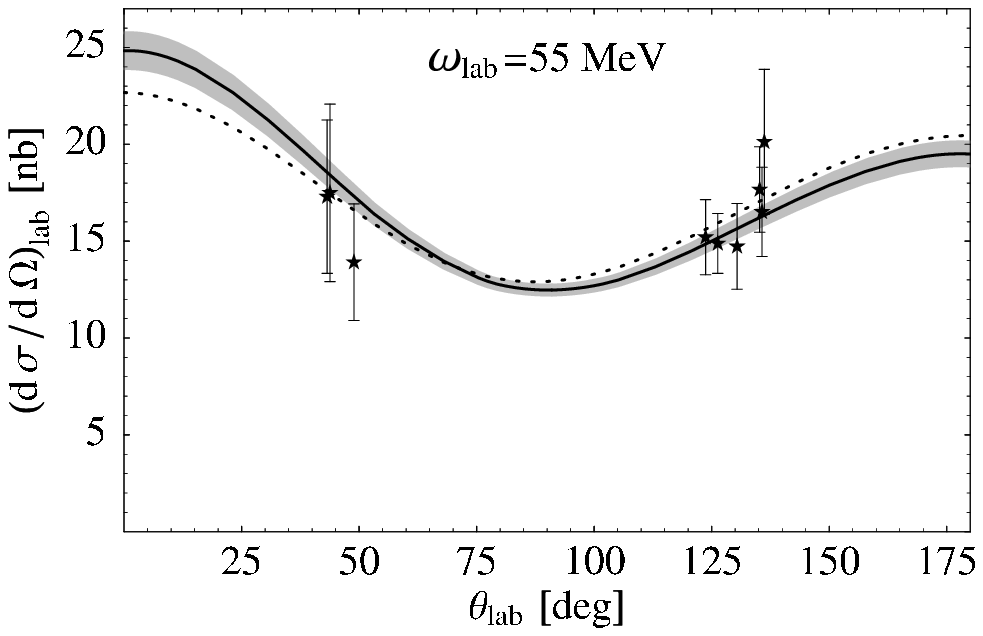}
    \\
    \includegraphics*[width=.48\linewidth]{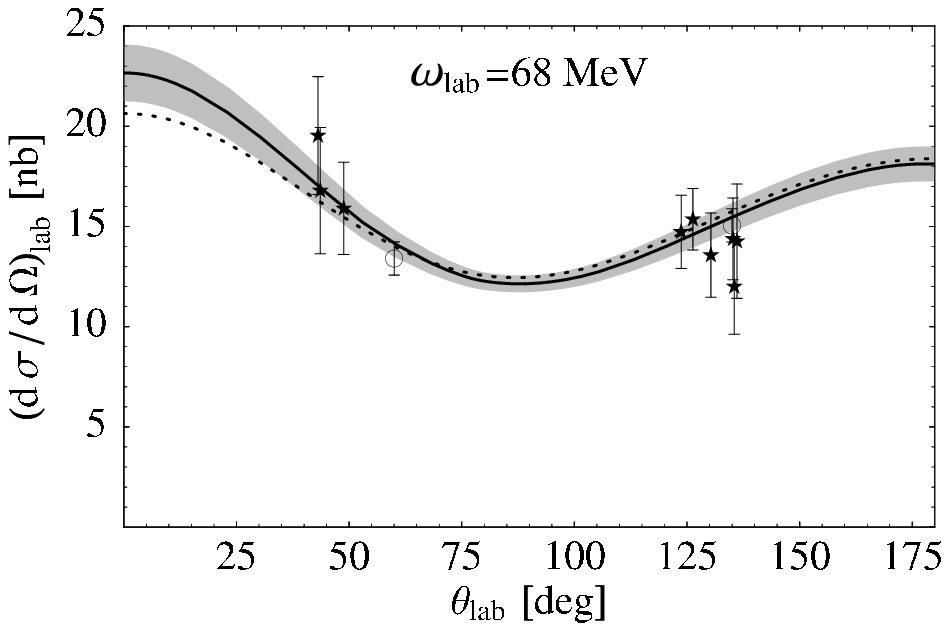}
    \hfill
    \includegraphics*[width=.48\linewidth]{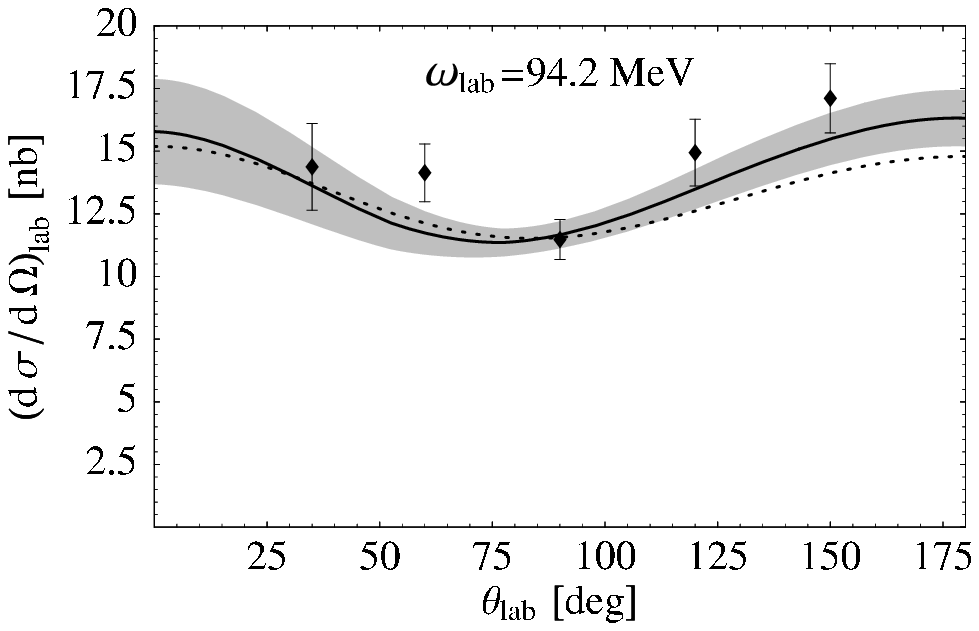}
    \\
    \caption{Results from a global fit of $\alpha_{E}^s$ and $\beta_{M}^s$ to
      all existing elastic $\gamma d$ data (solid). The grey bands are derived
      from our statistical errors. The dotted line represents ``fit IV'', one
      of the $\mathcal{O}(q ^4)$-HB$\chi$PT fits from Ref.~\cite{McGPhil1,McGPhil2},
      with central values $\alpha_{E}^s=11.5\cdot10^{-4}\;\mathrm{fm}^3$,
      $\beta _{M}^s= 0.3\cdot10^{-4}\;\mathrm{fm}^3$.  For the $\mathcal{O}(q
      ^4)$ calculation the NLO chiral wave function of Ref.~\cite{NLO} has
      been used, whereas our results were derived with the NNLO-version of
      this wave function~\cite{Epelbaum}.}
    \label{fig:2parameterfit2}
  \end{center}
\end{figure}

\begin{figure}[!htb]
  \begin{center}
    \includegraphics*[width=.48\linewidth]{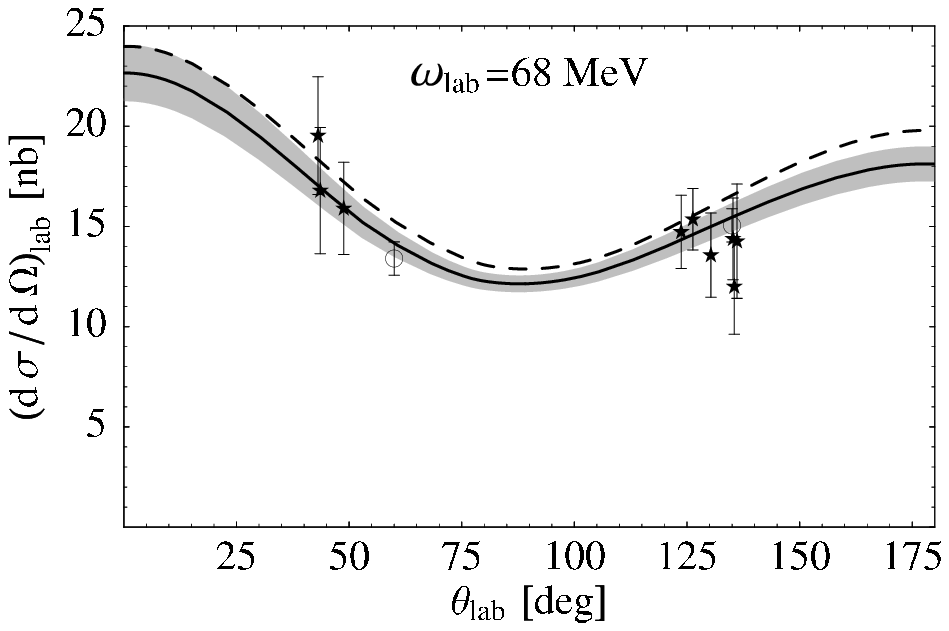}
    \hfill
    \includegraphics*[width=.48\linewidth]{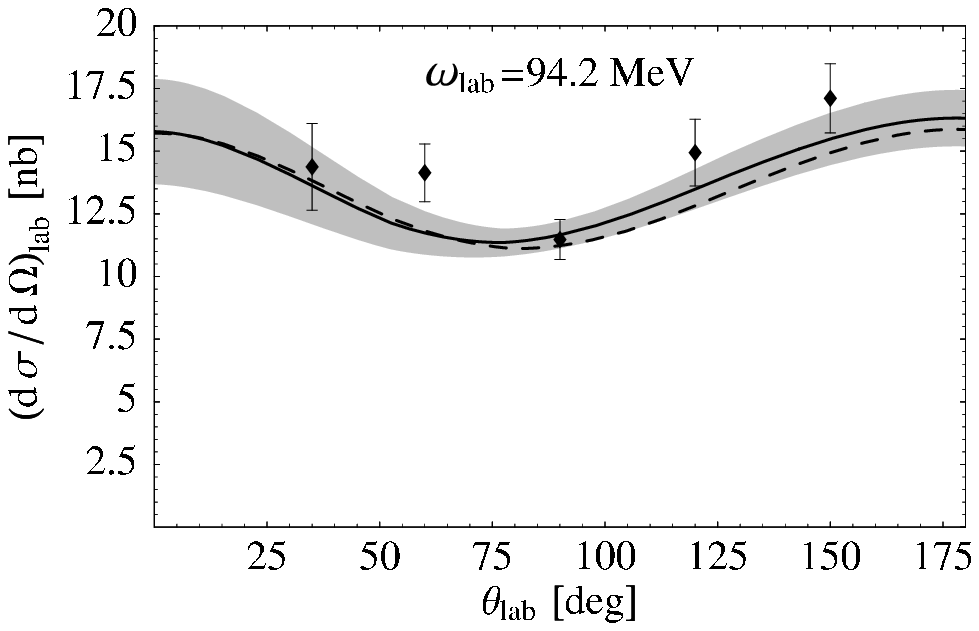}
    \caption{Results from a global fit of $\alpha_{E}^s$ and $\beta_{M}^s$ to
      all existing elastic $\gamma d$ data (solid), obtained with the NNLO
      chiral wave function~\cite{Epelbaum}. The grey bands are derived from
      our statistical errors. We compare here our results to the
      2-parameter-fit results from~\cite{deuteronpaper}, using the same chiral
      wave function (dashed).}
    \label{fig:2parameterfit3}
  \end{center}
\end{figure}

The plots corresponding to Eq.~(\ref{eq:globalfit}), together with the
(statistical) error bands, are shown in Fig.~\ref{fig:2parameterfit2}. The
predictions within our Green's-function hybrid approach, using the results for
$\alpha_{E}^p$ and $\beta_{M}^p$ from Eq.~(\ref{eq:exppHGHP}) for the proton
\textit{and} the neutron polarizabilities, describe the data already well, see
Fig.~\ref{fig:comparisonpertnonpert}.  It is therefore no surprise that also
the fitted curves are in good agreement with experiment. We compare our fit
results to ``fit IV'' from Ref.~\cite{McGPhil1,McGPhil2}, which is the $\mathcal{O}(q
^4)$ HB$\chi$PT fit to all data with central values
$\alpha_{E}^s=11.5\cdot10^{-4}\;\mathrm{fm}^3$,
$\beta_{M}^s=0.3\cdot10^{-4}\;\mathrm{fm}^3$. As explained in detail
in~\cite{deuteronpaper}, the only sizeable deviations are observed at 94.2~MeV
in the backward direction, due to the $\Delta$-resonance diagram,
Fig.~\ref{fig:SSEsingle}(a), which is not included in the calculation of
Ref.~\cite{McGPhil1,McGPhil2}.  In Fig.~\ref{fig:2parameterfit3} and
Table~\ref{tab:fits}, we also compare to the 2-parameter fit from
Ref.~\cite{deuteronpaper}, which was performed with the $\gamma d\rightarrow
\gamma d$ kernel according to third-order SSE, using the chiral wave
function~\cite{Epelbaum}. These curves correspond to the fits performed with
the ``effective'' data set, cf.~Table~3 and Fig.~11 in~\cite{deuteronpaper}.
Here we observe a constant offset in the differential cross section at 68~MeV,
whereas at 94.2~MeV the two curves are quite close to each other, similarly to
Fig.~\ref{fig:comparisonpertnonpert}.

\begin{figure}[!htb]
  \begin{center}
    \includegraphics*[width=.48\linewidth]{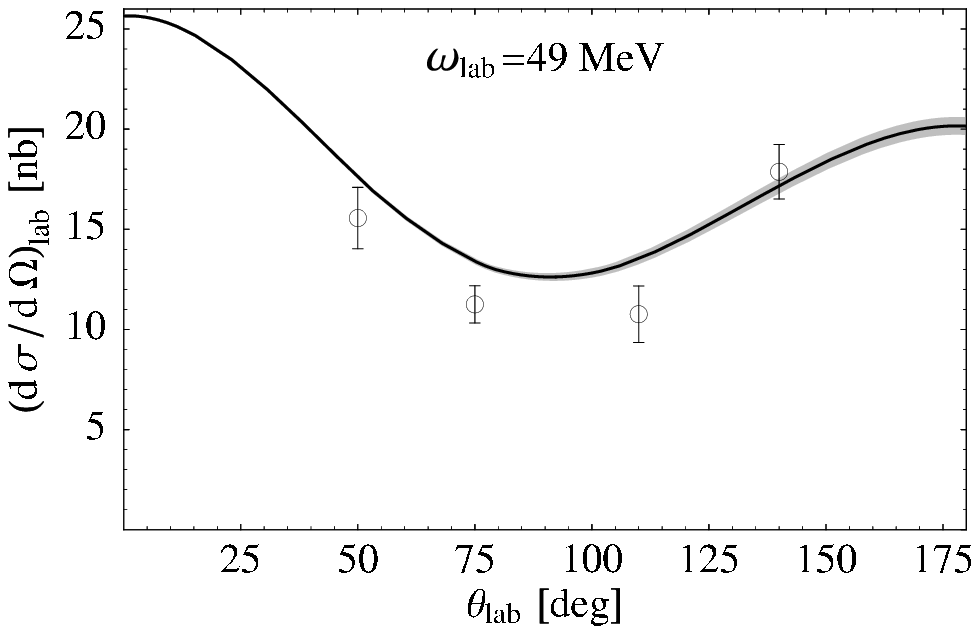}
    \hfill
    \includegraphics*[width=.48\linewidth]{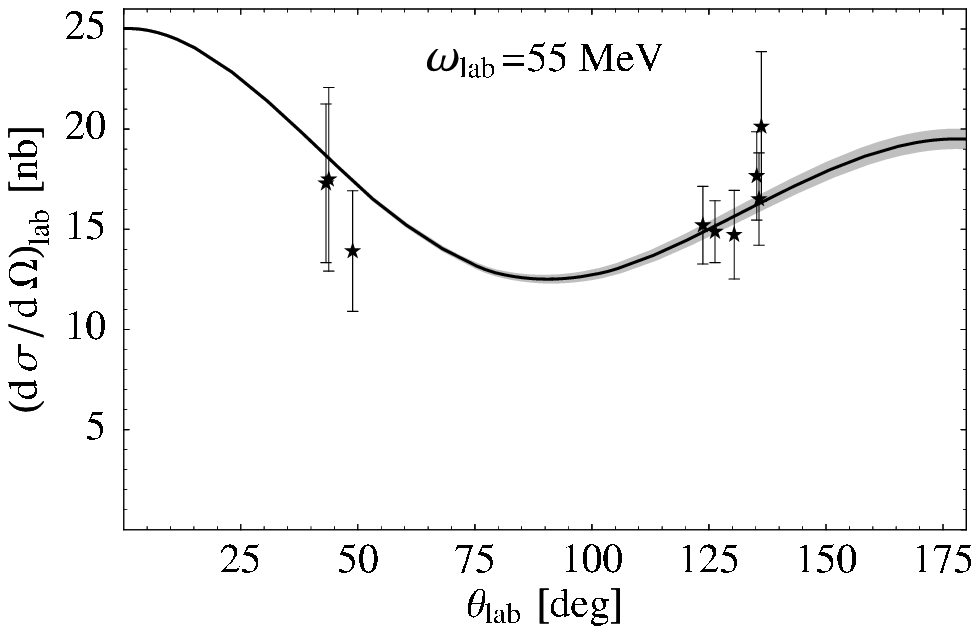}
    \\
    \includegraphics*[width=.48\linewidth]{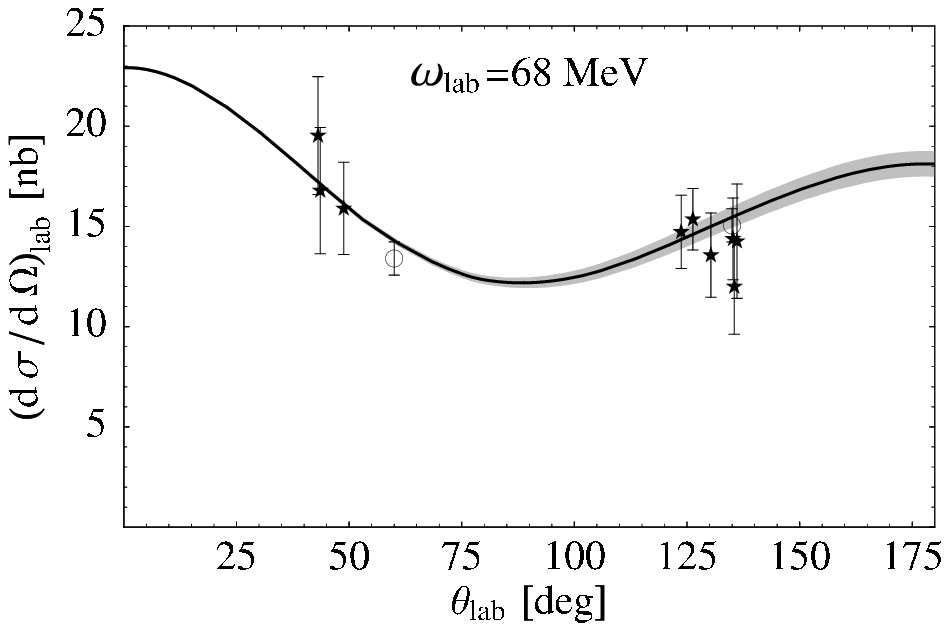}
    \hfill
    \includegraphics*[width=.48\linewidth]{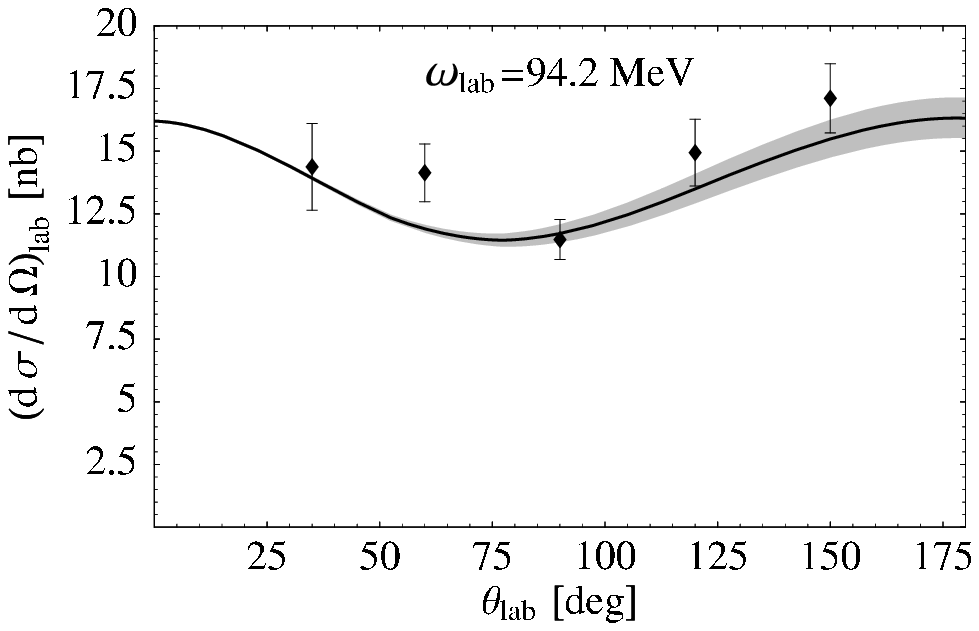}
    \caption{Results from a global fit of $\alpha_{E}^s$ to all existing
      elastic $\gamma d$ data, using the chiral wave function~\cite{Epelbaum}.
      $\beta_{M}^s$ is fixed via the Baldin sum rule, Eq.~(\ref{eq:Baldin}).
      The grey bands are derived from our statistical errors.}
    \label{fig:2parameterfitBaldin2}
  \end{center}
\end{figure}

The value of our global fit for $\alpha_{E}^s$ is slightly smaller, the 
one for $\beta_{M}^s$ slightly larger than the fit results of  
Ref.~\cite{deuteronpaper}, given in Eq.~(\ref{eq:deuteron1fits}) and in 
Table~\ref{tab:fits}, respectively. Nevertheless, both extractions agree 
well with each other within their error bars, and there is also very good 
agreement of Eq.~(\ref{eq:globalfit}) with the values 
given in~\cite{Schumacher}, 
see Eqs.~(\ref{eq:reviewp}) and~(\ref{eq:reviewn}). 
Furthermore, we find that the numbers given in Eq.~(\ref{eq:globalfit}) add up
nearly exactly to the isoscalar Baldin sum rule (see~\cite{deuteronpaper} 
for the proton and neutron sums used as input),
\begin{equation}
  \left.\phantom{\PCsq}\alpha_{E}^s+\beta_{M}^s
  \right|_{\mathrm{world}\;\, \mathrm{av.}}=
  (14.5\pm0.6)\cdot10^{-4}\;\mathrm{fm}^3,
  \label{eq:Baldin}
\end{equation}
which has been a serious problem in former extractions~\cite{Lvov,McGPhil1,McGPhil2}.
Therefore, in order to reduce the statistical error, we repeat our global fit,
using the central sum-rule value as an additional fit constraint. The results
are
\begin{align}
  \left.\phantom{\PCsq}\alpha_{E}^s\right|_\mathrm{Baldin}&=
  (11.3\pm0.7\,(\mathrm{stat})\pm0.6\,(\mathrm{Baldin}))\cdot10^{-4}\;\fm^3,
  \nonumber\\
  \left.\phantom{\PCsq}\beta_{M}^s \right|_\mathrm{Baldin}&= (
  3.2\mp0.7\,(\mathrm{stat})\pm0.6\,(\mathrm{Baldin}))\cdot10^{-4}\;\fm^3
  \label{eq:globalfitBaldin}
\end{align}
with $\chi^{2}/d.o.f.=0.95$.  Of course, the central values of
Eq.~(\ref{eq:globalfitBaldin}) are very similar to the ones of
Eq.~(\ref{eq:globalfit}), due to the nearly perfect agreement of the
2-parameter-fit result with the sum-rule value. However, the statistical error
of the Baldin-constrained fit is reduced by about 50\%. The plots arising from
the global, Baldin-constrained fit, together with the corresponding error
bars, are shown in Fig.~\ref{fig:2parameterfitBaldin2}.  The central curves
are nearly indistinguishable from the ones of Fig.~\ref{fig:2parameterfit2}.
In order to simplify comparison, we sum up our fit results, together with
those from Refs.~\cite{deuteronpaper,McGPhil1,McGPhil2} in Table~\ref{tab:fits}.
Obviously, all three extractions of the electric polarizability agree with
each other, whereas determining $\beta_M$ without the explicit inclusion of
$\Delta$ resonance degrees of freedom yields a negative central
value~\cite{McGPhil1,McGPhil2}, in contradiction to~\cite{Schumacher,deuteronpaper} and
the present work, which show also good agreement in this quantity.

\begin{table}[!htb] 
  \begin{center}
    \begin{tabular}{|c||l|l||l|l|}
      \hline 
      \rule[-1.2ex]{0ex}{3.5ex}
      &\multicolumn{2}{|c||}{$\alpha_E^s\;[10^{-4}\;\mathrm{fm}^3]$}
      &\multicolumn{2}{|c|}{$\beta_M^s\;[10^{-4}\;\mathrm{fm}^3]$}\\
      \hline
      \rule[-1.2ex]{0ex}{3.5ex}
      &2-parameter fit&1-parameter fit&2-parameter fit&1-parameter fit\\
      \hline
      \hline
      \rule[-1.2ex]{0ex}{3.5ex}
      Ref.~\cite{McGPhil1,McGPhil2}
      &$13.0\pm1.9^{+3.9}_{-1.5}$&
      &$-1.8\pm1.9^{+2.1}_{-0.9}$&
      \\
      \hline 
      \rule[-1.2ex]{0ex}{3.5ex}
      Ref.~\cite{deuteronpaper}
      &$     12.8\pm1.4\pm1.1$&$     12.6\pm0.8\pm0.7$
      &$\;\;\:2.1\pm1.7\pm0.1$&$\;\;\:1.9\mp0.8\mp0.7$
      \\
      \hline
      \hline
      \rule[-1.2ex]{0ex}{3.5ex}
      this work
      &$     11.5\pm1.4$&$     11.3\pm0.7$
      &$\;\;\:3.4\pm1.6$&$\;\;\:3.2\mp0.7$
      \\
      \hline
    \end{tabular}
  \end{center}
  \caption{Comparison of the fit results for the isoscalar nucleon
    polarizabilities $\alpha_E^s$ and $\beta_M^s$, achieved within
    $\calO(\epsilon^3)$ SSE~\cite{deuteronpaper} (Weinberg hybrid approach),
    $\calO(p^4)$ HB$\chi$PT~\cite{McGPhil1,McGPhil2} (Weinberg hybrid approach) and the
    Green's-function hybrid approach presented in this work, respectively. The
    first error-bar is statistical; a second error bar denotes the systematic
    uncertainty from the wave-function dependence~\cite{deuteronpaper,McGPhil1,McGPhil2}
    and the arbitrariness as to which data are included in the
    fit~\cite{McGPhil1,McGPhil2}. In the one-parameter fits, the error induced by the
    Baldin sum rule is not shown.}
  \label{tab:fits}
\end{table}

Combining Eqs.~(\ref{eq:globalfit}) or~(\ref{eq:globalfitBaldin}), 
respectively, with the (Baldin-constrained) fit results of 
Eq.~(\ref{eq:exppHGHP}), which we extracted from proton data using the SSE
framework, we calculate the values for the neutron polarizabilities as 
\begin{align}
  \alpha_{E}^n&=
  (12.0\pm2.0\,(\mathrm{stat})\pm0.4\,(\text{Baldin}))\cdot10^{-4}\;\fm^3,
  \nonumber\\
  \beta_{M}^n &= (
  4.0\pm2.1\,(\mathrm{stat})\pm0.4\,(\text{Baldin}))\cdot10^{-4}\;\fm^3
  \label{eq:globalfitneutron}
\end{align}
for the 2-parameter fit and
\begin{align}
  \left.\phantom{\PCsq}\alpha_{E}^n\right|_\mathrm{Baldin}&=
  (11.6\pm1.5\,(\mathrm{stat})\pm0.6\,(\mathrm{Baldin}))\cdot10^{-4}\;\fm^3,
  \nonumber\\
  \left.\phantom{\PCsq}\beta_{M}^n \right|_\mathrm{Baldin}&= (
  3.6\mp1.5\,(\mathrm{stat})\pm0.6\,(\mathrm{Baldin}))\cdot10^{-4}\;\fm^3
  \label{eq:globalfitBaldinneutron}
\end{align}
for the fit of the isoscalar polarizabilities including the Baldin constraint.
We consider these values to be as reliable as those from the quasi-free
Compton experiment of Ref.~\cite{Kossert}, as the isoscalar polarizabilities,
from which they are derived, have been determined by fitting our deuteron
Compton calculation, which fulfills the low-energy theorem, to all existing
elastic deuteron Compton-scattering data. That means there is no restriction
on either the energy, as in Ref.~\cite{deuteronpaper}, or on the angle, like
in Ref.~\cite{McGPhil1,McGPhil2}. From these values we deduce that the magnetic response
of the neutron is comparable to that of the proton and that nucleons are
paramagnetic. We also conclude that the isovector polarizabilities are
considerably smaller than the isoscalar ones. In other words, our analysis
shows that within the precision of the currently existing data, elastic
Compton scattering from the proton and the deuteron is in agreement with
\begin{align}
  \alpha_{E}^p\approx\alpha_{E}^n,\qquad\qquad \beta_{ M}^p\approx\beta_{
    M}^n.
\end{align}
These findings agree with those of Refs.~\cite{Schumacher,deuteronpaper}.

\section{Conclusion} 
\setcounter{equation}{0}
\label{sec:conclusion}

In this work we examined elastic Compton scattering from the deuteron.
Diagrams without an intermediate two-nucleon state have been calculated up to 
next-to-leading order in the Small Scale Expansion, an Effective Field Theory 
with nucleons, pions and the $\Delta(1232)$ resonance as explicit degrees of 
freedom. Those diagrams including the propagation of the two-nucleon system 
between the two photon vertices have been calculated using Green's-function 
methods. Therefore, and because of the fact that we use deuteron wave 
functions that have been derived from state-of-the-art 
$NN$-potentials, we refer to this approach 
as the ``Green's-function hybrid approach''.
For the photon coupling we make use of Siegert's theorem~\cite{Siegert}, 
which is well-known to guarantee the exact static limit, see 
e.g.~\cite{Karakowski1,Karakowski2,Arenhoevel,ArenhoevelII}.

In Section~\ref{sec:comparison}, we show that we have achieved a consistent
description of elastic deuteron Compton scattering which is valid from 0~MeV
up to $\w\sim 100$~MeV. The advantage of our calculation with respect to
Refs.~\cite{deuteronpaper,Phillips,McGPhil1,McGPhil2} is that we obtain a good
description of the data published below 60~MeV. This improvement is of course
connected to the correct static limit, which had not been obtained in those
publications. Therefore, other calculations reaching this limit and thus being
manifestly gauge invariant are also able to describe the low-energy data well,
see e.g.~\cite{Lvov,Karakowski1,Karakowski2,Rupak1,Rupak2,Rupak3}. However, we achieve good agreement
also with high-energy data, measured around 94.2~MeV~\cite{Hornidge}.  In this
energy regime the work of Ref.~\cite{Karakowski1,Karakowski2} fails and the ``pion-less''
Effective Field Theory used in \cite{Rupak1,Rupak2,Rupak3} is inapplicable.  The authors of
\cite{Lvov} have also problems to describe the data in the backward direction
without introducing surprisingly large isovector polarizabilities. It was
already shown in~\cite{deuteronpaper} that the main difference between their
approach and ours is in the energy dependence of the resonant Compton
multipoles, which is well captured in our calculation due to the inclusion of
explicit $\Delta$-resonance degrees of freedom, whereas the authors of
Ref.~\cite{Lvov} include the polarizabilities only via the leading and
subleading terms of a Taylor expansion in the photon energy.

There are a number of ways to check the theoretical uncertainties of our
calculation.  As demonstrated in Section~\ref{sec:wavefunctiondep} and
Fig.~\ref{fig:wavefunctiondep}, the error induced by dependence on the
deuteron wave function is tiny and may well be set to zero, while it is
sizeable in the extractions of Refs.~\cite{deuteronpaper,McGPhil1,McGPhil2}. In
Section~\ref{sec:potentialdep} and Fig.~\ref{fig:LOchiPT}, we compared the
cross-sections derived using either the AV18-potential, or the much cruder
one-pion exchange potential, regulated to reproduce the deuteron binding
energy and ${}^1S_0$-scattering length. This may for the time being serve as
LO potential of $\chi$EFT. We find that the magnetic polarizability using this
potential is less than $1\cdot10^{-4}\;\fm^3$ larger than using AV18, and the
electric polarizability decreases by practically the same amount. As it is
safe to assume that AV18 is a better representative of a fully self-consistent
$\chi$EFT potential at the level of our calculation (next-to-leading order)
than one-pion exchange, we classify uncertainties induced by the potential as
negligible. A third way to estimate the errors from the theory side is to
compare to our previous calculation~\cite{deuteronpaper}. With the
single-nucleon sector treated identically to the approach presented here, its
only difference is that the interaction kernel is expanded perturbatively.  As
argued in Section~\ref{sec:comparison}, this approximation is justified in the
r\'egime $\omega\sim\mpi$. There, both results should thus coincide up to
higher-order corrections stemming from the different treatment of the
two-nucleon sector.  Figure~\ref{fig:2parameterfit3} and in particular the
extracted polarizabilities in Table~\ref{tab:fits} show that both approaches
agree well within the statistical error, and do not deviate by more than
$1.3\cdot10^{-4}\;\fm^3$. Finally, higher-order corrections to the
energy-dependence of the polarizabilities themselves were in SSE estimated by
na\"ive dimensional analysis to be less than
$1\cdot10^{-4}\;\mathrm{fm}^3$~\cite{deuteronpaper}.  All these interdependent
estimates suggest a theory uncertainty of our extraction of
$\pm1\cdot10^{-4}\;\mathrm{fm}^3$ as appropriate. This is comparable with the
statistical error.

Having achieved a good description of all elastic deuteron Compton-scattering
data enabled us therefore to perform a global fit of the isoscalar
polarizabilities to all existing data points, published in
\cite{Lucas,Lund,Hornidge}. Our 2-parameter-fit results are (see
Table~\ref{tab:fits})
\begin{align}
  \alpha_{E}^s&=(11.5\pm1.4\,(\mathrm{stat})\pm1\,(\mathrm{theory})
)\cdot10^{-4}\;\fm^3,\nonumber\\
  \beta_{M}^s &=( 3.4\pm1.6\,(\mathrm{stat})\pm1\,(\mathrm{theory})
)\cdot10^{-4}\;\fm^3.
  \label{eq:globalfitconclusion}
\end{align}
We note that the numbers of Eq.~(\ref{eq:globalfitconclusion}) are very close
to the fit results (\ref{eq:exppHGHP}) for the proton polarizabilities,
determined in~\cite{HGHP} within the SSE framework. This leaves little space
for large isovector polarizabilities. Furthermore they agree extraordinarily
well with the isoscalar Baldin-sum-rule value, Eq.~(\ref{eq:Baldin}).
Therefore, in order to reduce the statistical error, we repeated our fits
including this constraint:
\begin{align}
  \left.\phantom{\PCsq}\alpha_{E}^s\right|_\mathrm{Baldin}&=
  (11.3\pm0.7\,(\mathrm{stat})\pm0.6\,(\mathrm{Baldin})\pm1\,(\mathrm{theory})
)\cdot10^{-4}\;\fm^3,
  \nonumber\\
  \left.\phantom{\PCsq}\beta_{M}^s \right|_\mathrm{Baldin}&= (
  3.2\mp0.7\,(\mathrm{stat})\pm0.6\,(\mathrm{Baldin})\pm1\,(\mathrm{theory})
)\cdot10^{-4}\;\fm^3.
  \label{eq:globalfitBaldinconclusion}
\end{align}
Combining Eqs.~(\ref{eq:globalfitconclusion})
or~(\ref{eq:globalfitBaldinconclusion}), respectively, with the proton numbers
of Eq.~(\ref{eq:exppHGHP}), we obtained the neutron polarizabilities
\begin{align}
  \alpha_{E}^n&=
  (12.0\pm2.0\,(\mathrm{stat})\,\pm0.4\,(\text{Baldin})\pm1\,(\mathrm{theory})
)\cdot10^{-4}\;\fm^3,
  \nonumber\\
  \beta_{M}^n &= (
  4.0\pm2.1\,(\mathrm{stat})\,\pm0.4\,(\text{Baldin})\pm1\,(\mathrm{theory})
)\cdot10^{-4}\;\fm^3
  \label{eq:globalfitneutronconclusion}
\end{align}
for the 2-parameter fit and
\begin{align}
  \left.\phantom{\PCsq}\alpha_{E}^n\right|_\mathrm{Baldin}&=
  (11.6\pm1.5\,(\mathrm{stat})\pm0.6\,(\mathrm{Baldin})\pm1\,(\mathrm{theory})
)\cdot10^{-4}\;\fm^3,
  \nonumber\\
  \left.\phantom{\PCsq}\beta_{M}^n \right|_\mathrm{Baldin}&= (
  3.6\mp1.5\,(\mathrm{stat})\pm0.6\,(\mathrm{Baldin})\pm1\,(\mathrm{theory})
)\cdot10^{-4}\;\fm^3
  \label{eq:globalfitBaldinneutronconclusion}
\end{align}
for the 1-parameter fit including the Baldin constraint.  In both fits the
fact that our deuteron Compton-scattering calculation is applicable in the
whole energy range from 0~MeV to 100~MeV enables us to include all data into
our fit of the isoscalar polarizabilities. From these results we deduce that
the neutron is paramagnetic and that proton and neutron behave rather
similarly when exposed to external electromagnetic fields. In both points our
deuteron Compton calculation agrees with
Refs.~\cite{Schumacher,Kossert,deuteronpaper} and with the Chiral Perturbation
Theory prediction that contributions to the isovector polarizabilities only
start beyond leading-one-loop order. This also justifies again that the two
iso-scalar short-distance contributions to the electric and magnetic scalar
polarizabilities in Sec.~\ref{sec:nonresonant} are the smallest set which
needs to be promoted by one order as in Refs.~\cite{deuteronpaper,HGHP} to
describe the data, given present uncertainties.

Our results for $\alpha_E$ and $\beta_M$ are in excellent agreement with those
from a comprehensive, experimental analysis of all quasi-free
data~\cite{Schumacher}.  That analysis did, however, not include elastic
deuteron Compton scattering due to the alleged discrepancy between theory and
elastic $\gamma d$ experiments.  This discrepancy has now been resolved, both
in~\cite{deuteronpaper} and in the Green's-function hybrid approach presented
in this work.

Finally, we strongly advocate enlarging the data base for elastic Compton
scattering on the deuteron.  If further experiments, as planned or running at
TUNL/HI$\gamma$S~\cite{Weller:2009zza,Ahmed}, at the S-DALINAC~\cite{Richter}
and at MAXlab~\cite{Feldman:2008zz,Feldman2}, provide additional quality data
below the pion threshold, the increased statistics would reduce the
statistical error in our fit of the isoscalar polarizabilities.  Recall that
the statistical errors of the two-parameter fit \eqref{eq:globalfitconclusion}
are larger than our theoretical uncertainty, and that the combined statistical
and Baldin-sum-rule error of the one-parameter fit
\eqref{eq:globalfitBaldinconclusion} is comparable to it. These errors are
increased by the systematic experimental uncertainties which may be estimated
more comfortably if more experiments using different methods were available.
On the theory side, a wider effort is under way to describe elastic Compton
scattering on the proton, deuteron and ${}^3$He from the Thomson limit to well
into the $\Delta$-resonance region in one model-independent, unified
framework. We focus at present on improving the accuracy by a full
next-to-next-to-leading order calculation with nucleons, pions and the
$\Delta(1232)$ as dynamical, effective degrees of
freedom~\cite{McGovern:2009sw,allofus}. Clearly, reduced error bars in
coherent Compton scattering from the deuteron would enable us to quantify by
how much proton and neutron polarizabilities differ.


\section*{Acknowledgments}

The authors acknowledge helpful discussions with Wolfram Weise, Norbert
Kaiser, Daniel R. Phillips and Andreas Nogga.  We are also grateful to G.
Feldman, J. Friar, G. Miller and D.-O. Riska for useful comments made on
several points. We thank D.R. Phillips for providing us with his deuteron
Compton code and E. Epelbaum and V. Stoks for their deuteron wave functions. A
number of our colleagues have encouraged us to publish these results even
after a substantial temporal hiatus which is entirely our fault.  We are
grateful to the ECT* for its hospitality and financial support during the
workshops ``Two-photon physics'' (RPH and HWG) and ``Nuclear forces and QCD:
never the twain shall meet?'' (HWG and TRH).  HWG also acknowledges support
and discussions during the TUNL/HI$\gamma$S mini-workshop on Compton
scattering.  This work has been supported in part by the Bundesministerium
f\"ur Forschung und Technologie, by Deutsche Forschungsgemeinschaft under
contract GR1887/2-2, by a National Science Foundation \textsc{Career} award
PHY-0645498 (HWG) and US-Department of Energy grant DE-FG02-95ER-40907 (HWG).

\newpage

\appendix
\section{Multipole Expansion of the Photon Field}
\setcounter{equation}{0}
\label{app:multipoleexp}
For the photon field $\vec{A}\ofxi$, we use the multipole expansion derived in
\cite{Karakowski1,Karakowski2} in analogy to~\cite{Rose}, see also
Ref.~\cite{PHD}. The result is
\begin{align}
\label{eq:multipoleexp}
\hat{\epsilon}_\lambda\,\e^{i\vec{k}\cdot\vec{\xi}}&=
\sum_{L=1}^\infty\sum_{M=-L}^L
\wignerd{L}{M}{\lambda}\,i^L\,\sqrt{\frac{2\pi\,(2L+1)}{L\,(L+1)}}\\
&\times\left\{-\frac{i}{\w}\,\nab_\xi\,\psi_L(\w\xi)\,Y_{L\,M}(\hat{\xi})-
i\,\w\,\vec{\xi}\,j_L(\w\xi)\,Y_{L\,M}(\hat{\xi})-
\lambda\,\vec{L}\,Y_{L\,M}(\hat{\xi})\,j_L(\w\xi)\right\}.\nonumber
\end{align}
$\hat{\epsilon}_\lambda$ denotes the photon polarization vector in the
spherical (helicity) basis.  The functions $\wignerd{L}{M}{\lambda}$ are the
Wigner $D$-functions $D_{M,\lambda}^L(\alpha,\theta,\gamma)$ for
$\alpha=\gamma=0$, angular momentum $L$ and magnetic quantum number $M$;
$Y_{L\,M}$ denote the spherical harmonics and $\psi_L(\w
r)=(1+r\,\frac{d}{dr})\,j_L(\w r)$ with the spherical Bessel functions
$j_L(z)=\sqrt{\frac{\pi}{2z}}J_{L+\frac{1}{2}}(z)$ and $J_L(z)$ the Bessel
functions of the first kind.  In the static (long-wavelength) limit only the
gradient term in Eq.~(\ref{eq:multipoleexp}) survives, as for
$\w\rightarrow0$, $j_ 1(\w r)\rightarrow\frac{1}{3}\,\w r$ and $\psi_1(\w
r)\rightarrow\frac{2}{3}\,\w r$.  This term turns out to be the dominant part
of the photon field for all energies under consideration. Therefore, like in
Ref.~\cite{Karakowski1,Karakowski2}, we define two scalar functions
\begin{align}
\label{eq:phidefinition}
\phi_i(\vec{r})&=-\sum_{L=1}^\infty\sum_{M=-L}^L\delta_{M,\lambda_i}\,
\frac{i^{L+1}}{\w}\,\sqrt{\frac{2\pi\,(2L+1)}{L\,(L+1)}}\,\psi_L(\w r)\,
Y_{L\,M}(\hat{r}),\nonumber\\
\phi_f(\vec{r})&= \sum_{L'=1}^\infty\sum_{M'=-L'}^{L'}(-1)^{L'-\lambda_f}\,
\wignerd{L'}{M'}{-\lambda_f}\,\frac{i^{L'+1}}{\w}\,\sqrt{\frac{2\pi\,(2L'+1)}
{L'\,(L'+1)}}\,\psi_{L'}(\w r)\,Y_{L'\,M'}(\hat{r}),\nonumber\\
\end{align}
which allow us to write
\begin{align}
\label{eq:gradphi}
\eps\,\e^{i\vec{k}_i\cdot\vec{\xi}}   &\approx\nab\phi_i\ofxi,\nonumber\\
\epspr\,\e^{-i\vec{k}_f\cdot\vec{\xi}}&\approx\nab\phi_f\ofxi.
\end{align} 

We want to decompose the photon field in its electric and magnetic part. 
Therefore, we write Eq.~(\ref{eq:multipoleexp}) 
in Eq.~(\ref{eq:schematically}) schematically as
$\vec{A}=\nab\phi+\Aone+\Atwo$ with
\begin{align}
  \label{eq:Aone}
  \Aone\ofxi&=-\sum_{\vec{k},\lambda=\pm1}\sum_{L=1}^\infty\sum_{M=-L}^L\lambda
  \,\sqrt{\frac{2\pi\,(2L+1)}{L\,(L+1)}}\,i^L\,j_L(\w\xi)\,\vec{L}\,
  Y_{L\,M}(\hat{\xi})\nonumber\\
  &\times\left[a_{\vec{k},\lambda}\,\delta_{M,\lambda}-
    a_{\vec{k},\lambda}^\dagger\,(-1)^{L+\lambda}\,\wignerd{L}{M}{-\lambda}\right],
  \\
  \Atwo\ofxi&=-\sum_{\vec{k},\lambda=\pm1}\sum_{L=1}^\infty\sum_{M=-L}^L
  \sqrt{\frac{2\pi\,(2L+1)}{L\,(L+1)}}\,i^{L+1}\,\w\,\vec{\xi}\,j_L(\w\xi)\,
  Y_{L\,M}(\hat{\xi})\nonumber\\
  &\times\left[a_{\vec{k},\lambda}\,\delta_{M,\lambda}-
    a_{\vec{k},\lambda}^\dagger\,(-1)^{L+\lambda}\,\wignerd{L}{M}{-\lambda}\right].
  \label{eq:Atwo}
\end{align}
$\Aone$ constitutes the magnetic part of the photon field, $\nab\phi+\Atwo$ is
the electric field \cite{Rose}.  The operators $a_{\vec{k},\lambda}^\dagger$
($a_{\vec{k},\lambda}$) create (destroy) a photon with momentum $\vec{k}$ and
polarization $\lambda$.

\section{Calculation of Diagrams with Intermediate\\$NN$-Scattering}
\setcounter{equation}{0}
\label{app:resonant}
Here we give details about the calculation of the diagrams sketched in  
Fig.~\ref{fig:displabel}. Special emphasis is put on the construction of the 
two-nucleon Green's function, drawing substantially from 
Ref.~\cite{Karakowski1,Karakowski2}.

$\Mfi{\phi\phi 1}$ (Eq.~(\ref{eq:Mfiphiphi1added})) will be calculated first, 
in analogy to Ref.~\cite{Karakowski1,Karakowski2}. Defining the shortcut 
$E_0\equiv\w+\frac{\w^2}{2\md}-B$ 
and neglecting prefactors for the moment we can write this amplitude as 
\begin{equation}
\Mfi{\phi\phi 1}\propto\sum_C\frac{\mx{d_f}{\psi_{L'}\,Y_{L'\,M'}}{C}
\mx{C}{\psi_L\,Y_{L\,M}}{d_i}}{E_0-\EC},
\end{equation}
where we suppressed the sums over $L,M$ and $L',M'$ for brevity, cf.
Eq.~(\ref{eq:phidefinition}).  Each wave function can be separated into a
radial part, denoted by the index '$\mathrm{rad}$', and an angular part,
denoted by a hat.  Furthermore, $\ket{C}$ is an eigenstate to $H^{np}$ with
eigenvalue $\EC$, so we can write
\begin{equation}
  \Mfi{\phi\phi 1}\propto\sum_{C_\mathrm{rad}\,\hat{C}}
  \mx{d_{f\mathrm{rad}}\,\hat{d}_f}{\psi_{L'}\,Y_{L'\,M'}\,\frac{1}
    {E_0-H_{\hat{C}}^{np}}}{C_\mathrm{rad}\,\hat{C}}
  \mx{C_\mathrm{rad}\,\hat{C}}{\psi_L\,Y_{L\,M}}{d_{i\,\mathrm{rad}}\,\hat{d}_i}.
\end{equation}
$\hat{C}$ is used as a shorthand notation for all angular quantum numbers of
the intermediate state, i.e. \mbox{$\ket{\hat{C}}=\ket{L_C\,S_C\,J_C\,M_C}$}.
Separating the radial from the angular part of $\ket{C}$ and inserting two
complete sets of radial states $\ket{r}$ and $\ket{r'}$ we get
\begin{align}
  \label{eq:separation}
  \Mfi{\phi\phi 1}&\propto\sum_{C_\mathrm{rad}\,\hat{C}}
  \int_0^\infty\int_0^\infty r^2 dr\,r'^2
  dr'\mx{\hat{d}_f}{Y_{L'\,M'}}{\hat{C}}
  \mx{d_{f\mathrm{rad}}}{\psi_{L'}}{r'}\\
  &\times\mx{r'}{\frac{1}{E_0-H_{\hat{C}}^{np}}}{C_\mathrm{rad}}
  \mxemp{C_\mathrm{rad}}{r}\mx{r}{\psi_L}{d_{i\,\mathrm{rad}}}
  \mx{\hat{C}}{Y_{L\,M}}{\hat{d}_i}.\nonumber
\end{align}
For the deuteron ($J=1$) wave function, we use the notation
$\Psi_{1m}(\vec{r}\,)=\sum_{l=0,2}\frac{u_l(r)}{r}\mathcal{Y}_m^{l11}(\hat{r})$
with $u_0(r)$ ($u_2(r)$) denoting the usual radial wave functions $u(r)$
($w(r)$) in position space, cf. e.g. Ref.~\cite{Ericson}.  The indices of the
angular wave functions $\mathcal{Y}$ are $l11$ for orbital angular momentum,
spin and total angular momentum of the deuteron state, $m\in\{-1,0,1\}$. Now
we can write Eq.~(\ref{eq:separation}) as
\begin{align}
  \label{eq:Mfiwithint}
  \Mfi{\phi\phi 1}&\propto\sum_{\hat{C}}\sum_{l=0,2}\sum_{l'=0,2}\,
  \mx{l'\,1\,1\,M_f}{Y_{L'\,M'}}{\hat{C}}\mx{\hat{C}}{Y_{L\,M}}{l\,1\,1\,M_i}\\
  &\times\doubleint r dr\,r' dr' u_{l'}(r')\,\psi_{L'}(\frac{\w r'}{2})\,
  \mx{r'}{\frac{1}{E_0-H_{\hat{C}}^{np}}}{r}\, \psi_L(\frac{\w
    r}{2})\,u_l(r),\nonumber
\end{align}
where we have removed the sum over $C_\mathrm{rad}$. Integrals without limits
are always integrated from $0$ to infinity throughout this work.

We now have to evaluate the double integral in Eq.~(\ref{eq:Mfiwithint}), 
including the Green's function 
\begin{equation}
  \green=\mx{r'}{\frac{1}{E_0-H_{\hat{C}}^{np}}}{r}.
  \label{eq:greendefinition}
\end{equation}
However, we need to evaluate the integral for arbitrary functions of $r$.
Therefore we describe how to calculate
\begin{equation}
  \mathcal{I}_{fi}^{ll'\hat{C}}=
  \doubleint r dr\,r' dr'\,u_{l'}(r')\,J_f(r')\,\green\,J_i(r)\,u_l(r).
  \label{eq:doubleint}
\end{equation}
We do so in two steps and define
\begin{equation}
  \chi_f^{l'\hat{C}}(r)\equiv\int r' dr'\,u_{l'}(r')\,J_f(r')\,\green.
\end{equation}
Once we have solved this first part, it is easy to numerically calculate the
remaining integral
\begin{equation}
  \mathcal{I}_{fi}^{ll'\hat{C}}=
  \int r dr\,u_l(r)\,J_i(r)\,\chi_f^{l'\hat{C}}(r).
\end{equation}
In order to find the function $\chi_f^{l'\hat{C}}(r)$~-- in the following we
use the abbreviation $\chi_{\hat{C}}(r)$ for brevity~-- we first note that
\begin{equation}
  \left(E_0-H_{\hat{C}}^{np}\right)\,\green=\mxemp{r'}{r}=
  \frac{\delta(r'-r)}{r^2}.
  \label{eq:Greensequation}
\end{equation}
Eq.~(\ref{eq:Greensequation}) defines the Green's function corresponding to
Schr\"odinger's equation with a central potential and the Hamiltonian
\begin{equation}
  H_{\hat{C}}^{np}=\frac{\vec{p}^{\,2}}{m_N}+V_{\hat{C}}(r)=-\frac{1}{m_N}\,
  \left(\frac{1}{r}\frac{d^2}{dr^2}r\right)+\frac{L_C\,(L_C+1)}{m_N\,r^2}+
  V_{\hat{C}}(r).
  \label{eq:HhatCnp}
\end{equation}
The dependence of the potential on the quantum numbers of the interim state
$\ket{C}$ is shown explicitly.  Note that the neutron-proton potential
contains a tensor part and therefore not only depends on the distance $r$ but
on the vector $\vec{r}$.  The tensor force mixes e.g. the deuteron $s$- and
$d$-states in Schr\"odinger's equation. Nevertheless, on the level of the
Green's function this matrix equation decouples, cf.~Ref.~\cite{Ericson}.  The
decoupling of Eq.~(\ref{eq:Greensequation}) guarantees that only the diagonal
terms of the tensor force contribute. Therefore, the orbital angular momentum
is well defined, which allows us to replace $\vec{L}^2\rightarrow
L_C\,(L_C+1)$ in Eq.~(\ref{eq:HhatCnp}).

Eqs.~(\ref{eq:Greensequation}) and~(\ref{eq:HhatCnp}) combine to
\begin{equation}
  \left[E_0+\frac{1}{m_N}\frac{d^2}{dr^2}-
    \frac{L_C\,(L_C+1)}{m_N\,r^2}-V_{\hat{C}}(r)\right]\,r\,\green=
  \frac{\delta(r'-r)}{r},
  \label{eq:radialoperator}
\end{equation}
which e.g. in the deuteron case reduces to the two differential equations
\begin{equation}
  \left[E_0+\frac{1}{m_N}\frac{d^2}{dr^2}-V_\mathrm{cent}(r)\right]
  \,r\,G_{0}(r,r';E_0)=
  \frac{\delta(r'-r)}{r},
  \label{eq:Greens}
\end{equation}
\begin{equation}
  \left[E_0+\frac{1}{m_N}\frac{d^2}{dr^2}-
    \frac{6}{m_N\,r^2}-V_\mathrm{cent}(r)+2\,V_\mathrm{ten}(r)\right]
  \,r\,G_{2}(r,r';E_0)=
  \frac{\delta(r'-r)}{r}.
  \label{eq:Greend}
\end{equation}
$V_\mathrm{cent}(r)$ denotes the central part of the potential,
$V_\mathrm{ten}(r)$ the tensor potential. The indices of the Green's functions
in Eqs.~(\ref{eq:Greens},~\ref{eq:Greend}) reflect the orbital angular
momentum state, whereas $J=1,\;S=1$ is not written down explicitly.  Acting
with the operator given in square brackets in Eq.~(\ref{eq:radialoperator}) on
$\chi_{\hat{C}}(r)$, the integral over $r'$ collapses and we find
\begin{equation}
  \left[\frac{d^2}{dr^2}+m_N\,\left(E_0-V_{\hat{C}}(r)\right)-
    \frac{L_C\,(L_C+1)}{r^2}\right]\,r\,\chi_{\hat{C}}(r)=m_N\,u_{l'}(r)\,J_f(r).
  \label{eq:diffeq}
\end{equation}
This is a second-order differential equation in $r$ with an inhomogeneity,
which can be interpreted as a source term. Its solutions are real for $E_0<0$
and complex for $E_0>0$. The latter case corresponds to $\w>B$, i.e.  the
photon carries enough energy to break up the deuteron into its two
constituents. Obviously, an imaginary part only appears in the $s$-channel
diagrams, where the incoming photon is absorbed before the other one is
emitted. In Appendix~\ref{app:photodisintegration} we will use the imaginary
part of the amplitudes to derive the total deuteron-photodisintegration cross
section via the optical theorem.

For $r\rightarrow\infty,\;$
$u_{l'}(r)J_f(r)\rightarrow 0$ due to $u_{l'}(r)\rightarrow 0$, i.e. 
Eq.~(\ref{eq:diffeq}) reduces to a homogeneous differential equation. 
Furthermore, $V_{\hat{C}}(r)\rightarrow 0$ for 
$r\rightarrow\infty$. Therefore, we are for large distances left with
\begin{equation}
  \left[\frac{d^2}{dr^2}+E_0\,m_N-\frac{L_C\,(L_C+1)}{r^2}\right]\,r\,
  \chi_{\hat{C}}(r)=0.
\end{equation} 
This equation is known to be solved by a linear combination of the spherical
Bessel functions of the first and second kind, $j_{L_C}(Qr)$ and $n_{L_C}(Qr)$
with $Q=\sqrt{m_N\,E_0}$. Note that $Q$ can be real or imaginary\footnote{The
  sign of the imaginary solution is determined by adding an infinitesimal
  imaginary part to $-B$, i.e. $B\rightarrow B-i\epsilon$.}, depending on
$E_0$.  In our case the boundary condition is that $\chi_{\hat{C}}(r)$ must be
an outgoing spherical wave for large $r$, cf.  e.g.~\cite{Sakurai}. Therefore
we may write
\begin{equation}
  \lim_{r\rightarrow\infty}\chi_{\hat{C}}(r)\propto h_{L_C}^{(1)}(Qr),
  \label{eq:condition}
\end{equation} 
with $h_{L_C}^{(1)}(Qr)$ the spherical Hankel function of the first kind,
defined as
\begin{equation}
  h_{L_C}^{(1)}(Qr)=j_{L_C}(Qr)+i n_{L_C}(Qr).
\end{equation}

Once we have numerical solutions for the homogeneous and the inhomogeneous
differential equation, we need to find the correct linear combination which
satisfies the condition~(\ref{eq:condition}). In other words we have to
determine the coefficient $\lambda$ which fulfills
\begin{equation}
  \lim_{r\rightarrow\infty}\left\{\chi_{\hat{C}}^\mathrm{in}(r)+
    \lambda\,\chi_{\hat{C}}^\mathrm{hom}(r)\right\}\propto h_{L_C}^{(1)}(Qr),
\end{equation}
where $\chi_{\hat{C}}^\mathrm{in}(r)$ ($\chi_{\hat{C}}^\mathrm{hom}(r)$)
denote the solution to the inhomogeneous (homogeneous) differential equation.
In the asymptotic limit, $\chi_{\hat{C}}(r)$ must be a linear combination of
$j_{L_C}(Qr)$ and $n_{L_C}(Qr)$ or, equivalently, of $j_{L_C}(Qr)$ and
$h_{L_C}^{(1)}(Qr)$. Therefore we can write the general solutions in the
following way:
\begin{align}
  \label{eq:chihom}
  \chi_{\hat{C}}^\mathrm{hom}(r)&=C_{\hat{C}}^\mathrm{hom}(r)\,
  \left[j_{L_C}(Qr)+t_{\hat{C}}^\mathrm{hom}(r)\,h_{L_C}^{(1)}(Qr)\right],\\
  \chi_{\hat{C}}^\mathrm{in }(r)&=C_{\hat{C}}^\mathrm{in }(r)\,
  \left[j_{L_C}(Qr)+t_{\hat{C}}^\mathrm{in }(r)\,h_{L_C}^{(1)}(Qr)\right]
  \label{eq:chiin}
\end{align}
with functions $C_{\hat{C}}^\mathrm{in/hom}(r)$,
$t_{\hat{C}}^\mathrm{in/hom}(r)$ which become the constants
$C_{\hat{C}}^\mathrm{in/hom}$, $t_{\hat{C}}^\mathrm{in/hom}$ for large $r$.
With the choice $\lambda=-C_{\hat{C}}^\mathrm{in}/C_{\hat{C}}^\mathrm{hom}$ we
find
\begin{equation}
  \lim_{r\rightarrow\infty}\left\{\chi_{\hat{C}}^\mathrm{in}(r)+
    \lambda\,\chi_{\hat{C}}^\mathrm{hom}(r)\right\}=C_{\hat{C}}^\mathrm{in}\,
  \left(t_{\hat{C}}^\mathrm{in}-t_{\hat{C}}^\mathrm{hom}\right)\,
  h_{L_C}^{(1)}(Qr),
\end{equation}
which satisfies the condition~(\ref{eq:condition}). Therefore we need to
determine the coefficients $C_{\hat{C}}^\mathrm{in}$,
$C_{\hat{C}}^\mathrm{hom}$. This has to be done in the region where
$C_{\hat{C}}(r)$, $t_{\hat{C}}(r)$ are constant, i.e. their derivatives
vanish. In this region
\begin{equation}
  \ln\chi_{\hat{C}}^\mathrm{in/hom}(r)=\ln C_{\hat{C}}^\mathrm{in/hom}+
  \ln\left[j_{L_C}(Qr)+t_{\hat{C}}^\mathrm{in/hom}\,h_{L_C}^{(1)}(Qr)\right].
\end{equation}
Defining $D^\mathrm{in/hom}=\frac{d}{dr}\ln\chi_{\hat{C}}^\mathrm{in/hom}=
\frac{{\chi_{\hat{C}}'}^\mathrm{in/hom}}{\chi_{\hat{C}}^\mathrm{in/hom}}$ we
find
\begin{equation}
  D^\mathrm{in/hom}=\frac{d}{dr}\ln\left[j_{L_C}(Qr)+
    t_{\hat{C}}^\mathrm{in/hom}\,
    h_{L_C}^{(1)}(Qr)\right]=\frac{\frac{d}{dr}j_{L_C}(Qr)+
    t_{\hat{C}}^\mathrm{in/hom}
    \,\frac{d}{dr}h_{L_C}^{(1)}(Qr)}
  {j_{L_C}(Qr)+t_{\hat{C}}^\mathrm{in/hom}\,h_{L_C}^{(1)}(Qr)}.
\end{equation}
From this equation we can easily determine $t_{\hat{C}}^\mathrm{in/hom}$,
which we use to solve Eqs.~(\ref{eq:chihom},~\ref{eq:chiin}) for
$C_{\hat{C}}^\mathrm{in/hom}$ and thus to determine $\lambda$.

Numerically, this is one of the most involved parts of this work. 
Fortunately, a nice 
and valuable cross-check to the routine can be performed. For this we consider
again the double integral to be calculated, Eq.~(\ref{eq:doubleint}).
This integral is obviously invariant under the interchange 
$r\leftrightarrow r'$. 
A general feature of Green's functions is that they are symmetric 
under $r\leftrightarrow r'$, i.e. $G_{\hat{C}}(r',r;E_0)=\green$. 
Therefore,
\begin{equation}
  \mathcal{I}_{fi}^{ll'\hat{C}}=
  \doubleint r' dr'\,r dr\,u_{l'}(r)\,J_f(r)\,\green\,J_i(r')\,u_l(r').
\end{equation}
This expression is identical to $\mathcal{I}_{fi}^{ll'\hat{C}}$ with
$i\leftrightarrow f$, $l\leftrightarrow l'$, i.e. our results must be
symmetric under $i\leftrightarrow f$, $l\leftrightarrow l'$. This is a
non-trivial check, because for $J_f(r)\neq J_i(r)$ completely different
functions $\chi_f^{l'\hat{C}}(r)$ are generated. Our routine agrees well with
this symmetry~-- the deviation caused by numerical uncertainties is less than
1\%.

Now all tools to calculate $\Mfi{\phi\phi 1,2}$ are prepared. However, as the
algebraic manipulations are not too complicated, we refer the interested
reader to Ref.~\cite{PHD} for further evaluation and the analytic results.
There we also compute $\Mfi{\phi\phi 3}$ and $\Mfi{\phi\phi 4}$.

We turn now to the evaluation of the subleading terms of
Section~\ref{sec:subleading}, where we make use of the continuity
equation~(\ref{eq:continuity}) at only one or even none of the two vertices.
Therefore, we have to specify the current $\vec{J}\ofxi$ and the relevant
parts of the photon field $\vec{A}\ofxi$. These are the non-gradient terms in
Eq.~(\ref{eq:schematically}), i.e. $\Aone$~(Eq.~(\ref{eq:Aone})) and
$\Atwo$~(Eq.~(\ref{eq:Atwo})).  The one-body current is considered first. It
consists of two parts, $\Jsigma$ and $\Jp$, cf.
Eqs.~(\ref{eq:Jsigma},~\ref{eq:Jp}).  All possible combinations of
$\Aone,\;\Atwo$ and $\Jsigma,\;\Jp$ have been calculated in \cite{Karakowski1,Karakowski2}.
We also evaluated all these amplitudes, however we found that only $\Jsigma$
gives visible contributions to the deuteron Compton cross sections.
Therefore, we may restrict ourselves to the following combinations:
($\Jsigma,\,\Aone$), denoted by $\sigma1$, and ($\Jsigma,\,\Atwo$), denoted by
$\sigma2$.

We now calculate $\int\Jsigma\ofxi\cdot\vec{A}^{(1,2)}\ofxi\,d^3\xi$.  We
start with the derivation for $\Aone$, writing only the $\xi$-dependent terms
for simplicity.
\begin{align}
  \int\Jsigma\ofxi\cdot\Aone\ofxi\,d^3\xi&\propto\int\sum_{j=n,p}
  \left[\nab_\xi\times\mu_j\,\vec{\sigma}_j\,\delta(\vec{\xi}-\vec{x}_j)\right]\,
  j_L(\w \xi)\,\vec{L}\,Y_{L\,M}(\hat{\xi})\,d^3\xi\nonumber\\
  &=\int\sum_{j=n,p}\mu_j\,\vec{\sigma}_j\,\delta(\vec{\xi}-\vec{x}_j)\,\nab_\xi
  \times \left(j_L(\w\xi)\,\vec{L}\,Y_{L\,M}(\hat{\xi})\right)\,d^3\xi,
\end{align}
where one partial integration has been performed. Now we evaluate the integral
and afterwards replace $\vec{x}_p\rightarrow \frac{\vec{r}}{2},\;
\vec{x}_n\rightarrow-\frac{\vec{r}}{2}$, cf. Eq.~(\ref{eq:cmvariables}),
yielding
\begin{equation}
  \int\Jsigma\ofxi\cdot\Aone\ofxi\,d^3\xi\propto2\left[\nab_r\times 
    \left(j_L(\frac{\w r}{2})\,\vec{L}\,Y_{L\,M}(\hat{r})\right)\right]
  \left(\mu_p\,\vec{\sigma}_p-(-1)^L\,\mu_n\,\vec{\sigma}_n\right),
  \label{eq:replacexpxnsigma1}
\end{equation}
where we used $Y_{L\,M}(-\hat{r})=(-1)^L\,Y_{L\,M}(\hat{r})$.  By the help of
the identity
$\vsh{L}{L}{M}=\frac{1}{\sqrt{L\,(L+1)}}\cdot\vec{L}\,Y_{L\,M}(\hat{r})$, see
e.g.~\cite{Edmonds}, this becomes
\begin{equation}
  \int\Jsigma\ofxi\cdot\Aone\ofxi\,d^3\xi\propto2\sqrt{L\,(L+1)}
  \left[\nab_r\times j_L(\frac{\w r}{2})\,\vsh{L}{L}{M}\right]
  \left(\mu_p\,\vec{\sigma}_p-(-1)^L\,\mu_n\,\vec{\sigma}_n\right).
\end{equation}
Now we can use the curl formula~\cite{Edmonds} for simplifying
$\left[\nab_r\times j_L(\frac{\w r}{2})\,\vsh{L}{L}{M}\right]$ and the
recursion relations for spherical Bessel functions to write
\begin{align}
  \int\Jsigma\ofxi\cdot\Aone\ofxi\,d^3\xi&\propto
  \left[i\,\w\,j_{L-1}(\frac{\w r}{2})\,\sqrt{\frac{L\,(L+1)^2}{2L+1}}\,
    \vsh{L}{L-1}{M}\right.\\
  &-\left.i\,\w\,j_{L+1}(\frac{\w r}{2})\,\sqrt{\frac{L^2\,(L+1)}{2L+1}}\,
    \vsh{L}{L+1}{M}\right]
  \left(\mu_p\,\vec{\sigma}_p-(-1)^L\,\mu_n\,\vec{\sigma}_n\right).\nonumber
\end{align}
We found, like the authors of Ref.~\cite{Karakowski1,Karakowski2}, that the numerical
importance of the various contributions rapidly decreases with increasing
multipolarity $L$. Therefore, the term proportional to $\vsh{L}{L+1}{M}$ may
be neglected. Defining $\vec{S}=\frac{\vec{\sigma}_p+\vec{\sigma}_n}{2}$ and
$\vec{t}=\frac{\vec{\sigma}_p-\vec{\sigma}_n}{2}$ and including all
prefactors, we get the result
\begin{align}
  \label{eq:intJsigmaAonescalar}
  \int\Jsigma&\ofxi\cdot\Aone\ofxi\,d^3\xi=\nonumber\\
  &-\sum_{\vec{k},\lambda=\pm1}\sum_{L=1}^\infty\sum_{M=-L}^L\lambda\,
  \sqrt{2\pi\,(L+1)}\,\frac{e\,\w}{2m_N}\,i^{L+1}\,j_{L-1}(\frac{\w r}{2})\,
  \vsh{L}{L-1}{M}\\
  &\times\left[\left(\mu_p-(-1)^L\,\mu_n\right)\,\vec{S}+
    \left(\mu_p+(-1)^L\,\mu_n\right)\,\vec{t}\,\right] \cdot
  \left[a_{\vec{k},\lambda}\,\delta_{M,\lambda}-
    a_{\vec{k},\lambda}^\dagger\,(-1)^{L+\lambda}\,\wignerd{L}{M}{-\lambda}\right].
  \nonumber
\end{align}
The scalar products are replaced using the relation
\begin{equation}
  \vsh{J}{L}{M}\cdot\vec{V}=\left[Y_L\otimes V\right]_{J\,M},
  \label{eq:TdotV}
\end{equation}
which holds for any vector (rank 1) operator ($\otimes$ denotes the
irreducible tensor product). An explicit proof of the
relation~(\ref{eq:TdotV}) is given e.g. in Ref.~\cite{PHD}. We use it to
finally rewrite Eq.~(\ref{eq:intJsigmaAonescalar}) as
\begin{align}
  \int\Jsigma\ofxi\cdot\Aone\ofxi\,d^3\xi&=-\sum_{\vec{k},\lambda=\pm1}
  \sum_{L=1}^\infty\sum_{M=-L}^L\lambda\,
  \sqrt{2\pi\,(L+1)}\,\frac{e\,\w}{2m_N}\,i^{L+1}\,j_{L-1}(\frac{\w r}{2})
  \nonumber\\
  &\times \left\{\left(\mu_p-(-1)^L\,\mu_n\right)\,\left[Y_{L-1}\otimes
      S\right]_{L\,M}+ \left(\mu_p+(-1)^L\,\mu_n\right)\,\left[Y_{L-1}\otimes
      t\right]_{L\,M}\,
  \right\}\nonumber\\
  &\times\left[a_{\vec{k},\lambda}\,\delta_{M,\lambda}-
    a_{\vec{k},\lambda}^\dagger\,(-1)^{L+\lambda}\,\wignerd{L}{M}{-\lambda}\right].
  \label{eq:intJsigmaAone}
\end{align}

We turn now to the calculation of $\int\Jsigma\ofxi\cdot\Atwo\ofxi\,d^3\xi$.
Again we restrict ourselves in the derivation to the $\xi$-dependent terms,
finding 
\begin{align}
  \int\Jsigma\ofxi\cdot\Atwo\ofxi\,d^3\xi&\propto\int\sum_{j=n,p}
  \left[\nab_\xi\times\mu_j\,\vec{\sigma}_j\,\delta(\vec{\xi}-\vec{x}_j)\right]
  \,\vec{\xi}\,j_L(\w\xi)\,Y_{L\,M}(\hat{\xi})\,d^3\xi\nonumber\\
  &=\nab_r\times\left(\vec{r}\,j_L(\frac{\w r}{2})\,Y_{L\,M}(\hat{r})\right)\,
  \left(\mu_p\,\vec{\sigma}_p+(-1)^L\,\mu_n\,\vec{\sigma}_n\right),
  \label{eq:intJsigmaAtwo}
\end{align}
where we have performed the same steps as in the derivation of
Eq.~(\ref{eq:replacexpxnsigma1}). Using the relation
\begin{equation}
  \vec{r}\,j_L(\frac{\w r}{2})\,Y_{L\,M}(\hat{r})=
  \sqrt{L\,(L+1)}\,r\,j_L(\frac{\w r}{2})\,
  \left[\frac{\vsh{L}{L-1}{M}}{\sqrt{(L+1)\,(2L+1)}}-
    \frac{\vsh{L}{L+1}{M}}{\sqrt{L\,(2L+1)}}\right],
  \label{eq:rjYequal}
\end{equation}
which is derived e.g. in~\cite{PHD}, and the curl formula we find
\begin{equation}
  \nab_r\times\left(r\,j_L(\frac{\w r}{2})\,
    \left[\frac{\vsh{L}{L-1}{M}}{\sqrt{(L+1)\,(2L+1)}}-
      \frac{\vsh{L}{L+1}{M}}{\sqrt{L\,(2L+1)}}\right]\right)=
  -i\,j_L(\frac{\w r}{2})\,\vsh{L}{L}{M}.
  \label{eq:usingcurl}
\end{equation}
Combining Eqs.~(\ref{eq:intJsigmaAtwo}-\ref{eq:usingcurl}), together with the
definitions of $\vec{S}$ and $\vec{t}$, cf.
Eq.~(\ref{eq:intJsigmaAonescalar}), yields
\begin{align}
  \int\Jsigma\ofxi\cdot\Atwo\ofxi\,d^3\xi&\propto
  -i\,\sqrt{L\,(L+1)}\,j_L(\frac{\w r}{2})\,\vsh{L}{L}{M}\nonumber\\
  &\times\left[(\mu_p+(-1)^L\,\mu_n)\,\vec{S}
    +(\mu_p-(-1)^L\,\mu_n)\,\vec{t}\right].
\end{align}
Including all prefactors, we get
\begin{align}
  &\int\Jsigma\ofxi\cdot\Atwo\ofxi\,d^3\xi=
  -\sum_{\vec{k},\lambda=\pm1}\sum_{L=1}^\infty\sum_{M=-L}^L\sqrt{2\pi\,(2L+1)}\,
  \frac{e\,\w}{2m_N}\,i^L\,j_L(\frac{\w r}{2})
  \left\{ (\mu_p+(-1)^L\,\mu_n)\right.\nonumber\\
  &\times\left. [Y_L\otimes S]_{L\,M} +(\mu_p-(-1)^L\,\mu_n)\,[Y_L\otimes
    t]_{L\,M}\right\}\cdot
  \left[a_{\vec{k},\lambda}\,\delta_{M,\lambda}-a_{\vec{k},\lambda}^\dagger\,
    (-1)^{L+\lambda}\,\wignerd{L}{M}{-\lambda}\right],
  \label{eq:intJsigmaAtwofinal}
\end{align}
making use of Eq.~(\ref{eq:TdotV}) once more.

We are now ready to calculate the amplitudes $\Mfi{\phi\,\sigma 1}$ and
$\Mfi{\phi\,\sigma 2}$. The results are given in Ref.~\cite{PHD}, together
with the amplitudes $\Mfi{\sigma 1\,\sigma 1}$ and $\Mfi{\sigma 2\,\sigma 2}$,
which do not contain the gradient part of the photon field. Nevertheless,
these contributions are strong, cf.  Fig.~\ref{fig:separation}, due to the
numerically large factor $(\mu_p-\mu_n)^2\approx22$ which appears in those
amplitudes.  As the terms with both photons coupling to the current $\Jp$,
Eq.~(\ref{eq:Jp}), are not supported by this factor, these contributions are
tiny and are not considered in our work. The mixed amplitudes $\Mfi{\sigma
  1\,\sigma 2}$, $\Mfi{\sigma 2\,\sigma 1}$ also turn out negligibly small,
cf. Ref.~\cite{PHD}.

So far we (explicitly) only considered one-body currents. However, there are
also non-negligible contributions from pion-exchange currents,
Fig.~\ref{fig:mesonexchange}. The corresponding expressions in
coordinate-space representation are (with $\vec{x}_1$,~$\vec{x}_2$ the
position of nucleon~1 and~2, respectively)
\begin{equation}
  \vec{J}\,^\mathrm{KR}_\mathrm{stat}(\vec{\xi};\vec{x}_1,\vec{x}_2)=
  \frac{e\,f^2}{m_\pi^2}\,
  \left(\vec{\tau}_1\times\vec{\tau}_2\right)_z\,
  \left[
    \vec{\sigma}_1\,\delta(\vec{x}_1-\vec{\xi})\,(\vec{\sigma}_2\cdot\hat{r})+
    \vec{\sigma}_2\,\delta(\vec{x}_2-\vec{\xi})\,(\vec{\sigma}_1\cdot\hat{r})
  \right]\,\frac{\partial}{\partial r}\frac{\e^{-m_\pi r}}{r}
  \label{eq:KR}
\end{equation}
for the Kroll-Ruderman (pair) current (Fig.~\ref{fig:mesonexchange}(a)), and
\begin{equation}
  \vec{J}\,^\mathrm{pole}_\mathrm{stat}(\vec{\xi};\vec{x}_1,\vec{x}_2)=-
  \frac{e\,f^2}{4\pi}\,
  \left(\vec{\tau}_1\times\vec{\tau}_2\right)_z\cdot
  \left(\nab_1-\nab_2\right)\,
  (\vec{\sigma}_1\cdot\nab_1)\,
  (\vec{\sigma}_2\cdot\nab_2)\,
  \frac{\e^{-m_\pi|\vec{x}_1-\vec{\xi}|}}{m_\pi\,|\vec{x}_1-\vec{\xi}|}\,
  \frac{\e^{-m_\pi|\vec{x}_2-\vec{\xi}|}}{m_\pi\,|\vec{x}_2-\vec{\xi}|}
  \label{eq:pole}
\end{equation}
for the so-called pion-pole current (Fig.~\ref{fig:mesonexchange}(b)), cf.
e.g. \cite{Ericson}\footnote{Our convention for the pion-nucleon coupling
  $f^2$ differs by a factor $4\pi$ from that used in \cite{Ericson}.}.  The
relative vector $\vec{r}$ is defined as $\vec{r}=\vec{x}_1-\vec{x}_2$.

Our numerical evaluations show that the explicit inclusion of the pole current
is well negligible in the process and energies under consideration. Therefore
we are only concerned with the Kroll-Ruderman current, Eq.~(\ref{eq:KR}). This
expression, however, is derived in the limit of static nucleons (denoted by
the index 'stat'), i.e. the correction due to the photon energy is neglected.
This being a rather crude approximation for $\w\sim100$~MeV, which is close to
the pion mass, we use
\begin{equation}
  \vec{J}\,^\mathrm{KR}(\vec{\xi};\vec{x}_1,\vec{x}_2)=
  \frac{e\,f^2}{m_\pi^2}\,
  \left(\vec{\tau}_1\times\vec{\tau}_2\right)_z\,
  \left[
    \vec{\sigma}_1\,\delta(\vec{x}_1-\vec{\xi})\,(\vec{\sigma}_2\cdot\hat{r})+
    \vec{\sigma}_2\,\delta(\vec{x}_2-\vec{\xi})\,(\vec{\sigma}_1\cdot\hat{r})
  \right]\,\frac{\partial}{\partial r}f^\mathrm{KR}(r)
  \label{eq:KRreal}
\end{equation}
instead of Eq.~(\ref{eq:KR}). The function $f^\mathrm{KR}(r)$ depends on the
photon energy and is defined as
\begin{equation}
  f^\mathrm{KR}_s(r)=\frac{\e^{-m_\pi r}}{2\,r}-2\pi\int
  \frac{d^3q}{(2\pi)^3}\,\frac{\e^{i\vec{q}\cdot(\vec{x}_2-\vec{x}_1)}}
  {\sqrt{m_\pi^2+\vec{q}^{\,2}}\,( \w-\sqrt{m_\pi^2+\vec{q}^{\,2}})}
  \label{eq:fKRs}
\end{equation}
for an $s$-channel diagram and
\begin{equation}
  f^\mathrm{KR}_u(r)=\frac{\e^{-m_\pi r}}{2\,r}-2\pi\int
  \frac{d^3q}{(2\pi)^3}\,\frac{\e^{i\vec{q}\cdot(\vec{x}_2-\vec{x}_1)}}
  {\sqrt{m_\pi^2+\vec{q}^{\,2}}\,(-\w-\sqrt{m_\pi^2+\vec{q}^{\,2}})}
  \label{eq:fKRu}
\end{equation}
for the $u$-channel, cf. Ref.~\cite{PHD}. Note that for $\w=0$,
Eqs.~(\ref{eq:fKRs}) and~(\ref{eq:fKRu}) reduce to the above expression
$f^\mathrm{KR}(r)=\frac{\e^{-m_\pi r}}{r}$.

Now we have another current at hand, which we can use to replace
$\vec{J}\ofxi$ in Eq.~(\ref{eq:disp}). However, one has to be careful in order
not to double-count certain contributions; e.g. it is not allowed to combine
$\Hint=-\int\vec{J}^\mathrm{\,KR}\ofxi\cdot\vec{A}^\mathrm{full}\ofxi\,d^3\xi$
at one vertex with $\Hint=-\int\vec{J}\ofxi\cdot\nab\phi\ofxi\,d^3\xi$ at the
other, as this contribution is already included~-- at least partly~-- in the
dominant terms, due to the use of Siegert's theorem, cf. discussion around
Eq.~(\ref{eq:implicit}).  The Kroll-Ruderman current changes isospin, i.e.
$\Hint=-\int\vec{J}^\mathrm{\,KR}\ofxi\cdot\vec{A}\ofxi\,d^3\xi$ transforms
the isospin-0 deuteron to an isospin-1 object.  Therefore we need another
isospin-changing interaction at the second vertex.  Pauli's principle
guarantees that the total wave function of the two-nucleon system has to be
antisymmetric under the exchange of the two constituents.  Stated differently,
the wave function has to fulfill $(-1)^{S+L+T}=-1$, i.e. in order to have
$T=1$ we need $S+L$ even.  The operator that turned out to be the most
important one numerically is $[Y_0\otimes t]_1$, cf.
Eq.~(\ref{eq:intJsigmaAone}).  The same observation has been made in
\cite{Karakowski1,Karakowski2}. Nevertheless, also the operator $Y_1$, which stems from
$\hat{\phi}_i$, $\hat{\phi}_f$, gives non-negligible contributions.  However,
in the amplitudes including $\hat{\phi}_i$ or $\hat{\phi}_f$ at the non-KR
vertex, one is not allowed to use the full photon field in
$\Hint=-\int\vec{J}^\mathrm{\,KR}\ofxi\cdot\vec{A}\ofxi\,d^3\xi$, as explained
after Eq.~(\ref{eq:KRreal}). Therefore we are left with
$\Mfi{\mathrm{KR}\,\mathrm{full}\,\sigma1}$, $\Mfi{\phi\,\mathrm{KR}1}$ and
$\Mfi{\phi\,\mathrm{KR}2}$, where 'KR$\,$full' denotes the integral over the
Kroll-Ruderman current, multiplied by the full photon field. There is no
danger of double-counting $\Mfi{\mathrm{KR}\,\mathrm{full}\,\sigma1}$, as we
only take into account the operator $[Y_{L-1}\otimes t]_L$.  This operator,
however, changes the deuteron spin, whereas the matrix elements arising from
$\phi_{i,f}$ are spin-conserving.  Further contributions, like the one where
$\vec{J}\ofxi=\vec{J}^\mathrm{\,KR}\ofxi$ at both vertices, turned out to be
small. The evaluation of the Kroll-Ruderman terms is given in Ref.~\cite{PHD},
along with further details and the analytic results for all amplitudes.

\section{Total Deuteron-Photodisintegration Cross Section}
\setcounter{equation}{0}
\label{app:photodisintegration}

Besides complying with the low-energy theorem, cf.  Section~\ref{sec:Thomson},
another important check on our calculation is the extraction of the total
deuteron-photodisintegration cross section from the Compton amplitude via the
optical theorem. This process has been studied more extensively~--
experimentally as well as theoretically, see e.g.~\cite{Partovi}~-- than
elastic deuteron Compton scattering and there is plenty of data below 100~MeV
to compare with.

The optical theorem in our normalization reads
\begin{equation}
  \sigma^\mathrm{tot}=
  \frac{1}{\w}\cdot\frac{1}{6}\,\sum_{i=f}\,\mathrm{Im}[\Mfi{}(\theta=0)],
\end{equation}
i.e. the total cross section is the sum over the imaginary part of all
deuteron Compton amplitudes in the forward direction with identical initial
and final photon and deuteron states ($\lf=\li$, $M_f=M_i$), divided by the
photon energy $\w$. Like the elastic deuteron Compton cross section, this sum
is divided by 6, as we have to average over the initial states.

We calculate this cross section in the lab frame, in order to be able to
compare to data. The rest of our work is performed in the $\gamma d$-cm frame,
which has also been chosen in Ref.~\cite{PHD}, and we refer to this reference
for the final results of our amplitudes.  Fortunately, these amplitudes are
easily transformed into the lab frame.  First we note that we only need to sum
over the $s$-channel diagrams, as only they become complex for photon energies
above the deuteron binding energy $B$, while the $u$-channel amplitudes stay
real for all photon energies, cf.  Section~\ref{sec:dominant}. As the authors
of Ref.~\cite{Karakowski1,Karakowski2} calculate in the lab frame, we convert our
calculation according to their work. In the $s$-channel the only change is
$\w+\frac{\w^2}{2m_d}\leftrightarrow\w-\frac{\w^2}{2m_d}$, because in the lab
frame, the deuteron's initial kinetic energy vanishes, whereas the total
intermediate momentum is $\vec{P}_C=\vec{k}_i$. In the cm frame we have
$\vec{P}_i=-\vec{k}_i$ and $\vec{P}_C=\vec{0}$ in the $s$-channel.

\begin{figure}[!htb]
  \begin{center}
    \includegraphics*[width=.6\linewidth]{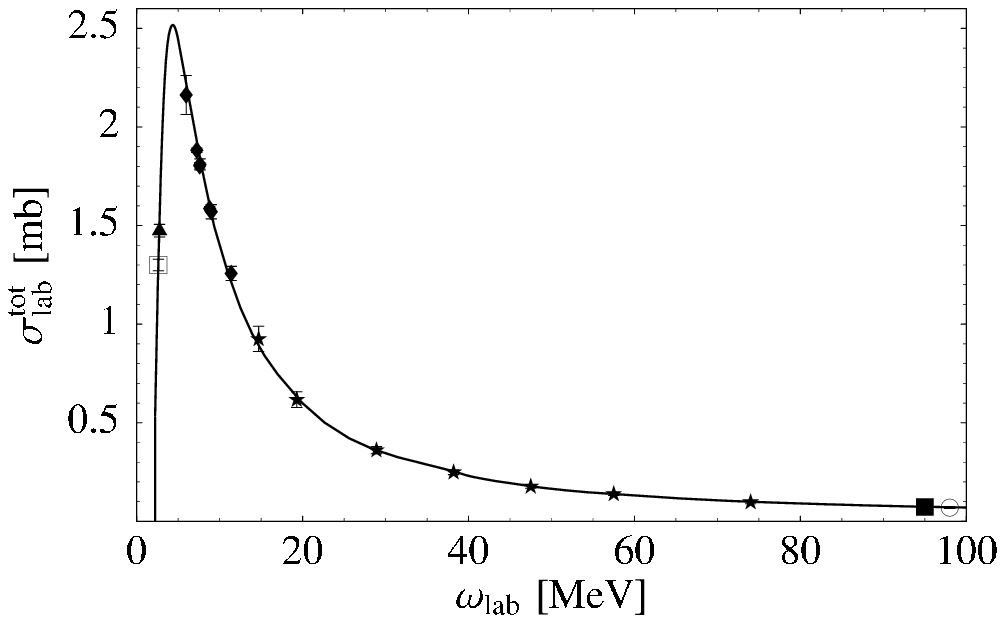}\\
    \includegraphics*[width=.48\linewidth]{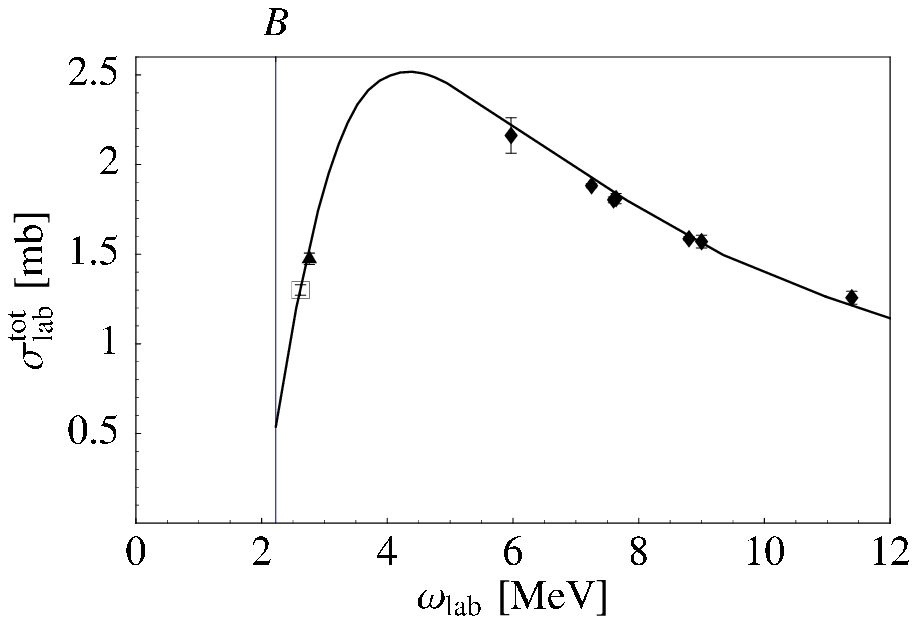}
    \hfill
    \includegraphics*[width=.48\linewidth]{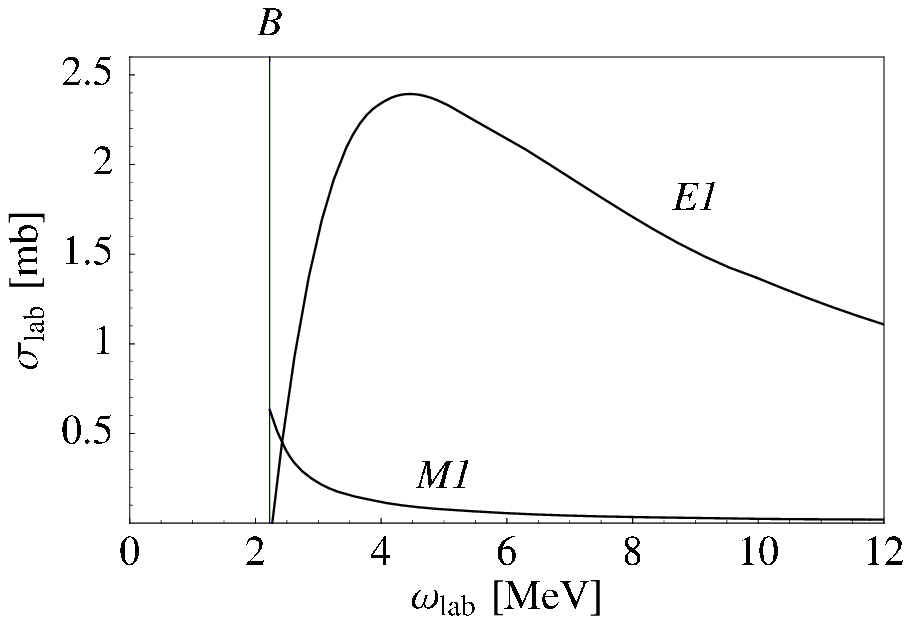}
    \caption{Total deuteron-photodisintegration cross section derived from our
      deuteron Compton amplitudes, together with data from \cite{MaB} (open
      box), \cite{Birenbaum} (diamond), \cite{Bernabei} (star), \cite{Meyer}
      (box), \cite{Sanctis} (circle).  The triangle corresponds to the
      weighted average of the data measured at 2.76~MeV \cite{Moreh,SnB,MMC},
      as determined in~\cite{Arenhoevelreview}.  '$E1$', '$M1$' denotes the
      contributions from the $E1$- and the singlet $M1$-transition,
      respectively. $B$ is the binding energy of the deuteron.}
    \label{fig:photodisintegration}
  \end{center}
\end{figure}

Our result for the total deuteron-photodisintegration cross section is shown
in Fig.~\ref{fig:photodisintegration}, together with data from
\cite{MaB,Birenbaum,Bernabei,Meyer,Sanctis,Moreh,SnB,MMC} which are described well by our calculation. In the
lower left panel the low-energy regime is enlarged, in order to emphasize the
non-vanishing value at threshold.
The by far most important contribution at threshold stems from the singlet
$M1$-transition of $\Mfi{\sigma1\,\sigma1}$, which is the amplitude with the
magnetic part of the photon field $\Aone$, Eq.~(\ref{eq:Aone}), coupling to
the spin current at both vertices. It corresponds to the operator
$[Y_{0}\otimes t]_1$, cf. Eq.~(\ref{eq:intJsigmaAone}), which transforms the
deuteron into a (singlet) $S_C=0$-state. $M1$ is the shorthand notation for
the magnetic coupling of a photon with $L=1$. Nevertheless, as is well-known,
already 1~MeV above threshold the cross section is completely dominated by the
amplitude $\Mfi{\phi\phi\,1}$ (Eq.~(\ref{eq:Mfiphiphi1added})), where for
$L,L'=1$ we have an $E1$-interaction at each vertex, and this dominance holds
for all higher energies.  In the lower right panel of
Fig.~\ref{fig:photodisintegration}, we show these two (most important)
contributions to the total photodisintegration cross section, denoted as
'$E1$' and '$M1$'. We observe the well-known rise of '$M1$', as $\w$
approaches the breakup threshold, cf. e.g.  \cite{Arenhoevelreview,Brown},
whereas '$E1$' is zero for $\w=B$. Note that $E1$ not only consists of
$\Mfi{\phi\phi\,1}$ but of all amplitudes with an $E1$-interaction at the
vertex of the incoming photon.

Strictly speaking there are also contributions from the one-body current
$\Jp\ofxi$, cf. Eq.~(\ref{eq:Jp}). The corresponding amplitudes are given in
\cite{Karakowski1,Karakowski2} but are not included in this work, as we found that their
contributions to the elastic deuteron Compton cross sections are tiny (of the
order of 1\%) and so is their effect on the total disintegration cross
section. Nevertheless, in the high-energy regime of our calculation, say for
$\w\sim 100$~MeV, they do give visible contributions to $E1$, but these cancel
nearly exactly against other terms which also contain $\Jp$. Therefore, when
we only look at the sum of all amplitudes contributing to
$\sigma^\mathrm{tot}$, we may well neglect the current $\Jp$.

We also compare our results with predictions for the strengths of 
electric and magnetic transitions close to threshold from the 
\textit{Effective Range Expansion}~\cite{Bethe1,Bethe2} given by 
\begin{equation}
  \sigma_\mathrm{ER}^\mathrm{el}=\frac{2}{3}\,\frac{e^2}{\gamma^2}\,
  \frac{(\frac{\w}{B}-1)^{3/2}}{(\frac{\w}{B})^3\,(1-\gamma\,r_t)},
  \label{eq:EREel}
\end{equation}
\begin{equation}
  \sigma_\mathrm{ER}^\mathrm{mag}=\frac{1}{6}\,\frac{e^2}{m_N^2}\,
  (\mu_p-\mu_n)^2\,\frac{k\,\gamma}{k^2+\gamma^2}\,\frac{(1-\gamma\,a_s+
    \frac{1}{4}\,a_s\,(r_s+r_t)\,\gamma^2-\frac{1}{4}\,a_s\,(r_s-r_t)\,k^2)^2}
  {(1+k^2\,a_s^2)\,(1-\gamma\,r_t)}
  \label{eq:EREmag}
\end{equation}
with $\gamma=\sqrt{m_N\,B}$.  The final-state relative momentum is
$k=|\vec{p}_p-\vec{p}_n|/2=\sqrt{m_N\,(\w-B)}$, and for the singlet scattering
length $a_s$ and the singlet (triplet) effective range $r_s$ ($r_t$) we use
$a_s=-23.749$~fm, $r_s=2.81$~fm and $r_t=1.76$~fm given in \cite{AV18}.  The
explicit form of Eqs.~(\ref{eq:EREel},~\ref{eq:EREmag}) is adopted from
\cite{Arenhoevelreview}.

\begin{figure}[!htb]
  \begin{center}
    \includegraphics*[width=.32\linewidth]{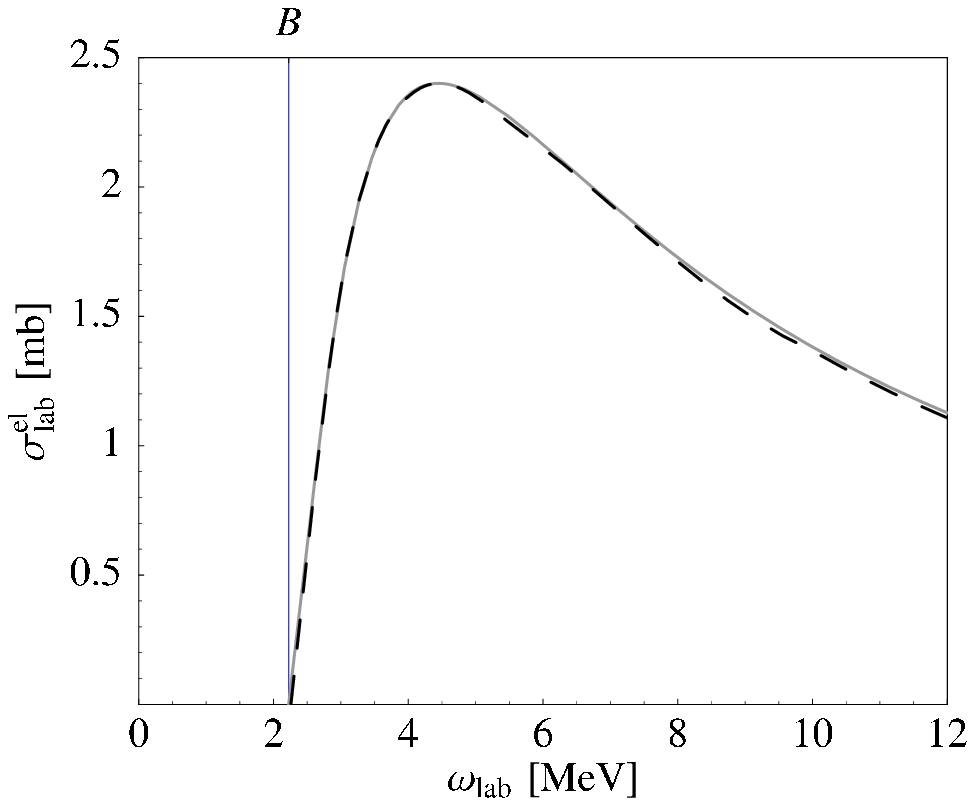}
    \includegraphics*[width=.32\linewidth]{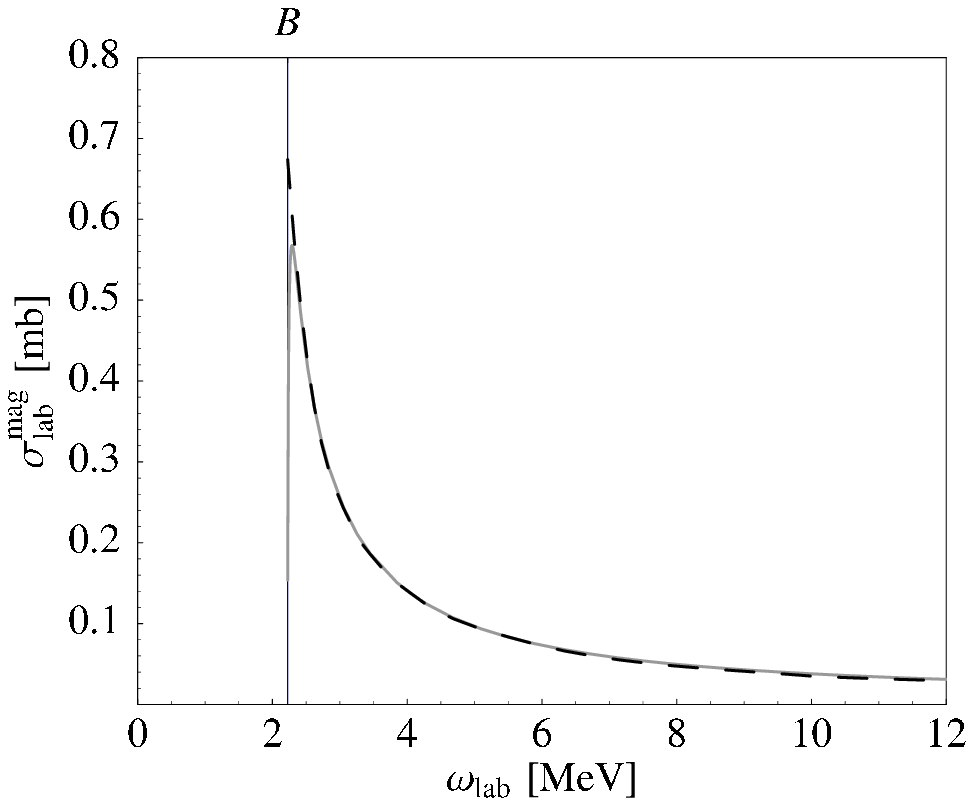}
    \includegraphics*[width=.32\linewidth]{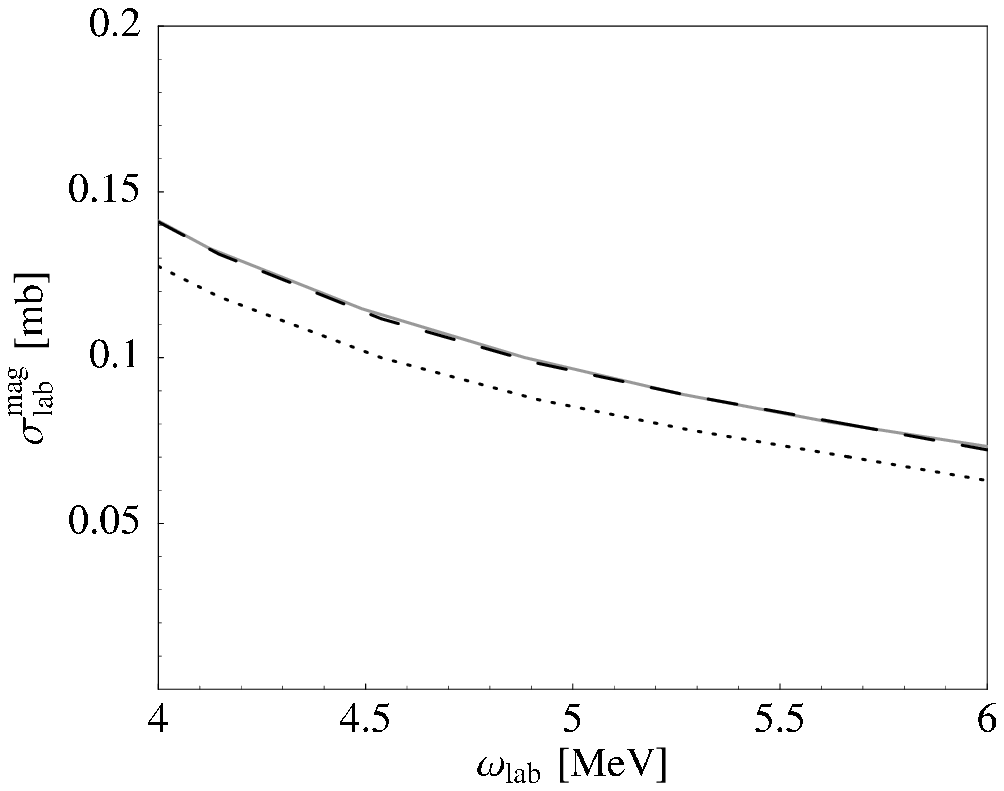}
    \caption{Comparison of our results (dashed) for the contributions of
      electric (left panel) and magnetic (middle and right panel) transitions
      to the total deuteron-photodisintegration cross section with predictions
      from the \textit{Effective Range Expansion} (grey). The dotted curve in
      the right panel does not include the Kroll-Ruderman diagrams.}
    \label{fig:photodisintegration2}
  \end{center}
\end{figure}

In Fig.~\ref{fig:photodisintegration2}, we compare our results with
Eqs.~(\ref{eq:EREel},~\ref{eq:EREmag}), finding excellent agreement between
both approaches.  We also demonstrate~-- in the right panel~-- the
non-negligible size of the KR~diagrams. The only visible contributions to the
total disintegration cross section at low energies including the
Kroll-Ruderman current are the amplitudes $\Mfi{\mathrm{KR}\,\sigma1}$, which
contribute to $\sigma^\mathrm{mag}$.

\newpage



\begin{thebibliography}{99}
\bibitem{GH01} H.W. Grie{\ss}hammer and T.R. Hemmert, Phys. Rev. C 65, 045207
        (2002).
\bibitem{HGHP} R.P. Hildebrandt, H.W. Grie{\ss}hammer, T.R. Hemmert and 
        B. Pasquini, Eur. Phys. J. A 20, 293 (2004).
\bibitem{Schumacher} M. Schumacher, Prog. Part. Nucl. Phys.~55, 567 (2005).
\bibitem{Olmos} Olmos de Leon et al., Eur. Phys. J. A 10, 207 (2001).
\bibitem{HHKLett1} T.R. Hemmert, B.R. Holstein and J. Kambor, Phys. Lett. 
        B 395, 89 (1997).
\bibitem{HHKLett2} T.R. Hemmert, B.R. Holstein and J. Kambor, 
        J.  Phys.  G 24, 1831 (1998).
\bibitem{Lucas} M. Lucas, Ph.D. thesis, University of Illinois (1994).
\bibitem{Lund} M. Lundin et al., Phys. Rev. Lett. 90, 192501 (2003).
\bibitem{Hornidge} D.L. Hornidge et al., Phys. Rev. Lett. 84, 2334 (2000).
\bibitem{Weller:2009zza}
  H.~R.~Weller, M.~W.~Ahmed, H.~Gao, W.~Tornow, Y.~K.~Wu, M.~Gai and R.~Miskimen,
  Prog.\ Part.\ Nucl.\ Phys.\  {\bf 62}, 257 (2009).
\bibitem{Weller} H.~Weller et al., HIGS-E-18-09; H.~Weller, private
  communication.
\bibitem{Miskimen} R.~Miskimen et al., HIGS-E-06-09; R.~Miskimen, private
  communication.
\bibitem{Miskimentalk} R.~Miskimen, \emph{Measuring the Spin-Polarizabilities
    of the Proton at \HIGS}, presentation at the INT workshop on Soft Photons
  and Light Nuclei, 17 June 2008, and private communication.
\bibitem{Gao} H.~Gao et al., HIGS-E-07-10; H.~Gao, private communication.
\bibitem{Ahmed} M.~Ahmed et al., HIGS-E-06-10; M.~Ahmed, private communication.
\bibitem{Feldman:2008zz}
  G.~Feldman {et al.},
  Few Body Syst.\ {\bf 44}, 325 (2008).
\bibitem{Feldman2}
  MAXlab experiment NP-006
  (http://www.maxlab.lu.se/kfoto/Experimental\linebreak{}Program/PAC\_info.html);
  G.~Feldman, K.~Fissum and L.~Myers, private communication.
\bibitem{Richter} A. Richter and P. von Neumann-Cosel, private communication 
        (2005).
\bibitem{AhrensBeckINT08} R.~Beck, \emph{Nucleon Compton Scattering at MAMI},
  talk at the INT workshop on Soft Photons and Light Nuclei, 17 June 2008;
  V.~Ahrens and.~R.~M.~Annand, private communication.
\bibitem{Lvov} M.I. Levchuk and A.I. L'vov, Nucl. Phys. A 674, 449 (2000).
\bibitem{Kossert} K. Kossert et al., Eur. Phys. J. A 16, 259 (2003).
\bibitem{deuteronpaper} R.P. Hildebrandt, H.W. Grie{\ss}hammer, T.R. Hemmert 
        and D.R. Phillips, Nucl. Phys. A 748, 573 (2005).
\bibitem{BKKM} V. Bernard, N. Kaiser, J. Kambor and U.-G. Mei{\ss}ner, Nucl.
        Phys. B 388, 315 (1992).
\bibitem{danielreview} D.R. Phillips, J. Phys. G 36, 104004 (2009).
\bibitem{Karakowski1} J.J. Karakowski and G.A. Miller, Phys. Rev. C 60, 014001 
        (1999).
\bibitem{Karakowski2} J.J. Karakowski, Ph.D. thesis, preprint nt@uw-99-6 (1999)
        {\tt [nucl-th/9901011]}.
\bibitem{Phillips} S.R. Beane, M. Malheiro, D.R. Phillips and U. van Kolck, 
        Nucl. Phys. A 656, 367 (1999).
\bibitem{McGPhil1} S.R. Beane, M. Malheiro, J.A. McGovern, D.R. Phillips and 
        U. van Kolck, Phys. Lett. B 567, 200 (2003); erratum ibid. B 607, 320 
        (2005).
\bibitem{McGPhil2} S.R. Beane, M. Malheiro, J.A. McGovern, D.R. Phillips and 
        U. van Kolck, Nucl. Phys. A 747, 311 (2005).
\bibitem{Weinberg1} S. Weinberg, Phys. Lett. B 251, 288 (1990).
\bibitem{Weinberg2} S. Weinberg, Nucl. Phys. B 363, 3 (1991).
\bibitem{Nijm} V.G. Stoks, R.A. Klomp, C.P. Terheggen and J.J. de Swart, Phys.
        Rev. C 49, 2950 (1994).
\bibitem{Bonn} R. Machleidt, F. Sammarruca and Y. Song, Phys. Rev. C 53, 1483 
        (1996).
\bibitem{AV18} R.B. Wiringa, V.G. Stoks and R. Schiavilla, Phys. Rev. C 51, 38
        (1995).
\bibitem{Beane} S.R. Beane, V. Bernard, E. Epelbaum, U.-G. Mei{\ss}ner and 
        D.R. Phillips, Nucl. Phys. A 720, 399 (2003).
\bibitem{Danieled} D.R. Phillips, Phys. Lett. B 567, 12 (2003).
\bibitem{Rholecture} M. Rho, in Proceedings of the 10th Taiwan Nuclear 
        Physics Spring School, Hualien, Taiwan, China (2002).
\bibitem{Nogga}  A. Nogga, R.G.E. Timmermans and U. van Kolck, 
        Phys.~Rev.~C~72, 054006 (2005).
\bibitem{Birse}  
  M.~C.~Birse,
  Phys.\ Rev.\  C 74, 014003 (2006)
  [arXiv:nucl-th/0507077].
\bibitem{PHD}  R.P. Hildebrandt, Ph.D. thesis, TU M\"unchen, 
        mediaTUM\_disshab\_000000000003041 (2005) {\tt [nucl-th/0512064]}.
\bibitem{hg1} H.W. Grie{\ss}hammer, in preparation.
\bibitem{chidyn2006} H.~W.~Grie\3hammer, [nucl-th/0611074].
\bibitem{menu07} H.~W.~Grie\3hammer, [arXiv:0710.2924 [nucl-th]].

\bibitem{Rupak1} S.R. Beane and M.J. Savage, Nucl. Phys. A 694, 511 (2001).
\bibitem{Rupak2} H.W. Grie{\ss}hammer and G. Rupak, Phys. Lett. B 529, 57 (2002).
\bibitem{Rupak3}  J. Chen, X. Ji and Y. Li, Phys. Rev. C~71, 044321 (2005).

\bibitem{Friar} J.L. Friar, Ann. of Phys. 95, 170 (1975).
\bibitem{Siegert} A.J.F. Siegert, Phys. Rev. 52, 787 (1937).
\bibitem{Arenhoevel} H. Arenh\"ovel, Z. Phys. A 297, 129 (1980).
\bibitem{ArenhoevelII}  M. Weyrauch and H. Arenh\"ovel, 
        Nucl. Phys. A 408, 425 (1983).
\bibitem{BKM} V. Bernard, N. Kaiser and U.-G. Mei{\ss}ner, Int. J. Mod. Phys. E
        4, 193 (1995).
\bibitem{HHKK} T.R. Hemmert, B.R. Holstein, J. Kambor and G. Kn{\"o}chlein:
        Phys. Rev. D 57, 5746 (1998).
\bibitem{Epelbaum} E. Epelbaum, W. Gl{\"o}ckle and U.-G. Meissner, 
        Eur. Phys. J. A 19, 125 (2004); ibid. 19, 401 (2004).
\bibitem{Rose} M.E. Rose, Elementary Theory of Angular Momentum, John Wiley \& 
        Sons, inc.,  New York$\:\cdot\:$London$\:\cdot\:$Sydney (1957). 
\bibitem{Ericson} T. Ericson and W. Weise, Pions and Nuclei, Clarendon Press, 
        Oxford, ISBN~0-19-852008-5 (1988).
\bibitem{danielpaper} D.R. Phillips, J. Phys. G 34, 365 (2007). 
\bibitem{Rho} T.-S. Park, K. Kubodera, D.-P. Min and M. Rho, Nucl. Phys. 
        A~646, 83 (1999).
\bibitem{NLO} E. Epelbaum, W. Gl{\"o}ckle and U.-G. Mei{\ss}ner, 
        Nucl. Phys. A 671, 295 (2000).
\bibitem{McGovern:2009sw}
  J.~A.~McGovern, H.~W.~Grie\3hammer, D.~R.~Phillips and D.~Shukla,
  [arXiv:0910.1184 [nucl-th]].
\bibitem{allofus} H.~Grie{\ss}hammer, J.~McGovern, D.~R.~Phillips, D.~Shukla,
  work in progress.
\bibitem{Sakurai} J.J. Sakurai, Modern Quantum Mechanics, Addison-Wesley
        Publishing Company, ISBN~0-201-53929-2 (1994).
\bibitem{Edmonds} A.R. Edmonds, Angular Momentum in Quantum Mechanics, 
        University Press, Princeton, ISBN~0-691-07912-9 (1996).
\bibitem{VMK} D.A. Varshalovich, A.N. Moskalev and V.K. Khersonskii, 
        Quantum Theory of Angular Momentum, World Scientific, Singapore, 
        ISBN~9971-50-107-4 (1988). 
\bibitem{Partovi} F. Partovi, Ann. Phys. 27, 79 (1964).
\bibitem{MaB} P. Marin, G.R. Bishop and H. Halban, Proc. Phys. Soc. 67A, 1113 
        (1954).
\bibitem{Birenbaum} Y. Birenbaum et al., Phys. Rev. C~32, 1825 (1985).
\bibitem{Bernabei} R. Bernabei et al., Phys. Rev. Lett. 57, 1542 (1986).
\bibitem{Meyer} H.O. Meyer et al., Phys. Rev. C~31, 309 (1985).
\bibitem{Sanctis} E. de Sanctis et al., Phys. Rev. C~34, 413 (1986).
\bibitem{Moreh} R. Moreh, T.J. Kennett and W.V. Prestwick, Phys. Rev. C 39, 
        1247 (1989); erratum ibid. C~40, 1548 (1989).
\bibitem{SnB} A.H. Snell, E.C. Barker and R.L. Sternberg, Phys. Rev. 80, 637 
        (1950).
\bibitem{MMC} W.R. McMurray and C.H. Collie, Proc. Phys. Soc. 68A, 181 (1955).
\bibitem{Arenhoevelreview} H. Arenh\"ovel and M. Sanzone, Photodisintegration
        of the Deuteron, Few Body Systems Suppl. 3, Springer-Verlag 
        Wien$\:\cdot\:$New York, ISBN~3-211-82276-3 (1991).
\bibitem{Brown} G.E. Brown and A.D. Jackson, The Nucleon-Nucleon Interaction,
        North-Holland Publishing Company, Amsterdam$\:\cdot\:$Oxford,
        ISBN~$0\:7204\:03359$ (1976).
\bibitem{Bethe1} H.A. Bethe, Phys. Rev. 76, 38 (1949); J. Schwinger, 
        hectographed notes on nuclear physics, Harvard University (1947).
\bibitem{Bethe2} H.A. Bethe and C. Longmire, Phys. Rev. 77, 647 (1950).
\end{thebibliography}
\end{document}